\documentclass[useAMS,usenatbib,onecolumn]{mn2e}
\usepackage{subfigure}
\usepackage{graphicx}
\usepackage{times}

\def\bI{{\bf I}}
\def\bw{{\bf w}}
\def\bl{{\bf l}}
\def\bO{{\bf \Omega}}
\def\dfl{{\tilde f}_\bl}
\def\ldw{\bl\cdot\bw}
\def\ldo{\bl\cdot\bO}
\def\ores{\bl\cdot\bO - m\Omega_p}
\def\sign{\mathop{\rm sign}\nolimits}

\def\erf{\mathop{\rm erf}\nolimits}

\def\spose#1{\hbox to 0pt{#1\hss}}
\def\lta{\mathrel{\spose{\lower 3pt\hbox{$\mathchar"218$}}
     \raise 2.0pt\hbox{$\mathchar"13C$}}}
\def\gta{\mathrel{\spose{\lower 3pt\hbox{$\mathchar"218$}}
     \raise 2.0pt\hbox{$\mathchar"13E$}}}
\let\simless=\lta

\def\aj{AJ}
\def\apj{ApJ}
\def\apjl{ApJL}
\def\mnras{MNRAS}

\title[Bar--halo interaction]{The Bar--Halo Interaction--I. From
Fundamental Dynamics to Revised N-body Requirements}

\author[Weinberg \& Katz]{Martin D. Weinberg\thanks{E-mail:
    weinberg@astro.umass.edu (MDW); nsk@kaka.astro.umass.edu (NK)} 
  and Neal Katz\footnotemark[1] \\
  Department of Astronomy, University of Massachusetts, Amherst,
}

\begin{document}

\label{firstpage}

\date{\today}
\pagerange{\pageref{firstpage}--\pageref{lastpage}} \pubyear{2005}

\maketitle
\begin{abstract}
  A galaxy remains near equilibrium for most of its history.  Only
  through resonances can non-axisymmetric features such as spiral arms
  and bars exert torques over large scales and change the overall
  structure of the galaxy.  In this paper, we describe the resonant
  interaction mechanism in detail and derive explicit criteria for the
  particle number required to simulate these dynamical processes
  accurately using N-body simulations, and illustrate them with
  numerical experiments.  To do this, we perform a direct numerical
  solution of perturbation theory, in short, by solving for each
  orbit in an ensemble and make detailed comparisons with N-body
  simulations.  The criteria include: sufficient particle coverage in
  phase space near the resonance and enough particles to minimise
  gravitational potential fluctuations that will change the dynamics
  of the resonant encounter.  These criteria are general in concept
  and can be applied to any dynamical interaction.  We use the
  bar--halo interaction as our primary example owing to its technical
  simplicity and astronomical ubiquity.
  
  Some of our more surprising findings are as follows.  First, the
  Inner-Lindblad-like resonance (ILR), responsible for coupling the
  bar to the central halo cusp, requires more than ${\cal O}(10^8)$
  equal mass particles within the virial radius for a Milky-Way-like
  bar in an NFW profile \citep{Navarro.Frenk.ea:97}. Second, orbits
  that linger near the resonance receive more angular momentum than
  orbits that move through the resonance quickly.  Small-scale
  fluctuations present in state-of-the-art particle-particle
  simulations can knock orbits out of resonance, preventing them from
  lingering and, thereby, decrease the torque.  This particularly
  affects the ILR.  However, noise from orbiting substructure remains
  at least an order of magnitude too small to be of consequence. The
  required particle numbers are sufficiently high for scenarios of
  interest that apparent convergence in particle number is misleading:
  the convergence is in the noise-dominated regime.  State-of-the-art
  simulations are not adequate to follow all aspects of secular
  evolution driven by the bar-halo interaction.  It is not possible to
  derive particle number requirements that apply to all situations,
  e.g. more subtle interactions may be even more difficult to
  simulate.  Therefore, we present a procedure to test the
  requirements for individual N-body codes to the actual problem of
  interest.

\end{abstract}


\section{Introduction}
\label{sec:intro}

During most of its lifetime, a galaxy undergoes long periods of
secular evolution.  The subtle dynamical effects driving this
evolution are best studied using analytic techniques that can
accurately follow the accumulation of weak perturbations for long
periods of time.  The mere existence of a present-day disk galaxy
selects against strong or frequent mergers since its formation owing
to the fragility of galactic disks.  The evolution of such galaxies
must be dominated by long-term secular changes, which are harder to
model using N-body simulations.  However, since galaxy evolution is
punctuated by epochs of violent nonlinear evolution, e.g. initial
formation, and major and minor mergers, modern researchers must rely
upon N-body simulations, which can easily follow strong perturbations
for short periods of time and adapt to evolving equilibria. Similarly,
modelling realistic astronomical scenarios that include collisionless
dark matter and star particles, gas, star formation and mass loss from
winds require simulations.  In the coming decade, simulations of
galaxies that accurately evolve the dynamics over many gigayears will
be possible, including a more physical treatment of the ISM, star
formation, and mass loss.  As researchers' reliance on N-body
simulation continues to grow and simulations continue to be used as
the gold standard for theoretical verification, it is important to
verify that N-body simulations truly capture the dynamical mechanisms
that drive quiescent galaxy evolution.

The secular evolution of quiescent galaxies is driven by structural
asymmetries, often triggered by environmental perturbations such as
satellites and group interactions or by local instabilities such as in
swing amplification \citep{Toomre:81,Jog:92,Fuchs:01}.  The response
of a galaxy to an asymmetry results in torques that globally
redistribute energy and angular momentum among the dark matter,
stellar, and gas components and thereby change the galaxy's
equilibrium mass distribution.  A barred galaxy is the simplest, most
well-defined example of strong inter-component evolution and,
therefore, a good litmus test for our understanding of long-term
galaxy evolution mediated by resonant interactions; more than half of
all galaxies are strongly barred in the near IR
\citep{Eskridge.Frogel.ea:00,Jogee.Barazza.ea:04}. We will emphasise
the bar--halo interaction as an example throughout this paper but one
should remember that the same dynamical arguments apply to any
evolving disturbance including interactions between the inner and
outer disk, the spheroid, and the dark matter halo.  A merging
satellite, for example, will be the subject of a forthcoming paper.
The bar--halo interaction has been studied recently by a large number
of groups with a variety of differing conclusions
\citep{Debattista.Sellwood:00, Sellwood:03, Athanassoula:03,
Valenzuela.Klypin:03}.  The goal of this paper is to provide a
detailed understanding of the dynamical mechanisms underlying this
simplest of intercomponent interactions and as a guide to the
requirements necessary to reproduce these dynamics accurately in
N-body simulations.

The bar-halo interaction is most often described in terms of dynamical
friction although, as we will see later in this paper many aspects of
this interaction are qualitatively different.  \citet[hereafter
TW]{Tremaine.Weinberg:84} and \citet{Weinberg:85} explained the bar
slow down observed in N-body simulations \citep{Sellwood:81} by using
dynamical friction as a paradigm and by deriving a formalism
appropriate for the quasi-periodic orbits typical of galaxies.  In
short, the bar interacts with the dark matter halo near
resonances. This induces a wake that lags the bar and, therefore,
torques the bar and slows its pattern speed.  The use of the
Chandrasekhar formula produces the correct scaling for the halo--bar
torque with rapid evolution, but does not properly represent the
underlying mechanism.  To see this, consider a sphere of orbits with a
rotating bar pinned to its centre.  If one stood on the rotating bar
and looked at the surrounding orbits in general they would execute
rosettes.  Because orbits spend more time near apocenter than
pericenter, orbits will torqued by the bar if their apocenters lead
the bar.  However, eventually the apocenters of the same orbit will
trail the bar as the rosette fills in.  If one waits long enough, the
apocenter will appear at every phase relative to the bar and the net
torque on the orbit will vanish.  If one applies this argument to
every orbit, the bar can never apply a torque!

What went wrong?  We made two related but inconsistent assumptions:
(1) we can ignore the closed periodic orbits because they are measure
zero in phase space; and (2) we can wait sufficiently long for the
orbits to look like filled in rosettes.  Consider an orbit that is not
quite closed.  This nearly closed orbit will have apsides that precess
so slowly that it will never look like a filled in rosette over an
astrophysically realistic time period because a galaxy is only a
finite number of bar periods old.  As one makes the time interval
shorter, more orbits will not look like filled in rosettes.  These
orbits {\em will} receive a net torque over this finite period, which
causes the bar to slow.  However, as the bar slows, these nearly
closed orbits no longer find themselves nearly closed and a new set of
orbits take their place.  This describes the essence of resonant
angular momentum transfer.  The resonance itself refers to the closed
orbit condition: frequencies of the orbit being commensurate with the
frequency of the bar pattern.  In other words, the angular momentum
exchange is caused by the breaking of adiabatic invariants near a
resonance.  Even though the periodic orbits have zero measure, they
influence the dynamics over a finite measure of phase
space\footnote{This is well-known in density-wave theory but less well
  appreciated in the current context.}.  The importance of resonances
in galactic disk dynamics was explored by \citet[hereafter
  LBK]{Lynden-Bell.Kalnajs:72}.  The dynamics of this process is
qualitatively different than the sum over scatterings that leads to
the Chandrasekhar formula.  This paper will describe these dynamics in
more detail, derive explicit conditions based on Hamiltonian
perturbation theory that must be satisfied before these resonant
dynamics can be obtained in N-body simulations, and demonstrate them
with N-body examples.

In an earlier paper, \citet[hereafter WK]{Weinberg.Katz:02}, we
described the interaction between a bar and a dark matter halo based
on a combination of perturbation theory and N-body simulations.  We
noted that in cuspy haloes the following low-order resonance extends
all the way to the centre:
\begin{equation}
-\Omega_r + 2\Omega_\phi = 2\Omega_p
\label{eq:ilr}
\end{equation}
where $\Omega_r$ ($\Omega_\phi$) is the frequency of the radial
(azimuthal) oscillation and $\Omega_p$ is the pattern frequency of the
bar.  This resonance for arbitrary eccentricity orbits is analogous to
the classical Inner Lindblad Resonance (ILR) for nearly circular
orbits.  We will call this resonance the ILR throughout this paper
although we really mean its {\em hot} analogue. The reason that the
ILR extends all the way to the centre owes to the relationship between
frequencies for radial orbits, $\Omega_r=2\Omega_\phi$.\footnote{For
density profiles less steep than singular isothermal
\citep{Touma.Tremaine:97}.} Therefore, in a cusp where
$\Omega_p\ll\Omega_r$ and $\Omega_p\ll\Omega_\phi$, there is always
some orbit in an isotropic system such that equation (\ref{eq:ilr})
holds.  Since the specific angular momentum in a central dark matter
cusp is very small, if the bar can torque orbits at this resonance, it
could make large changes to the inner density profile.  A linear
perturbation theory calculation that includes self-gravity suggested
that these changes might be significant and the predicted changes were
observed in an N-body simulation (WK).  Our results differed with the
conclusions of published simulations
\citep[e.g.][]{debattista.sellwood:98} only in that this central
evolution had not been previously examined.  Previous simulations
focused on the slowing of the bar, which is widely seen in simulations
\citep{Hernquist.Weinberg:92, debattista.sellwood:98,
Valenzuela.Klypin:03} but whose rate remains controversial.

WK offered some explanation for these differing results.  The elements
of our interpretation fell into two categories: (1) numerical
limitations: astronomically unrealistic (Poisson) noise disrupting the
quasi-periodic dynamics and (2) the sensitivity of the evolution to
the particular halo, disk and bar profiles.  This paper will explore
the first of these issues and the underlying dynamics in detail
beginning with an elaboration of the physical picture presented above
in \S\ref{sec:bardyn}. We will explore the second category in
\citet[hereafter Paper II]{Weinberg.Katz:05b}.  \citet{Debattista:02}
and \citet[hereafter S3]{Sellwood:03} have suggested that the these
differences are caused by the fixed pattern speed assumption in WK,
which makes the width of the resonance in frequency space narrow,
whereas real resonances in slowing bars are broad.  However, the
breadth in frequency space from the finite lifetime of the bar is
similar to the breadth from the slowing of the bar.  We will describe
why the breadth of the resonance in frequency space is a weak effect
in \S \ref{sec:bardyn} and show that the requirements to accurately
simulate this resonance are very stringent and likely explain most of
the differences.

Most of the comparisons between simulations and analytic theory have
examined the overall rate of angular momentum transferred between a
bar and a halo using an appropriately developed formula from LBK or
TW.  In contrast, we compare the analytic predictions from
perturbation theory and the results of N-body simulations by examining
the details of the dynamical mechanisms on small scales in phase
space.  We were surprised to find that the time scale of secular
evolution in galaxies, e.g. bar slow down, can be so fast that the LBK
formalism gives quantitatively inaccurate results \citep{Weinberg:04}.
We describe a second surprise in this paper: orbits may linger near
the resonance, which causes the change in conserved quantities to
scale as the square root of the perturbation strength (described in WK
as the {\em slow limit}) rather than as the square of the perturbation
strength (the {\em fast limit}).  Such a possibility was discussed in
TW but we find that in practice it is important for the ILR.  The
proper identification of the dynamical mechanisms and their regimes is
a necessary first step in being able to compare with simulations.

All this makes the astronomically relevant regimes for a slowing bar
of at least modest strength not easy to describe with analytic perturbation
theory.  One needs to include the direct time dependence as
described in \citep{Weinberg:04} and an interaction that may linger
near the resonance for an arbitrary time.  Using Hamiltonian perturbation
theory, we can reduce the exact solution to a series of
one-dimensional Hamiltonian problems.  We can then solve these problems
using a sequence of symplectic mappings or direct integrations.  This
brute-force perturbation technique allows us to obtain solutions for an
arbitrary amplitude and
time dependence while maintaining the well-understood aspects of
secular perturbation theory.  We describe this approach in
\S\ref{sec:bardyn}.

In \S\ref{sec:simres}, we discuss the requirements for an N-body
simulation to accurately follow these resonant dynamical processes and
identify three criteria that must be satisfied. The first criterion
requires that the phase space around the resonance be adequately
populated.  The finite number of particles used to trace the
gravitational field causes fluctuations on all scales.  These
fluctuations can change the dynamics of an orbit near resonance.  We
divide these into small- and large-scale fluctuations and derive two
additional particle number criteria.  The final criterion also
provides estimates for astronomical noise sources such as dark-matter
substructure. We illustrate the consequences of time-dependence
and multiple dynamical regimes in \S\ref{sec:simres} using the
generalisation of the familiar LBK formula for finite-time
interactions presented in \citet{Weinberg:04}.  We end with a
discussion in \S\ref{sec:disc} and summarise in \S\ref{sec:summ}.

\section{Basic principles}
\label{sec:bardyn}

There are two complementary ways to describe the dynamics of bar---halo
interactions: 1) consider the global macroscopic response of the halo
to the bar and compute the subsequent evolution; and 2) consider the
sum of each orbit's individual response to the perturbation and
compute the evolution as the net change in each orbit's conserved
quantities.  Each point of view provides a different insight but both
points of view are formally equivalent and will lead to identical
outcomes.  The former is natural for comparison with N-body
simulations and the latter with methods and results from nonlinear
dynamics.  Both require careful attention to the resonances but have
different virtues depending on the application.

\begin{figure}
  \centering
  \includegraphics[width=0.7\linewidth]{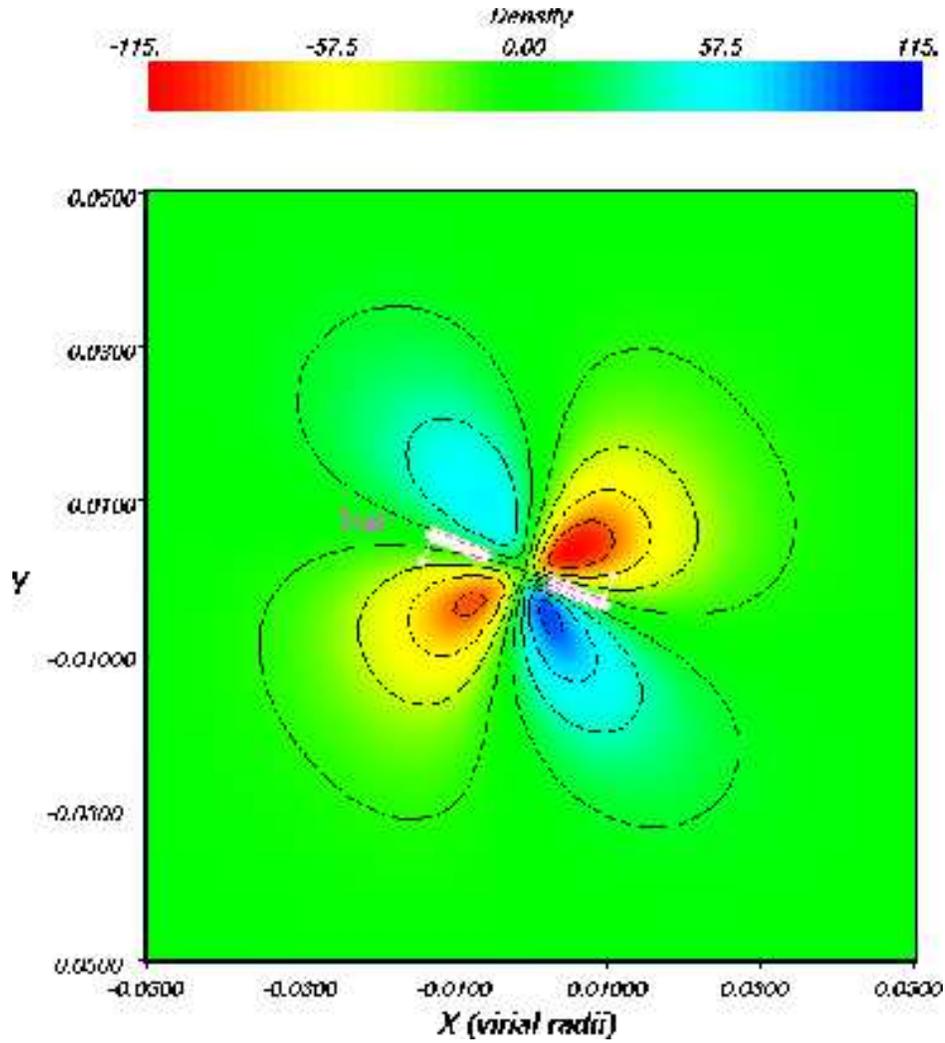}
  \caption{The wake in a halo caused by live disk bar in a live
    dark-matter halo. The contours and colour mapping show the dark
    matter wake density in a cut through the disk mid plane from blue
    (overdense) to red (underdense).  The bar position angle is shown
    in white with the direction of rotation indicated by the arrows.
    The live disk bar simulations are described in detail in
    \protect{\citet{Holley-Bockelmann.Weinberg.ea:05}}.}
  \label{fig:barwake}
\end{figure}

\begin{figure}
  \centering
  \includegraphics[width=0.3\linewidth]{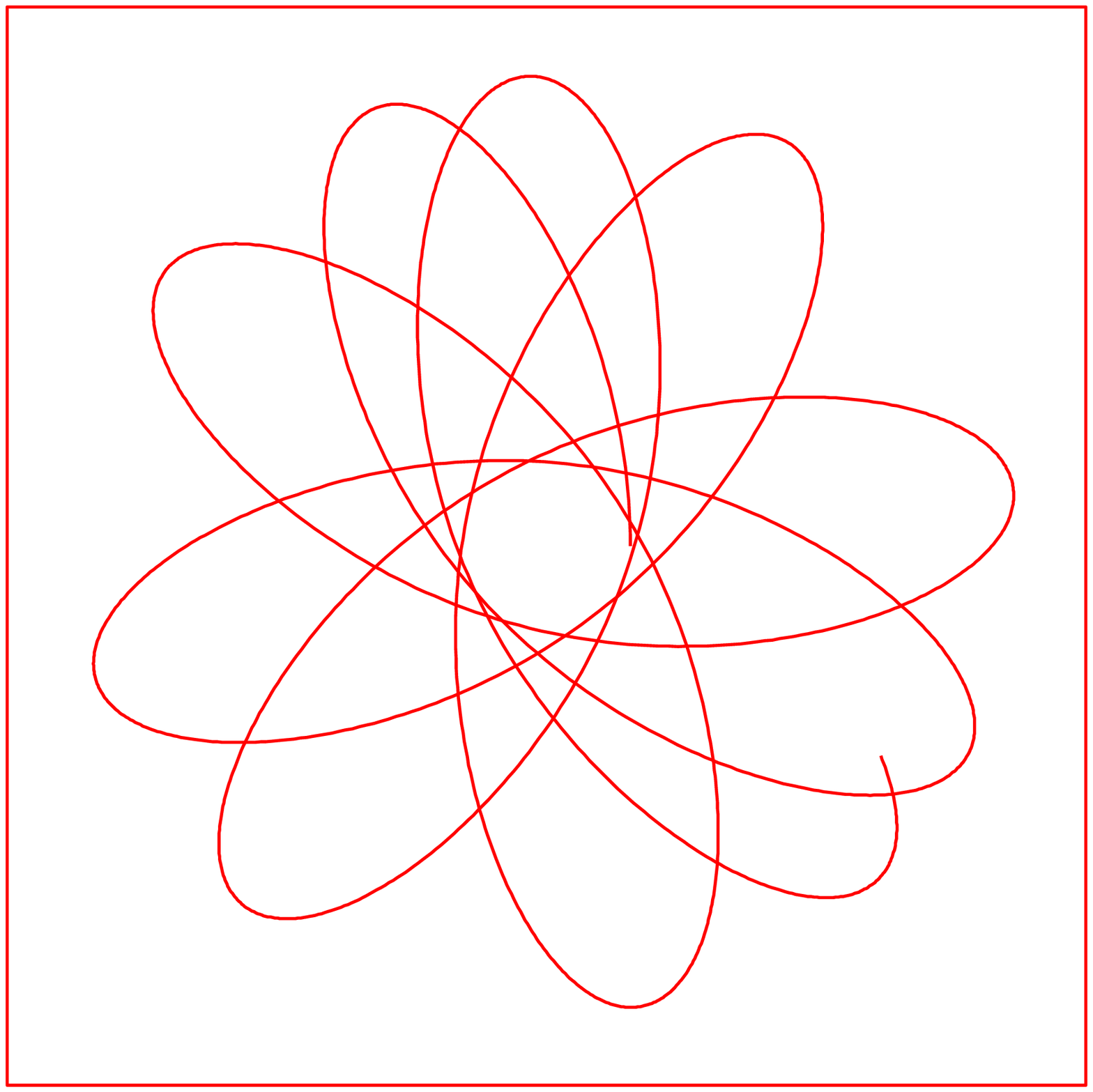}
  \includegraphics[width=0.3\linewidth]{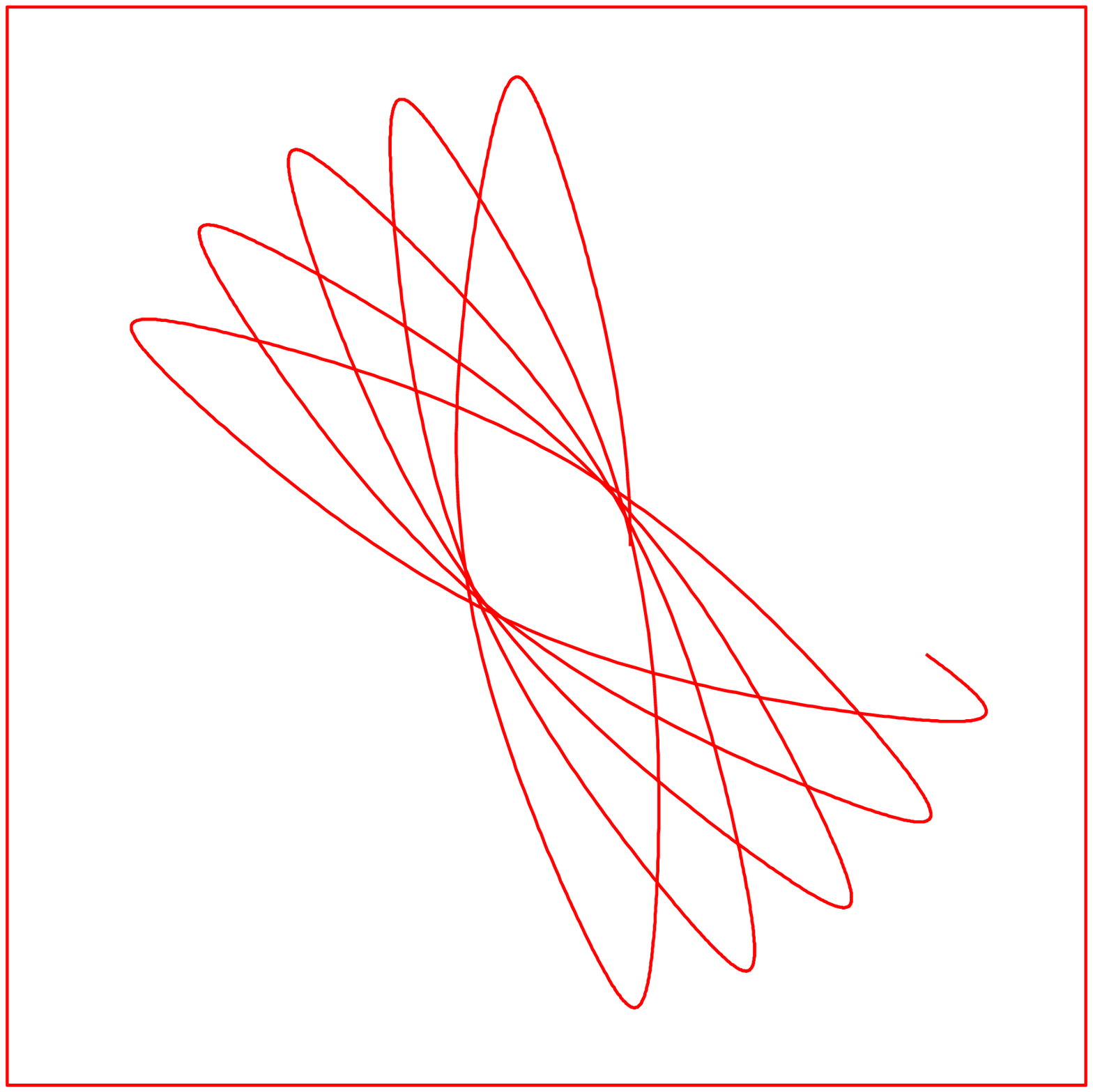}
  \includegraphics[width=0.3\linewidth]{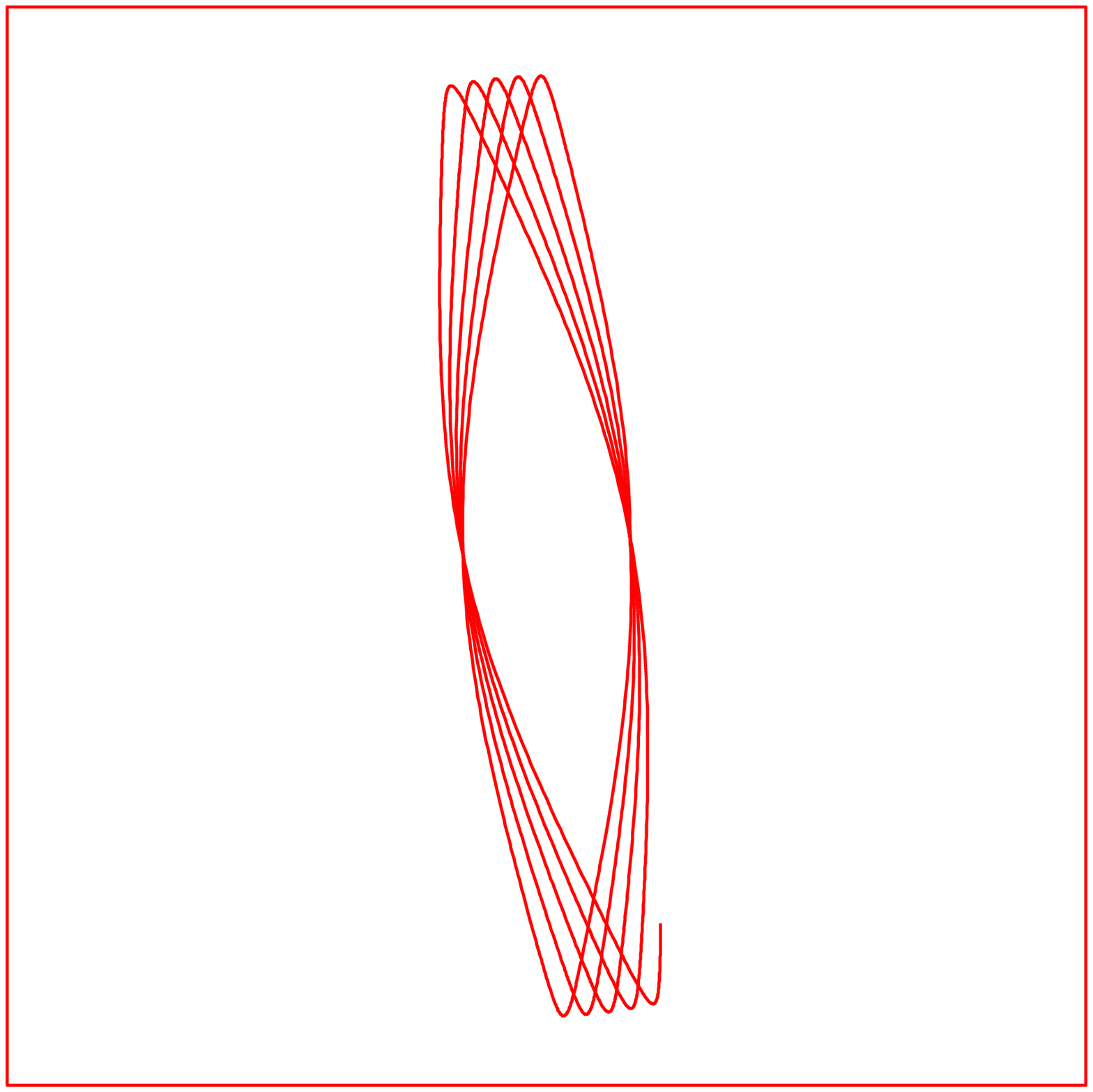}
  \caption{Orbit trajectories viewed the in the frame of a rotating
    bar.  An orbit far from a resonance has the standard ``rosette''
    appearance (Panel A, left).  An orbit near a resonance, in this
    case a 1:2 resonance, has slowly precessing apsides (Panel B,
    centre).  The precession rate decreases as the resonance is
    approached (Panel C, right).  An orbit at resonance is closed.}
  \label{fig:resorb}
\end{figure}

In the first point of view, the bar excites a wake in the halo.  An
example of such a wake in a self-consistent bar simulation is shown in
Figure \ref{fig:barwake}.  The excited dark matter halo wake lags the
bar, which causes a torque on the bar and removes angular momentum.
Naively one might expect the wake to be symmetric about the bar.  An
explanation for the lag requires us to consider the second point of
view: the response of individual orbits.  The basic physical picture
was outlined in \S\ref{sec:intro}: an arbitrary orbit has apsides that
precess either forward or backward in the rotating bar frame,
depending on the orbit's energy and angular momentum.  A
forward-precessing case is shown in the first panel of Figure
\ref{fig:resorb}.  Over short periods of time, the orbit may torque
the bar and later the bar may torque the orbit.  However, if we look
at this orbit averaged over some time interval $T$, long compared to
both its orbital and precession periods, the bar will see an
axisymmetric ring of mass density. Such an orbit, therefore, does not
change its angular momentum and presents {\sl no net torque} on the
bar.  However, there will always be some orbits that are very nearly
closed in the bar frame; these are the {\sl commensurate} or {\sl
resonant orbits}.  Near commensurabilities, the precession appears to
stop or slow down so much that it might as well be stopped.  More
precisely, for some fixed time interval $T$, the precession rate is
sufficiently slow that density of the orbit averaged over the
available time is not axisymmetric.  This situation is shown in Panels
(b) and (c) of Figure \ref{fig:resorb} for orbits successively closer
to resonance.  Resonant orbits feel a coherent forcing at the same
phase over many periods.  For these orbits, adiabatic invariance is
broken and the actions {\em can} change.  Therefore, the orbit can
exchange angular momentum with the bar, which causes both the bar and
orbit to evolve.

Figure \ref{fig:resorb} shows an orbit with prograde procession, but
there also exists a corresponding orbit with retrograde procession at
a slightly different (in this case larger) energy.  To lowest order,
these orbits cancel and there is no evolution.  Although the net
torque from the appropriately chosen pair may cancel, the phase-space
density will usually vary with energy and hence the average over phase
space will not generally cancel: there will either be more prograde or
retrograde orbits.  More precisely, the net torque caused by a
particular resonance will depend on the gradient of the phase-space
distribution function at that resonance (LBK, TW). At any one time,
halo orbits are gaining and losing angular momentum owing to all the
resonances but the actual net torque occurs as these first order
effects cancel.  If there were no phase-space density gradient near
the resonance, in many cases there would be no evolution.  In
addition, this net torque imposes a direction to the evolution and the
broken symmetry causes the response to either lag or lead the bar
position angle.  For a given bar perturbation, the response of the
halo and, therefore, the net torque on the bar will depend on the
phase-space structure of the dark halo.

From the point of view of an individual orbit, the bar induces a
periodic distortion in its trajectory, analogous to the modulation of
a pendulum by a sinusoidal force.  Averaged over an ensemble of orbits
with different phases, the sinusoidal response cancels.  However,
orbits that pass through resonance receive a permanent change that is
proportional to the amplitude of the perturbation.  The ongoing
secular evolution changes both the properties of the bar and the halo.
Therefore, the position of the resonance, defined by the closed
non-precessing orbit, slowly drifts through phase space (see
\S\ref{sec:calib}).  Hence, the time spent by any orbit near a
resonance is finite because the entire system evolves as a consequence
of the torque applied to these commensurate orbits (TW).  If the
secular evolution is rapid, the change in energy and angular momentum
(or actions) caused by a particular resonance is usually small for any
orbit \citep{Weinberg:85}.  The net change for these orbits do not
cancel since there are always some measure of orbits close enough to
being closed such that the first-order response does not cover all
phases.  The net angular momentum change then results from the
coherent interaction between the forced excitation of the orbit and
the slowly changing forcing potential, making the magnitude of the
response second-order in the bar perturbation amplitude.  In summary,
the net torque is proportional both to the phase space gradient of the
unperturbed phase space distribution and to the amplitude of the
changes in the angular momenta of individual orbits, which is
proportional to the square of the perturbation amplitude.  If the
secular evolution is slower, orbits may linger near the resonance.  In
this case, the interaction is nonlinear and no longer depends on
cancellation or the gradient of the phase-space density.  The change
in angular momentum for these interactions depends on the square root
of the perturbation amplitude (TW).  We will see that this regime is
important for the ILR.  Altogether the overall evolution of the galaxy
is driven by the net torque from all the resonances and affects a
significant fraction of all orbits.  Note that the evolution of the
galaxy halo is not caused by a dynamical instability but is {\em
secular} since it is driven by the exchange of angular momentum with
the disk bar.

A common misconception is that resonances are extremely ``thin'' and,
therefore, will only affect a set of orbits of measure zero.  Based on
the physical explanation above, this statement is incorrect for
several reasons.  First, one must be careful to distinguish between
the width of the resonance in phase space and the width of the
resonance in frequency space.  The width in phase space depends on the
integers defining the commensurability\footnote{The triple $(-1, 2,
2)$ in eq. \ref{eq:ilr}.  The mathematical definition will be
presented in \S\ref{sec:hamilton}.}, the amplitude of the perturbation
and the frequency of the perturbation, $\Omega_p$.  The {\em resonance
width} scales as the square root of the bar amplitude and inversely
with the second partial with respect to the resonant action (see eq.
\ref{eq:reswid} and associated discussion).  The sign and magnitude of
the torque depends on the phase of the apoapse as it passes through
the resonance and the net torque results from the sum over all possible
phases.  If there were insufficient orbital density in a resonance
width then one would not get the ensemble result but the contributions
from a few orbits at arbitrary phase, which would give a larger,
fluctuating contribution.  This leads to an important particle number
criterion (\S\ref{sec:calib}).

Secondly, for a real stellar system, the frequency spectrum of the
perturbation is not made of sharp lines but is broadened both by the
finite age of the galaxy and by the time-dependence of the driving bar
perturbation.  This is related to our consideration of finite time
intervals $T$ in the previous discussion.  As long as the integral
under each ``line'' in the spectrum is approximately the same, the
time-asymptotic result from secular perturbation theory remains valid.
In other words, nearly the same results obtain as long as the
resonances are not overlapping, which is true unless the bar slows is
so quickly that distinct resonances disappear, which we will
demonstrate does not occur in practise.  This approximation also
breaks down if the overall evolution of the bar pattern speed or the
density profiles is so slow that the changes in the individual orbits
near the resonance receive nonlinear perturbations.  This situation
can occur for some resonances even when the bar pattern speed changes
rapidly as the bar loses angular momentum
\citep{debattista.sellwood:98,Athanassoula:03}.  In these cases, the
perturbation equations may be solved directly, as we will describe
below.  Even if the bar pattern speed did not change or were held
fixed artificially, new orbits would still be affected as the
phase-space structure of the system itself evolved.  The orbits at or
near the resonance will change their actions and will, therefore, occupy
a different part of phase-space.  This in turn causes the halo
potential to change and reach a new equilibrium, moving fresh material
into the resonance.

Finally, even in the extremely artificial situation where both the
background potential and the bar pattern speed were held fixed, a
significant number of orbits would still be affected as long as the
system only existed for a finite time.  Orbits near a resonance
precess away from the resonance but they only do so very slowly: the
closer they are to the resonance, the slower their precession.  Given
an infinite time such orbits would precess through all angles and give
an axisymmetric time averaged orbit that would feel no net torque from
the bar. In such an eternal system, the orbits affected would be a set
a measure zero.  However, any astrophysical system only exists for a
finite time so the bar changes the actions of many orbits.  We
explicitly compute the extent in phase space of resonances for finite
time perturbations in \S\ref{sec:simres}.

These same arguments imply that there is no evolution without
resonances for a {\em collisionless} system in a near-equilibrium
state.  In the case of a rotating bar, for some fixed $T\gg t_{dyn}$,
the density of a time-averaged orbit sufficiently far from a resonance
is axisymmetric (see Fig. \ref{fig:resorb}, first panel) and, therefore,
applies no torque.  In the absence of all resonances, the conserved
actions, energy and angular momentum, are preserved for all time in
an axisymmetric collisionless stellar system.  In this sense,
post-formation near-equilibrium galaxy evolution is governed by the
resonant transfer of angular momentum. Resonances are not the
exception but are required for galaxy evolution!  For example, global
``modes'', i.e. a density wave that self-similarly evolves and whose
ensemble describes all possible excitations such as spiral arms, bars,
and halo modes, must dominate angular momentum transfer in the absence
of strong non-equilibrium perturbations such as mergers.  In the
absence of these ``modes'', secular evolution can only occur by local
{\em collisional} scattering, as in an accretion disk, but this
process has a characteristic time scale much longer than a Hubble time
in galaxies with the observed amount of substructure.

\subsection{Hamiltonian perturbation theory}
\label{sec:hamilton}

In this and the next several subsections, we convert this physical
picture to rigorous criteria for computing resonance phenomena in
particle simulations.  In later sections and in Paper II, we will
explore these dynamics in N-body simulations directly.

One can estimate the overall torque applied by a bar
analytically by summing the change in the action caused by the
perturbation after some period $T\gg t_{dyn}$ for each orbit in an
ensemble using the collisionless Boltzmann equation (CBE).  Although
more difficult, this approach does the averaging analytically. However, one
must be careful to treat the formal divergences that occur at
resonances.  These divergences are caused by the infinitely large
amplitude achieved by an infinitesimal number of particles, but are
not a cause for physical concern.
Alternatively, one may solve the perturbation theory equations
directly (see \S\ref{sec:solve}).  One must also take care to include
all the important resonances in this sum including those associated
with both discrete and continuous modes.  This is a straightforward if
somewhat complicated calculation if both self gravity and modes are
excluded, which we do here.  However, self-gravity can
enhance the response at large {\sl global} scales
\citep{Weinberg:98} and is also responsible for the existence of
weakly damped discrete modes \citep{Weinberg:94}, such as the $m=1$
sloshing mode.  These modes persist for astronomically relevant time
scales and some are sufficiently long-lived that they must be included
to follow the dynamics correctly.  Real astrophysical systems are not
eternal and damped modes are observationally evident in asymmetries
\citep[e.g.][]{Vesperini.Weinberg:00}.  With additional work, self gravity and
point modes can be accommodated by analytic perturbation theory,
although analytic estimates of complex realistic scenarios are
difficult.  (In \S\ref{sec:numsol}, we present a direct solution to
the perturbation theory problem that circumvents all of
these difficulties at the expense of CPU time.)  We begin by
presenting the solution of the CBE both to make contact with previous
work (e.g. LBK \& TW) and illustrate the difficulties.  This development
also includes all of the background needed for direct solution.

The torque on the halo may be computed analytically by expanding the
rotating bar potential in a Fourier series, where each angle
corresponds to a quasi-periodic degree of freedom for an orbit.  In a
spherical system, one degree of freedom corresponds to radial motion,
one to azimuthal motion in the orbital plane, and one to the
orientation of the orbital plane.  This last angle has zero
frequency.  The coefficients of the expansion will then depend only on
actions $\bI$:
\begin{equation}
  H_1({\bf r}, t) = \sum_{l_1,l_2,l_3=-\infty}^{\infty}
  H_{1\,\bl}(\bI) e^{i(l_1 w_1 + l_2 w_2 + l_3 w_3 - m\Omega_p t)}
  \label{eq:v1exp}
\end{equation}
where $\Omega_p$ is the bar pattern speed, $m$ is azimuthal wave
number defined by the spherical harmonic $Y_{lm}$, $\bI=(I_1, I_2,
I_3)$ are the actions with their corresponding angles $\bw=(w_1, w_2,
w_3)$, and $\bl=(l_1, l_2, l_3)$ is an integer vector describing each
term in the Fourier series.  We will use subscripts '0' and '1' to
denote terms that are zero- and first-order in the perturbation
amplitude.  The perturbed Hamiltonian includes both the imposed
gravitational potential of the bar and the gravitational potential
resulting from the response of the halo.  These actions follow
naturally from Hamilton-Jacobi theory; $I_1$, can be immediately
identified with the radial action $I_r$, $I_2$ with the angular
momentum in the orbital plane $J$ and $I_3$ with the projection of the
angular momentum along the $z$ axis $J_z$
\citep[e.g.][]{Goldstein:50}. From Hamilton's equations, we can obtain
the frequency of the angles: $\Omega_j=\partial H_0/\partial I_j$.
The subscripted $H_{1\,\bl}$ denotes a coefficient in the action-angle
series.

\subsection{Solving the perturbation theory}
\label{sec:solve}

\subsubsection{Canonical transformation to slow and fast variables}
\label{sec:canonical}

The natural decomposition of phase space into resonant variables
follows from an action-angle transform of the collisionless Boltzmann
equation (CBE) in phase space and a Laplace transform in time
\citep[e.g.][]{Weinberg:98a}.  To solve for the response, we begin
with the linearised CBE:
\begin{equation}
  {\partial f_1\over\partial t} + 
  {\partial H_0\over\partial\bI}\cdot
  {\partial f_1\over\partial\bw} - 
  {\partial H_1\over\partial\bw}\cdot
  {\partial f_0\over\partial\bI} = 0.
\label{eq:LCBE}
\end{equation}
The quantities $f_0$ and $H_0$ are the unperturbed phase space
distribution function and the unperturbed Hamiltonian and $f_1$ and
$H_1$ are the first-order terms.  The Fourier-Laplace transform of the
CBE is
\begin{equation}
  s\dfl + i\ldo\dfl - i\bl\cdot{\partial f_o\over\partial\bI}
  {{\tilde H}_{1\,\bl}} = 0
\label{eq:lcbe}
\end{equation}
where $s$ is the Laplace transform variable, the tilde indicates a
Laplace transformed quantity, the subscript $\bl=(l_1, l_2, l_3)$
indicates an action-angle transformed variable and $\bO=\partial
H_o/\partial\bI$.  Remember that the total perturbing potential $H_1$
is the response combined with the external perturbation.  The solution
is the inverse Laplace transform of the series:
\begin{equation}
  f_1(s) = \sum_{l_1,l_2,l_3=-\infty}^{\infty} \dfl(\bI, s)
  e^{i(l_1 w_1 + l_2 w_2 + l_3 w_3)}
  \label{eq:dfser}
\end{equation}
where ${\bf w}$ are the three angle variables and ${\bf I}$
are the three action variables.

Each term in the solution described by equation (\ref{eq:dfser}) is
oscillatory, proportional to $\exp(i\bl\cdot{\bf w})$.  
For a fixed perturbing frequency $\Omega_p$, the inverse Laplace
transform couples each term to the perturbing frequency and yields
oscillations of the form $\exp[i(\bl\cdot{\bf w} - m\Omega_p t)]$.
For example, assuming an $l=m=2$ quadrupole perturbation as we do for
a bar, the integers $(l_1, l_2, l_3)$ take the following values:
$l_1\in(-\infty, \infty)$, $l_2\in{-2,0,2}$ and $l_3=2$.  Orbits with
$\ldo - m\Omega_p \rightarrow 0$ define the closed {\em resonant}
orbits in the frame of reference rotating with the bar.  For the parts
of phase space very near a resonance described by a particular $\bl =
(l_1, l_2, l_3)$, the argument of the exponential will change very
slowly for one term and rapidly vary for most of the other terms.

The separation of these characteristic motions is easily performed by
making a canonical transformation to two new degrees of freedom where
one of the new coordinates corresponds to the angle of the
commensurability $w_s=\ldw - m\phi_p(t)$ and
\[
\phi_p(t) = \int^t dt^\prime\Omega_p(t^\prime)
\]
is the position angle of the perturbation.  This can be done with the
canonical transformation generating function (e.g. Goldstein Type 2,
op. cit.): $F=w_1 I_f + [\ldw - m\phi_p(t)] I_s$.  The new actions and
angles are therefore:
\begin{eqnarray}
  w_f &=& w_1 \\
  w_s &=& \ldw - m\phi_p(t) \\
  I_s &=& {1\over l_2}I_2 \\
  I_f &=& I_1 - {l_1\over l_2} I_2 \\
\end{eqnarray}
with $H_{new} = H_{old} - m\Omega_p I_s$ and with $w_3$ and $I_3$ as
before.  The angle $w_s$ is often called the ``slow'' angle because
its conjugate frequency vanishes at the resonance.  There is some
arbitrariness in the choice of the ``fast'' angle, $w_f$.  For
example, the choice $w_f=w_2$ yields the conjugate actions $I_s =
I_1/l_1$ and $I_f = I_2 - l_2I_2/l_1$.  This choice might be useful
when $l_2=0$.  In both cases, the angle $w_s$ varies very slowly and
$w_f$ varies rapidly relative to $w_s$ near the resonance defined by
the vector of integers $\bl$.  We can take advantage of this situation
by averaging over some time interval $T$ long enough so that all the
terms but the resonant one vanish owing to the rapid oscillation in
$w_f$ but that is small enough so that the argument of the exponent is
nearly unchanged.  In this way, we reduce the problem to a collection
of one-dimensional pendulum equations in $w_s$, one for each resonance
\citep[e.g. the {\em averaging theorem},][]{Arnold:78}.

The averaged Hamiltonian in these new coordinates takes the form
\begin{equation}
  H(I_s) = H_0(I_{s,r}) + 
  {1\over2}\left.{\partial^2H_0(I_s)\over\partial
  I_s^2}\right|_{I_{s,r}}(I_s - I_{s,r})^2 + H_{1\bl}\cos w_s
  \label{eq:hamprt}
\end{equation}
where $I_{s,r}$ is the slow action at resonance and $H_0$ is the
averaged unperturbed Hamiltonian with constant $\Omega_p$, up to some
arbitrary constant.  Equation (\ref{eq:hamprt}) is a pendulum equation
with arm length $M\equiv(\partial^2H_0(I_s)/\partial I_s^2)^{-1/2}$ and
acceleration $H_{1\,\bl}$.  The relative signs for the terms in
equation (\ref{eq:hamprt}) can be arranged by changing the phase of
$w_s$ as necessary; real positive values are assumed in the square
roots of the expressions that follow.  

The area defined by the infinite-period trajectory\footnote{The {\em
homoclinic} trajectory for the pendulum problem, i.e. the pendulum
orbit that results in the pendulum standing upside down on its pivot.}
is naturally identified as the {\it width} of the particular
resonance; this width is proportional to the square root of the
Fourier coefficient $H_{1\,\bl}(\bI)$ from equation (\ref{eq:v1exp})
(see also eq.  \ref{eq:reswid}).  A single resonance potential can
cover a significant volume in physical space.  Individual resonances
can overlap in phase space and will lead to chaos and diffusion in the
usual way \citep{Chirikov:79,Lichtenberg.Lieberman:83}.  However, in
galaxy dynamics on large scales, a few low-order resonances often
dominate the response in both physical and phase space.

Hamilton's equations in these new variables with the Hamiltonian from
equation (\ref{eq:hamprt}) explicitly represents the dynamics near
resonance.  These equations may include an arbitrary time dependence
in the perturbation consistent with the averaging.  However, these
equations only describe a single orbit.  A full description of the
secular evolution of a galaxy requires the sum of responses over the
entire phase space and motivates beginning with equation
(\ref{eq:LCBE}).  However, remember that there are
astronomical regimes where an orbit may linger close to the resonance
making weak nonlinearities important and in these cases we must resort
to direct solution over an ensemble of orbits, as we describe below.

The inverse Fourier-Laplace transform of equation (\ref{eq:lcbe}) is
straightforward albeit cumbersome  and does not
require the equivalent one-dimensional problem explicitly.  Having
solved for the self-consistent phase-space distribution function for
the response, we may then go back to the first order solution of the
collisionless Boltzmann equation for the distribution function and
integrate over velocities to get the shape of the gravitational
potential corresponding to a particular resonance, derive the overall
torque, or compute any other phase-space moment of interest.  For a
given orbit with actions $\bI$, the position ${\bf x}$ determines the
angle ${\bf w}$.  The integral of $\dfl$ over velocity then requires
an implicit solution of the equations defining the angles from
positions and momenta. This full, final solution is self-consistent in
the limit that the force from the perturbation and the response are
small compared to the background restoring force.  The result is a
second-order perturbation theory calculation.  Here, for simplicity,
we will eliminate
the self consistency by assuming that $H_1$ includes only the external
perturbation rather than both the external perturbation and the halo
response, which is of the same order of magnitude.

As an example of this whole procedure, we sketch the calculation of
the response of a homogeneous core to a bar or satellite rotating with
frequency $\Omega_p$ outside the core.  We focus on the ILR,
$\bl=(-1,2,2)$, although the basic features of the example apply to
any $\bl$.  To solve equation (\ref{eq:lcbe}), we need an expression
for $H_{1\,\bl}$.  We begin by relating the angle variables to
physical quantities.  Based on coordinate transformations and
geometry, the angle $w_1$ describes the radial oscillation of the
trajectory, $w_2$ describes the azimuthal oscillation of the
trajectory, and $w_3$ determines the orientation of the orbital plane,
which is invariant for a spherical potential.  On can express $w_3$ in
terms of the colatitude $\theta$ of the trajectory and the elevation
of the orbital plane $\beta$ (TW).  In addition, we find that
$l_2=-l,-l+2,\ldots,l-2,l$ and $l_3=m$ with
$l_1=-\infty,\ldots,\infty$.  The first contributing multipole to the
ILR is the quadrupole $l=2, |m|=2$.  Furthermore, we can write the
perturbing potential as an inner quadrupole $V_1(r,\phi,\theta,t) = c
Y_{lm}(\theta,\phi-\phi_p(t)) r^2$, where $c$ is a constant, because
we are considering the perturber to be outside the phase-space region
of interest and, since by symmetry $V_{l\,-m} = V_{lm}^\ast$, it is
sufficient to consider $m=2$ only.  With all of these identifications,
it is then straightforward to compute the necessary Fourier
coefficient:
\begin{equation}
H_{1\,\bl}(\bI, t) = {1\over(2\pi)^3}
\int\int\int d\bw e^{-i\ldw} c r^2(\bI,\bw)
Y_{22}(\theta,0)^{im(\phi-\phi_p(t))}.
\label{eq:fourier}
\end{equation}
The only explicit time dependence is a pure sinusoidal oscillation so
the Laplace transform yields:
\begin{eqnarray}
{\tilde H}_{1\,\bl}(\bI, s) &=& {1\over s+im\Omega_p}{1\over(2\pi)^3}
\int\int\int d\bw e^{-i\ldw} c r^2(\bI,\bw)
Y_{22}(\theta,0)^{im\phi} \nonumber \\
&\equiv& {1\over s+im\Omega_p} {\bar H}_{1\,\bl}(\bI)
\label{eq:htrans}
\end{eqnarray}
The integrals defining ${\bar H}_{1\,\bl}(\bI)$ can be performed
analytically because the motion in a homogeneous core is purely
harmonic with constant frequencies throughout the core, but its exact
functional form is not important here.

We now substitute equation (\ref{eq:htrans}) into equation
(\ref{eq:lcbe})
\begin{equation}
  {\tilde f}_{1\,\bl}(\bI, s) = i\bl\cdot{\partial f_0\over\partial\bI} {\bar
  H}_{1\,\bl}(\bI) {1\over s + i\ldo}{1\over s + im\Omega_p}
\end{equation}
and perform the inverse Laplace transform as follows:
\begin{eqnarray}
  f_{1\,\bl}(\bI, t) &=& {1\over2\pi i}\int^{c+i\infty}_{c-i\infty} ds e^{st}
{\tilde f}_{1\,\bl}(\bI, s)  \nonumber \\
&=& i\bl\cdot{\partial f_0\over\partial\bI} {\bar H}_{1\,\bl}(\bI) 
\left\{ 
  {e^{-i\ldo t}\over i(m\Omega_p-\ldo)} +
  {e^{-im\Omega_p t}\over i(\ldo-m\Omega_p)} 
\right\} \nonumber \\
  &=& -\bl\cdot{\partial f_0\over\partial\bI} {\bar H}_{1\,\bl}(\bI) 
e^{-i(\ldo+m\Omega_p)t/2} 
{ \sin[(\ldo-m\Omega_p)t/2]\over (\ldo-m\Omega_p)/2 }.
\end{eqnarray}
This equation has several interesting features.  First, the term
\[
{ \sin[(\ldo-m\Omega_p)t/2]\over (\ldo-m\Omega_p)/2 }
\]
oscillates rapidly and in the limit $t\gg1/\bO$, $\Omega_p$ approaches
$\pi\delta(\ldo-m\Omega_p)$.  This implies that after a long time, one
will only see a contribution for a commensurability
$\ldo-m\Omega_p=0$.  If the satellite or bar was within or nearly
within the homogeneous core, $\Omega_1\approx\Omega_2\approx\Omega_p$
and no ILR can possibly exist for small values of ${\bf l}$.  A small
decrease in $\Omega_p$ can remove most of the low-order resonances,
and outside of the core, one has
$\Omega_p\simless\Omega_1\approx\Omega_2$ and, therefore, no low-order
resonances exist.  Second, in a truly homogeneous core, $f_0(\bI)$ is
constant so that $\partial f_0/\partial\bI=0$ and therefore
$f_{1\,\bl}=0$ independent of the resonance $\bl$.  These features
imply that a bar deep inside of a halo core will have a dramatically
reduced torque.

\subsubsection{Derivation of first-order changes in angular momentum}

We may derive an expression for the change in angular momentum
similarly.  We begin by deriving the change in angular momentum for a
single orbit.  The Liouville theorem gives us the rate of change in
angular momentum for each orbit as:
\begin{equation}
  {dL_z\over dt} = {\partial L_z\over\partial t} + [H, L_z]
\end{equation}
where $[\cdot,\cdot]$ are Poisson brackets
\citep[e.g.][]{Goldstein:50}.  Because $L_z$ is a conserved quantity
in the absence of any perturbation, this equation becomes
\begin{equation}
  {dL_z\over dt} = {\partial H\over\partial\bI}\cdot{\partial
  L_z\over\partial\bw} - {\partial H\over\partial\bw}\cdot{\partial
  L_z\over\partial\bI} = -{\partial H_1\over\partial w_3}
\label{eq:torque1}
\end{equation}
where $\bw$, $\bI$ are action-angle variables and $H_1$ is the
first-order, perturbed Hamiltonian.

Now expanding $H_1$ as an action-angle expansion
\begin{equation}
  H_1(\bI, \bw, t) = \sum_\bl H_{1\,\bl, t}(\bI) e^{i\ldw},
\end{equation}
we can describe the evolution of a particular orbit's z angular
momentum component:
\begin{equation}
  {dL_z\over dt} = -\sum_\bl il_3 H_{1\,\bl}(\bI, t) e^{i\ldw}.
  \label{eq:dldt}
\end{equation}
We may now integrate over some time interval $T$, defined to be long
compared to an orbital time for our orbit with action $\bI$ but short
compared to the overall evolutionary time scale.  The angles now are
explicit functions of time: $\bw(t) = \bw_o + \bO t$, where
$\bO\equiv\partial H_o/\partial\bI$ and $\bw_o$ is the angle vector at
$t=0$.  The solution for $L_z(t)$ is particularly simple for the
sinusoidal time dependence in a simple rotating pattern
(e.g. eq. \ref{eq:fourier}).  Integrating equation (\ref{eq:dldt}),
one finds
\begin{eqnarray}
  \Delta L_z(T) &\equiv& \int^T_0 dt{dL_z\over dt}
  =
  -\sum_\bl il_3 H_{1\,\bl}(\bI,0)  e^{i(\ldw_0)}
  \left.\left[{e^{i(\ores)t}\over i(\ores)}\right]\right|^T_0
  \nonumber \\
  &=&
  -\sum_\bl il_3 H_{1\,\bl}(\bI,0)  e^{i(\ldw_0)} 
    \left\{e^{i(\ores)T/2} {\sin(\ores)T/2\over(\ores)/2} \right\},
\label{eq:deltalz0}
\end{eqnarray}
where we denote $H_{1\,\bl}(\bI, t) = H_{1\,\bl}(\bI,
0)\exp(-im\Omega_p t)$.  Equation (\ref{eq:deltalz0}) describes the
change in angular momentum $L_z$ for a single orbit.  To determine the
change in angular momentum for a particular portion of phase space, needed
to perform a statistical comparison to a N-body simulation, we
must average equation (\ref{eq:deltalz0}) over an orbit ensemble. To
do this, we multiply equation (\ref{eq:deltalz0}) by the phase-space
distribution function $f(\bI, \bw, t) = f_o(\bI) + f_1(\bI, \bw, t)$,
where $f_1(\bI, \bw, t) = \sum_{\bl^\prime} f_{\bl^\prime}(\bI,t)e^{i\bl^prime dw}$,
and average over the initial phase, $w_o$.

There are several subtleties in evaluating the average of equation
(\ref{eq:deltalz0}).  First, note that the bracketed term on the
right-hand side is steeply peaked about $\ores=0$ in the limit
$T\gg1/\Omega_{1,2}$; this, of course, owes to the resonance defined
by $\ores=0$.  Second, an average over angles is non-vanishing only
for the term $f_1$ ($\Delta L_z$) when $\bl= -\bl$.  This implies that
the net torque is second order in $H_{1\bl}$ and this procedure yields
a generalised form of the LBK formula \citep{Weinberg:04}.  However,
by performing the average over $\bw$, we are implicitly assuming a
time interval such that all phases are uniformly represented and this
ignores an important feature of the resonance under some
circumstances.  Recall that near the resonance, we can replace the
multidimensional problem with a one-dimensional one in the resonant
angle $w_s$.  Although $w_1$ and $w_2$ remain distributed in phase,
the distribution of $w_s$ is correlated by the dynamics of the
resonance.  To evaluate this average including the correlation, we can
replace the integral over phase with an integral over time,
\begin{equation}
  \int^{2\pi}_0dw_s \exp(iw_s) = \Omega_s\int^{T_s}_0 dt \exp(iw_s(t))
  \label{eq:phaseav1}
\end{equation}
and evaluate the integral using pendulum dynamics.  Far from the
resonance, $w_s(t)$ will be linear with time.  As the orbit approaches
the resonance defined by the infinite-period trajectory, the phases
will linger near the top pivot point of the pendulum, i.e. an upside
down pendulum, and the run of $w_s(t)$ will look like stairs with
horizontals at $-\pi$ and $\pi$ ($0$ and $2\pi$) if the coefficient of
the pendulum potential is negative (positive).  Near the
infinite-period trajectory, $T_s\rightarrow0$, $\Omega_s\rightarrow0$
with $T_s\Omega_s\rightarrow2\pi$.  Because the phase will accumulate
near the top pivot point in the one-dimensional potential,
$\exp(iw_s)$ will take on the value $1$ or $-1$, again depending on
the sign of the resonance potential.  Putting this together the phase
integral takes the value
\begin{equation}
  \int^{2\pi}_0dw_s \exp(iw_s) \rightarrow 2\pi \sign(H_{1\bl})
\label{eq:phaseav}
\end{equation}
near the resonance.  This bunching and break down of the phase
averaging is a consequence of broken adiabatic invariance at the
resonance.  In the limit of long time intervals, the contributing
orbits will be closer and closer to the infinite-period trajectory
and, therefore, more and more closely bunched in phase about the top
pivot point.  Our detailed investigation shows that resonances in
astronomically realistic systems can be bunched and require a more
subtle treatment than that described here.  In particular, one must
consider the simultaneous variation of $I_s$ with $w_s$ and this
correlates $H_{1\bl}$ with $w_s$.  In this case, $\langle\Delta
L_z(t)\rangle$ scales as $|H_{1\bl}|^{1/2}$ rather than
$|H_{1\bl}|^2$.  TW call the division between these regimes the {\em
slow} and {\em fast} limits and we describe them in the next section.
Rather than perform very complicated slow-limit phase averages and
time ordering, we will solve the equations of motion by direct
numerical integration for an ensemble of orbits.

\subsubsection{The slow and fast limits}

The rate of secular evolution determines whether or not an orbit {\em
lingers} near the resonance for many periods or quickly moves through
the resonance.  The rate of pattern speed evolution
characterises the secular change.
The natural width in frequency space
is the change in the commensurate frequency across the resonant width.
The ratio of the rate of pattern speed evolution to the square of the natural
width in frequency is a
dimensionless ratio that TW call the {\em speed} ($s$).  In their
perturbation calculation, values of $s\ll1$ give rise to changes in the
action that scale as the square root of the perturbation strength and
$s\gg1$ give changes that scale as the square of the perturbation
strength.  TW call these the {\em slow} and {\em fast} limits.  For a
perturbed system whose only time-dependence is its pattern speed,
TW defined the quantity {\em speed},
\begin{equation}
s = {|m{\dot\Omega}_p|\over\left|V_{1,{\bf l}}\,
  {\partial({\bf l}\cdot{\bf\Omega})/\partial I_s}\right|}
\end{equation}
where $s\gg1$ is {\em fast} and $s\ll1$ is {\em slow}.

For our purposes, these two regimes complicate the perturbation theory
but do not result in a dramatic qualitative change in the evolution of
a particular orbit.  However, the standard perturbation theory
(e.g. LBK) assumes the fast limit.  A set of exploratory simulations
that explicitly tracked $s$ for individual orbits show that the
dominant resonances have contributions from the slow limit unless the
bar slows down within a rotation period.  The ILR is {\em dominated}
by slow-limit encounters.  In practise, we find that $s$ is ${\cal
O}(1)$ for most low-order resonances.

\subsection{Numerical solution of the the averaged equations: tools
  and approach}
\label{sec:numsol}

Solving the linearised CBE for arbitrary time dependence in the
perturbation and for arbitrary values of the speed $s$ is difficult if
not intractable.  The added complication of time dependence and
contributions from both the slow and fast limits motivates us to
abandon explicitly solving the CBE by analytic phase averaging in
favour of direct integration of the perturbed Hamiltonian.  Averages
of phase-space ensembles are then performed by Monte-Carlo
integration.  Although more CPU intensive, this direct approach to
perturbation theory simplifies the complexity of tracking multiple
time scales, requires no differentiation between the fast and slow
limits, and better facilitates a comparison to N-body simulations.  In
principle, one can include the self gravity of the response using this
method although we do not do so here.

\subsubsection{Twist mapping}

We can look at the topology and effect of diffusion on the resonant
islands by converting the perturbed Hamiltonian near a resonance to a
twist mapping applying the now standard procedure
\citep[e.g.][]{Lichtenberg.Lieberman:83}.   Begin with the
one-dimensional averaged equations,
\[
H(I_s, w_s) = H_0(I_s) + H_{1{\bf l}}(I_s)e^{iw_s}
\]
where $H_0$ and $H_{1{\bf l}}$ are the zeroth-order and perturbed
  parts of the Hamiltonian, respectively.  By choosing the phase, with
  no loss in generality, we may keep only the real part to get
\begin{eqnarray}
H(I_s, w_s) = H_0(I_s) + H_{1{\bf l}}(I_s)\cos(w_s).
\end{eqnarray}
The surface of section implied by the averaging principle are the
successive returns to the $I_s, w_s$ plane separated by time intervals
$T_2 \equiv 2\pi/\Omega_f$, where $\Omega_f$ is the fast frequency.
>From Hamilton's equations, the change in action between two successive
periods $T_2$ denoted by subscripts $n$ and $n+1$ is
\begin{eqnarray}
  I_{s,n+1} &=& I_{s,n} -\int^{T_2}_0 dt {\partial H_{1{\bf l}}\over\partial
  w_s}(I_{s,n+1}, I_f, w_{s,n}+\Omega_{s,n}t, w_f + \Omega_{f,n}t)
  \nonumber \\
  &=& I_{s,n} - H_{1{\bf l}}({I_{s,n+1}}, I_f) \left[
    {\sin(w_{s,n}+\Omega_{s,n}T_2) - \sin(w_{s,n}) \over \Omega_{s,n}}
    \right]
  \nonumber \\
  \label{eq:twisteqm}
\end{eqnarray}
where the second equality assumes a simple oscillatory time dependence
for the perturbation.  To derive a symplectic mapping, we choose a
canonical generating function of the form
\begin{equation}
F_2 = I_{s,n+1}w_{s,n} + 2\pi A(I_{s,n+1}) + G(I_{s,n+1}, w_{s,n}).
\label{eq:genfct}
\end{equation}
The canonical transformation is then:
\begin{eqnarray}
  I_{s,n+1} &=& {\partial F_2\over\partial w_{s,n}} =  I_{s,n+1} +
  {\partial G\over\partial w_{s,n}} \nonumber \\
  w_{s,n+1} &=& {\partial F_2\over\partial I_{s,n+1}} = w_{s,n} + 2\pi\alpha +
  {\partial G\over\partial I_{s,n+1}}
  \label{eq:twistsym}
\end{eqnarray}
where $\alpha\equiv dA/dI_{s,n+1}$.  Comparing equations
(\ref{eq:twisteqm}) and (\ref{eq:twistsym}) we can deduce
that
\begin{equation}
G(I_{s,n+1}, w_{s,n}) = {H_{1{\bf l}}(I_{s,n},
  I_f)\over\Omega_s(I_{s,n}, I_f)} 
\left[\cos(w_{s,n}) - \cos(w_{s,n}+\Omega_{s,n}T_2)\right] +
  h(I_{s,n+1}).
\label{eq:genfct2}
\end{equation}
The integration constant $h(I_{s,n+1})$ yields a phase shift that can
be absorbed into $\alpha$.

The mapping defined by equations (\ref{eq:twistsym}) and
(\ref{eq:genfct2}) is well-defined and straightforwardly applied but,
for computational speed, the phase-space quantities must be tabled and
evaluated using interpolation methods.  Truncation error can then lead
to an imperfect symplectic mapping, but this situation can be improved
as follows.  On rearranging terms, equation (\ref{eq:twistsym}) has
the form
\begin{eqnarray}
  I_{s,n+1} &=& I_{s,n} + f(I_{s,n+1}, w_{s,n}) \nonumber \\
  w_{s,n+1} &=& w_{s,n} + g(I_{s,n+1}, w_{s,n}),
\label{eq:twist}
\end{eqnarray}
which implies that
\begin{equation}
  -{\partial g\over\partial w_{s,n}} =  {\partial f\over\partial
   I_{s,n+1}}
\end{equation}
must hold.
The function $g(I_{s,n+1}, w_{s,n})$ has only a sinusoidal dependence
on $w_{s,n}$ and this implies that the integral in the previous
equation can be performed exactly:
\begin{equation}
g(I_{s,n+1}, w_{s,n}) = -\int dw_{s,n}{\partial g\over\partial w_{s,n}}.
\end{equation}
 Deriving the mapping in this way, rather than explicitly from
equations (\ref{eq:genfct}) and (\ref{eq:genfct2}), guarantees the
symplectic condition.

Investigation of the orbit topology for the astronomically-motivated
bar perturbation used here shows that most resonances have the
standard pendulum topology with one important exception: the relative
amplitude of the ILR is sufficiently large that orbit families
bifurcate as has been described extensively elsewhere \citep[e.g.][for
a review]{Contopoulos.Grosbol:89}.  This does not present an
in-principle problem but, for simplicity, we restrict ourselves to
moderate to weak strength bars for comparison between perturbation
theory and simulations to remove the complications caused by orbit bifurcation.

The action-angle or Cartesian phase-space coordinates can be obtained
at any iteration $n$ (eq. \ref{eq:twist}) in the twist mapping by another
canonical
transformation.  We apply two-body interactions in the diffusive limit
by tabulating the standard orbit averaged diffusion coefficients
\citep{Spitzer:87}.  At each step, we transform from slow and fast
actions and angles to Cartesian coordinates, apply the change in
parallel and perpendicular velocities obtained from a Monte Carlo
realization, and then transform back to slow and fast variables.

\subsubsection{Numerical solution of the phase-averaged equations of motion}
\label{sec:numavg}

As previously described, the first step in canonical perturbation
theory is to expand the phase-space quantities in a Fourier series of
actions and angles:
\[
\Xi({\bf r}, t) = 
\sum_{l_1,l_2,l_3} \Xi_{l_1,l_2,l_3}({\bf I}({\bf r}, {\bf v}))
e^{i[l_1w_1({\bf r}, {\bf v}) + 
  l_2w_2({\bf r}, {\bf v}) + l_3w_3({\bf r}, {\bf v}) - m\phi(t)]}
\]
where we have assumed that the time dependence is periodic with phase
angle $\phi(t)$ as previously defined and
\begin{equation}
\Xi_{l_1,l_2,l_3}({\bf I}) = \oint d{\bf w} e^{i(l_1w_1 + l_2w_2 +
l_3w_3)} \Xi({\bf r}({\bf I}, {\bf r}), t).
\label{eq:angletrans}
\end{equation}
If $\Xi$ is a perturbation and we consider Hamilton's equations, we
may transform to new action and angle variables where one angle takes the
form
\[
w_s\equiv l_1w_1 + l_2w_2 + l_3w_3 - m\phi_p(t).
\]
As long as $|{\dot w}_s|$ is sufficiently large, the reaction of a
orbit to the perturbation will be oscillatory.  However, in the
limit that
\[
{\dot w}_s = l_1\Omega_1 + l_2\Omega_2 + l_3\Omega_3 - m\Omega_b(t)
\ll \min_j(\Omega_j)
\]
the period of the oscillations will be very long and in the limit
${\dot w}_s \rightarrow0$ the perturbation may cause a permanent shift
in the actions, i.e. a {\em resonance}.  Near a resonance, the
three-degree-of-freedom equations of motion can be transformed to a
one-degree-of-freedom problem as described in \S\ref{sec:canonical}.
The one-dimensional equations of motion take the following form:
\begin{eqnarray}
  {\dot w}_s &=& {\bf l}\cdot{\bf \Omega} - m\Omega_p  +
  {\partial W_{l_1,l_2,l_3}({\bf I})\over\partial I_s} e^{iw_s}
  \nonumber \\
  {\dot I}_s &=& -i W_{l_1,l_2,l_3}({\bf I})
  \label{eq:onedeqm}
\end{eqnarray}
where $W$ is the transform of the perturbation that follows from the
application of equation (\ref{eq:angletrans})

The validity of the averaging theorem is the only practical limitation
in this approximation.  It may break down for large amplitude
perturbations if resonances overlap \citep[e.g.][]{Chirikov:79} or if
the system loses or changes equilibrium.  The approximation has two
important advantages.  First, it separates the resonant variable from
the others and, thereby, allows the dynamics of the resonance to be
isolated from the remaining system,  allowing us to check the
one-dimensional problem against the full integration of the equations
of motion (examples below).  Second, because we have averaged over
the rapid oscillations, the one-dimensional problem can be solved
numerically with a larger time step.

However, unlike the N-body problem, forces in
the one-dimensional mapping depend on the velocity, i.e. $\dot{I}_s$,
and, therefore, we cannot use an explicit symplectic integrator
such as Leap Frog.  Rather, we use the semi-implicit symplectic RK2
method:
\begin{eqnarray}
  q_{n+1} = q_n + {\partial H\over\partial
    p}\left({p_n+p_{n+1}\over2}, {q_n+q_{n+1}\over2}\right), \nonumber \\
  p_{n+1} = p_n - {\partial H\over\partial
    q}\left({p_n+p_{n+1}\over2}, {q_n+q_{n+1}\over2}\right).
\end{eqnarray}
We apply this to equation (\ref{eq:onedeqm}) and iterate until it
converges using the algorithm:
\begin{eqnarray}
  w^{r+1}_{s,n+1} &=& w^r_{s,n} + {\partial H\over\partial
    I_s}\left({w^r_{s,n}+p^r_{s,n+1}\over2},
    {w^r_{s,n}+q^r_{s,n+1}\over2}\right), \nonumber \\ 
    I^{r+1}_{s,n+1} &=& I_{s,n} - {\partial H\over\partial
    w_s}\left({I^r_{s,n}+I^r_{s,n+1}\over2},
    {w^r_{s,n}+w^r_{s,n+1}\over2}\right)
\end{eqnarray}
where $r=1,2,\ldots$ until $\Delta w_s\equiv|w^{r+1}_{s,n+1} -
w^r_{s,n}|<\epsilon_1$ and $\Delta I_s|I^{r+1}_{s,n+1} -
I^r_{s,n}|<\epsilon_2$.  In practice, this iteration converges for
$r\simless10$ when $\epsilon_j=10^{-10}$

\subsubsection{Comparison with the direct ODE}
\label{sec:ODE}

\begin{figure}
\centering
\subfigure[non-resonant]{
  \includegraphics[width=0.49\textwidth]{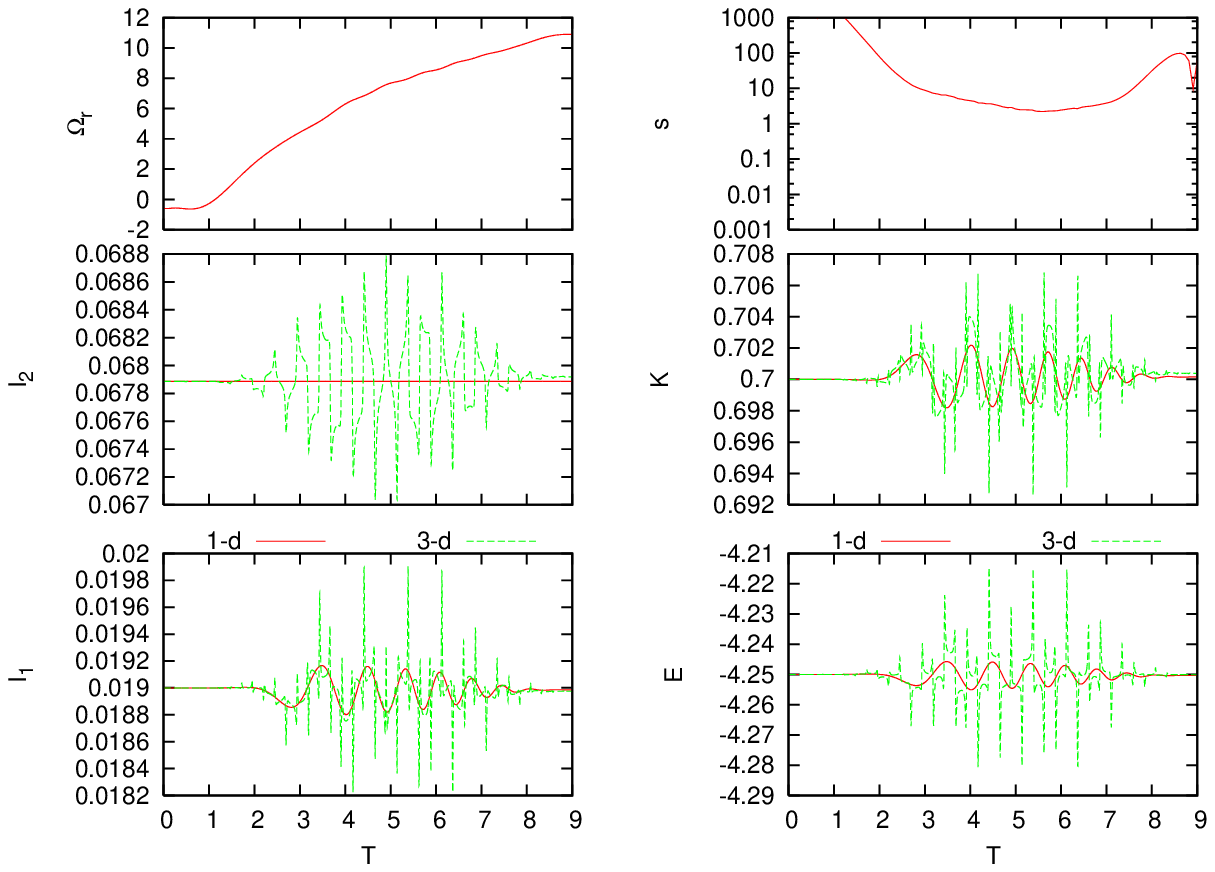}}
\subfigure[resonant]{
  \includegraphics[width=0.49\textwidth]{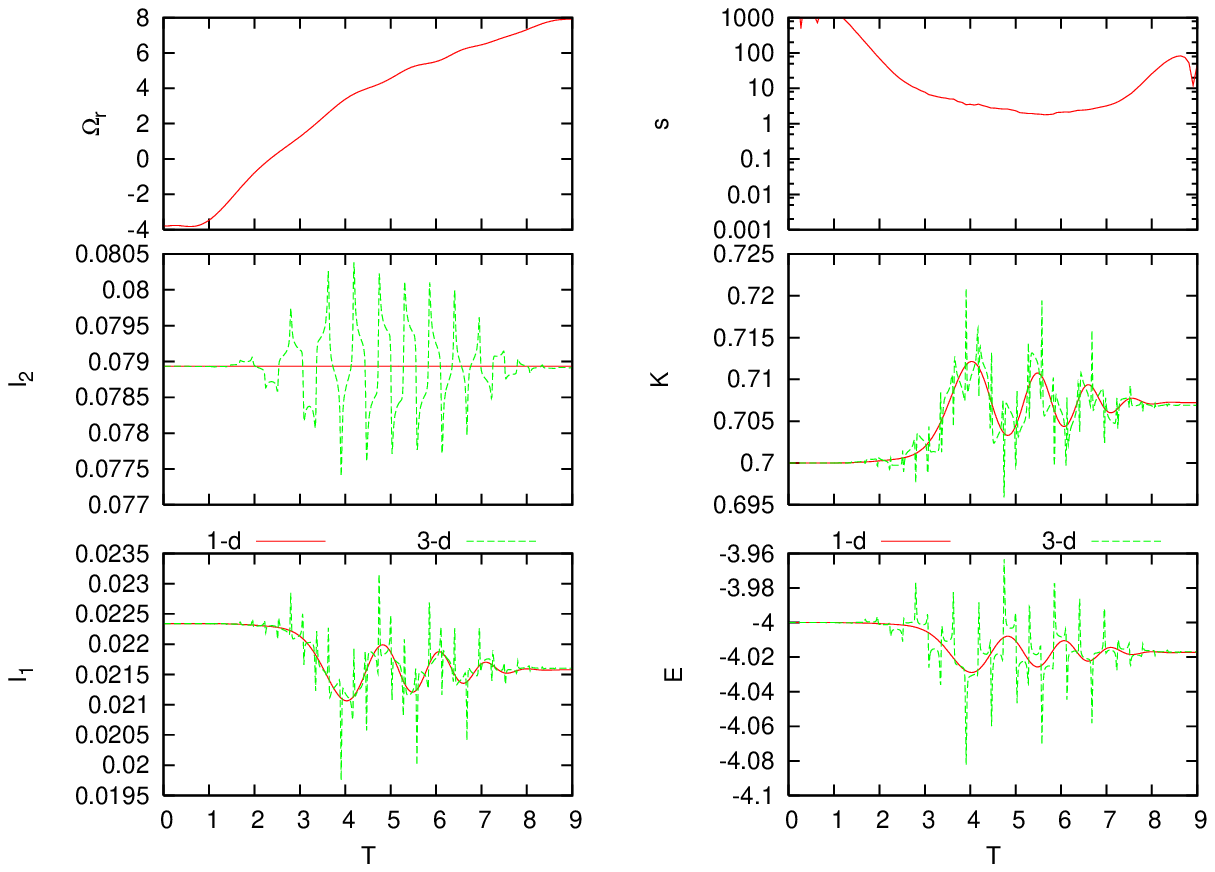}}
\caption{Panels show the evolution of an orbit during pattern speed
  evolution of a bar for (a) a non-resonant and (b) a resonant orbit.
  We plot the actions $I_1 \& I_2$, the energy $E$, the scaled angular
  momentum $\kappa=J/J_{max}$, the resonance quantity $\Omega_r\equiv
  l_1\Omega_1 + l_2\Omega_2 - 2\Omega_p(T)$, and the `speed' of the
  resonance $s$ versus time for the resonance $l_1=1$ and $l_2=0$
  (DRR).  The red line shows the one-dimension time averaged solution
  and the green line the full three-dimensional solution.}
\label{fig:comporb10}
\end{figure}

\begin{figure}
\centering
\subfigure[non-resonant]{
  \includegraphics[width=0.49\textwidth]{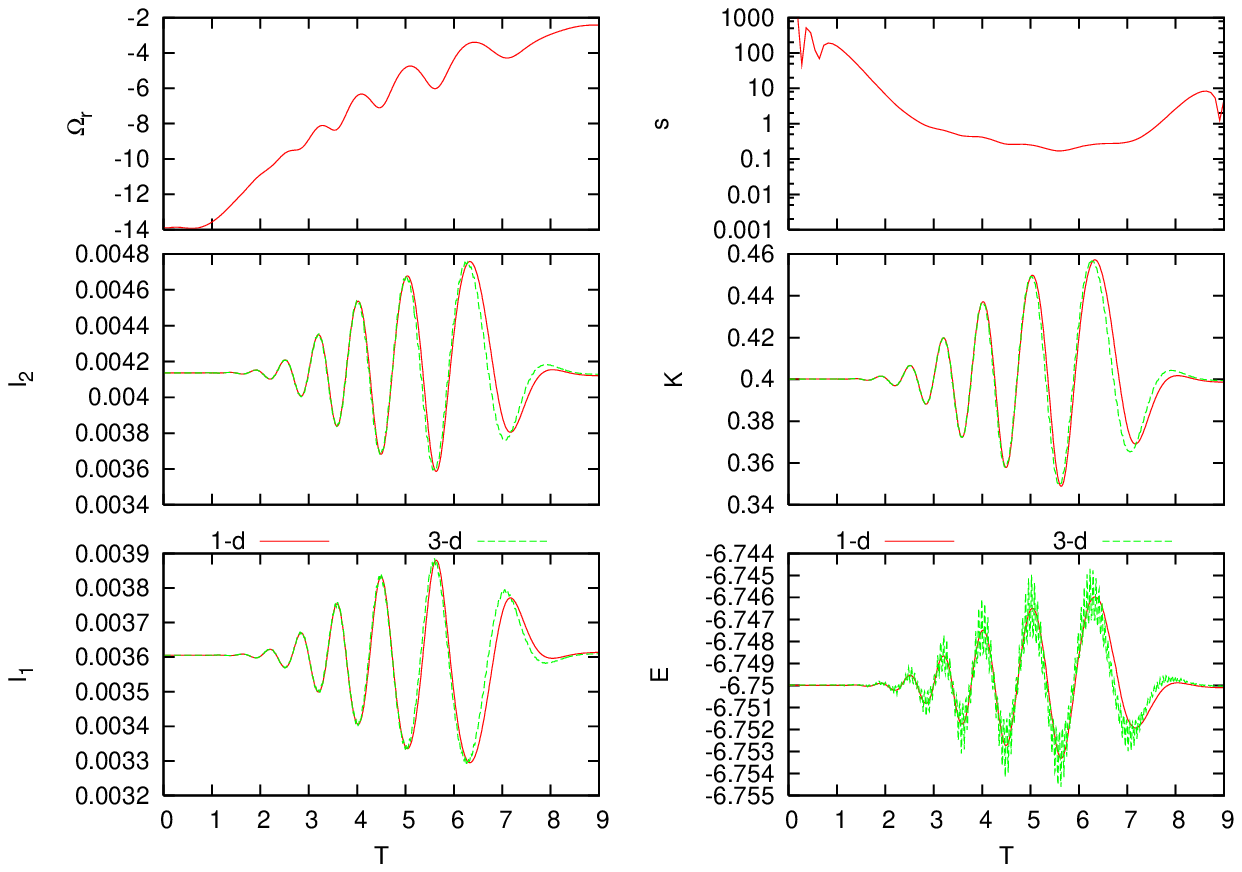}}
\subfigure[resonant]{
  \includegraphics[width=0.49\textwidth]{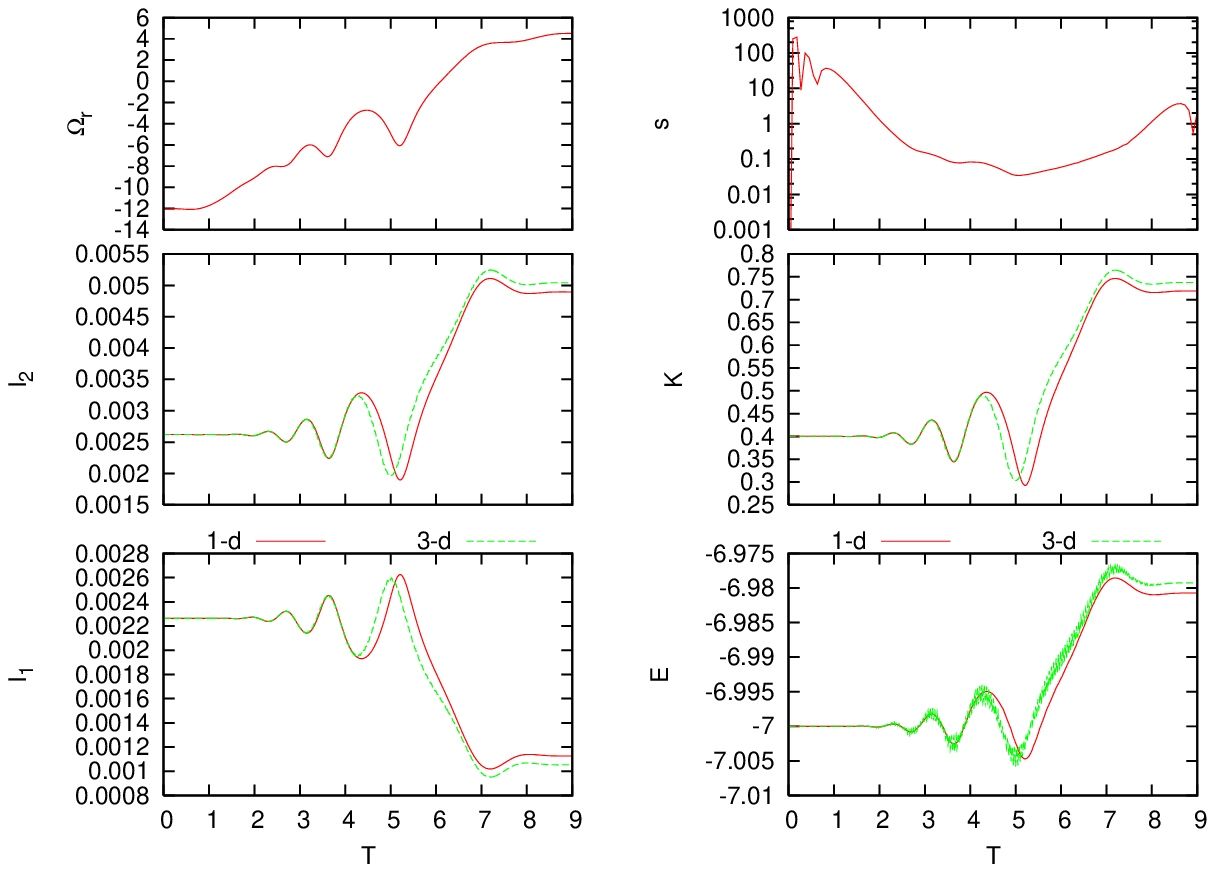}}
\caption{As in Figure \protect{\ref{fig:comporb10}} but for the
resonance $l_1=-1$ and $l_2=2$ (ILR).}
\label{fig:comporbILR}
\end{figure}

We want to compare orbit integrations using the time averaged
one-dimensional equations with those integrated directly using the
three-dimensional equations of motion in the bar quadrupole
gravitational potential.  For the one-dimensional case, we can perform
a canonical transformation to commensurate and non-commensurate
action-angle variables and average over the non-commensurate angles,
just as in the perturbation theory.  For a spherical system, this
leaves a reduced, two-degree of freedom problem: one corresponds to
the commensurate degree of freedom and one corresponds to the node or
{\em angle of ascension} of the orbital plane\footnote{This is
analogous to the {\em First point of Aries} in defining celestial
coordinates .  Clearly, this degree of freedom can not be averaged
because its unperturbed frequency is zero.}.  The conjugate action to
this angle is the $z$ component of the angular momentum.  The four
first-order ordinary differential equations, Hamilton's equations, can
then be integrated as an initial value problem independent of slow or
fast limit considerations.  For the three-dimensional case, we simply
integrate the orbit using the combined halo and quadrupole bar
potential using the quadrupole component of the bar perturbation as
described in \S\ref{sec:barpert}.

We assume an NFW dark matter halo and the bar length is the NFW halo
scale length $r_s$.  We impose a time-dependent pattern speed from an
N-body simulation with a slowing bar that has a mass that is 1\% of
the enclosed dark-matter halo mass within the bar radius, which has a
2\% force perturbation at maximum.  We choose such a weak perturbation
to make sure that the perturbation remains only weakly nonlinear in
the slow limit, where the strength scales as $|H_{1\bl}|^{1/2}$.  We
use a standard Leap Frog algorithm for the direct solution of
Hamilton's equations in Cartesian coordinates and use the implicit
symplectic algorithm to evolve the averaged equations (described in
\S\ref{sec:numavg}).  Because in the averaged equations the fast
oscillatory motion has been removed near the resonance, the time step
can be 100 times larger than for the ordinary Cartesian evolution.  At
each time step, the Cartesian phase space is transformed to the new
variables for comparison in the figures.

Figures \ref{fig:comporb10} and \ref{fig:comporbILR} compares resonant
and non-resonant orbits for the $l_1=1, l_2=0$ and the $l_1=-1, l_2=2$
(ILR) resonances integrated using the time averaged one-dimensional
perturbation theory and by directly integrating the full
three-dimensional equations of motion.  The $l_1=1, l_2=0$ resonant
orbit returns to apocenter twice during each bar period.  To simplify
the discussion, we will call this the {\em direct radial resonance}
(DRR).  An equilibrium phase-space distribution in a spherical halo
may be described by two conserved quantities.  For convenience we will
use the energy $E$ and the angular momentum scaled to the maximum for a
given energy $\kappa\equiv J/J_{max}(E) \in [0,1]$.

For DRR (Fig. \ref{fig:comporb10}), the {\em fast} variable is $w_2$
and, therefore, $I_2$ is conserved through the resonance as seen for
the one-dimensional averaged orbit.  The full three-dimensional
problem exhibits an oscillation in $I_2$ but there is no net change.
Outside of resonance, slower modulations by the resonance are seen in
both the three-dimensional and one-dimensional solutions for $I_1$.
The rapid oscillation from the fast degree of freedom is superimposed
on this slower motion in the three-dimensional solution.  Finally, the
net change $I_s=I_1$ is zero for the non-resonant orbit and non-zero for
the resonant orbit.

The overall behaviour for the ILR (Fig. \ref{fig:comporbILR})
resonance is similar although both $I_1$ and $I_2$ now both change
owing to resonance passage.  Note that both the ILR and DDR
transitions are not fully in the slow or fast regime.  Remember that
convenient analytic approximations only exist for $s\ll1$ (slow) or
$s\gg1$ (fast) and encounters with $s\approx1$, like these, cannot be
solved accurately by analytic perturbation theory but require a
numerical solution of the one-dimensional equations of motion.

\subsubsection{Comparison of one- and three-dimensional orbit
  integrations for a phase-space ensemble}
\label{sec:comp1d3d}

In this section, we compare the net changes in the halo phase space
caused by the rotating bar using both perturbation theory and the full
equations of motion.  The latter is performed using an N-body code
with a fixed halo potential.  We begin with a Monte-Carlo generated
phase space for the equilibrium distribution in Cartesian space: $x,
y, z, u, v, w$ and for each particle we perform the same orbit
calculations described in \S\ref{sec:ODE}.  The bar mass and length
and dark matter halo are as described in the previous section.
However, now the evolution of the bar pattern speed is determined by
conservation of the total angular momentum in the three-dimensional
calculation.  The pattern speed for this weak bar slows by 3\%
during the evolution.  The bar amplitude is slowly turned on and then
turned off to avoid transients.  Transients will not affect the total
torque significantly but may produce difficult to interpret
phase-space features.  The evolution of the bar pattern speed with
time is used as an input to the one-dimensional experiment.

\begin{figure}
\centering
\subfigure[1-d]{
  \includegraphics[width=0.49\textwidth]{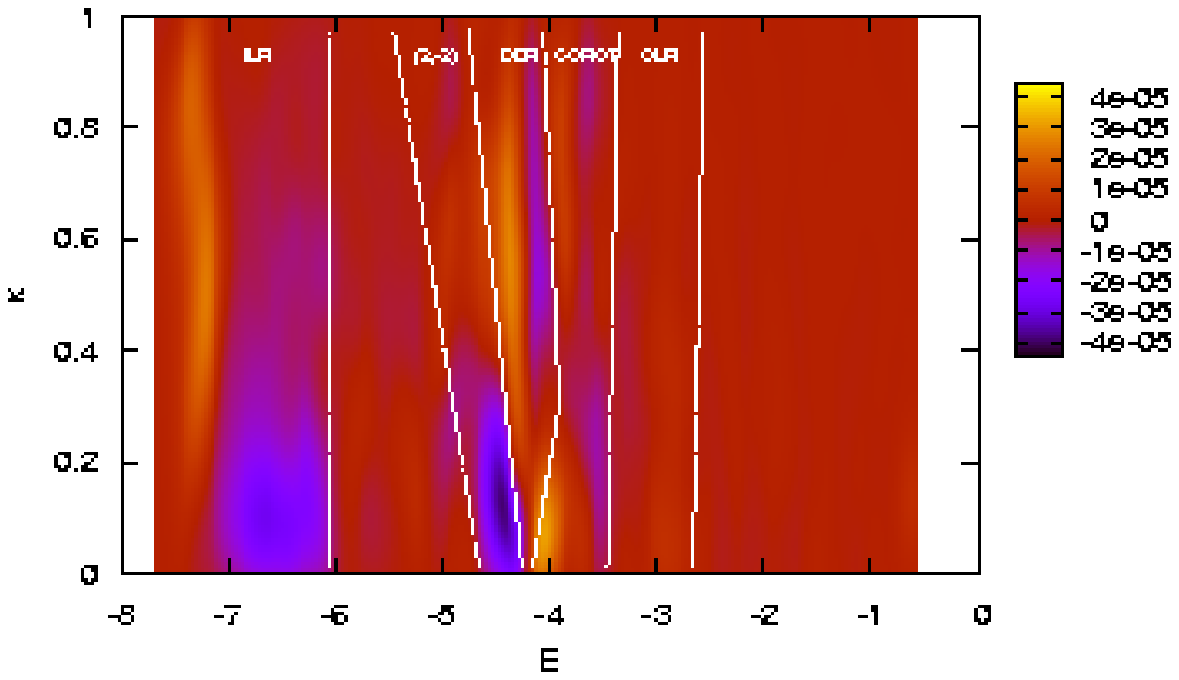}}
\subfigure[3-d]{
  \includegraphics[width=0.49\textwidth]{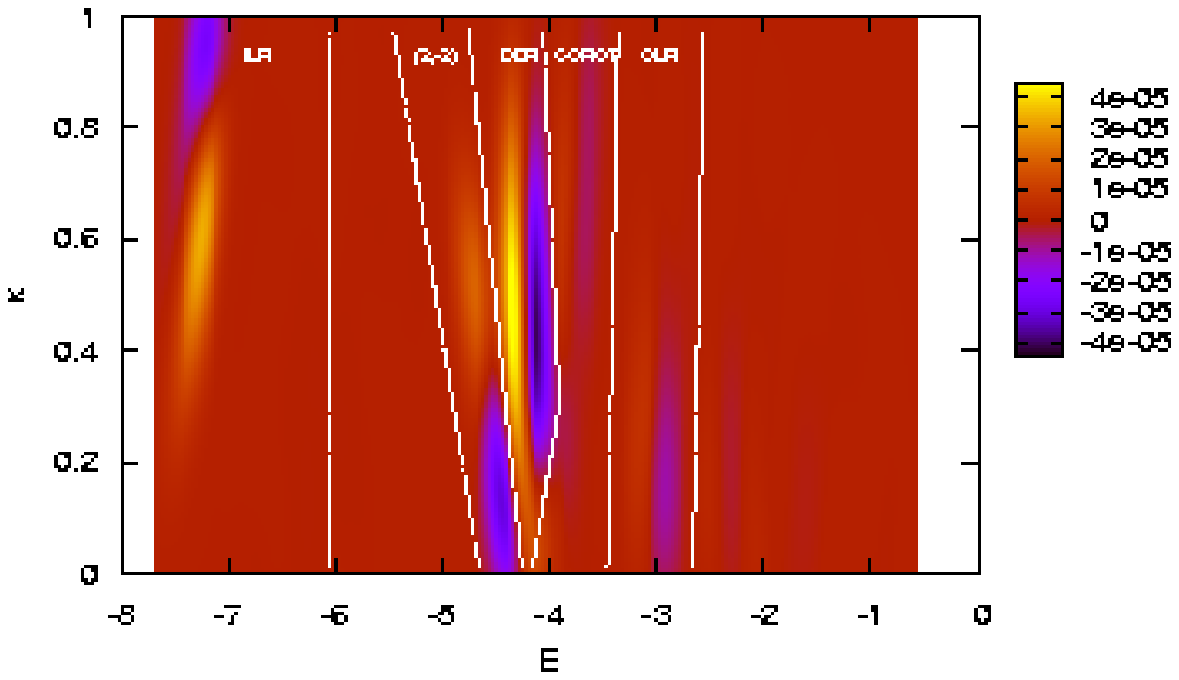}}
\caption{Distribution of changes in $L_z$ in phase space using the
  one-dimensional (a) and three-dimensional (b) equations of motion.}
\label{fig:comp1d3d}
\end{figure}

In Figure \ref{fig:comp1d3d}, we show the ensemble change in the $z$
component of the angular momentum $\Delta L_z$ during the bar evolution.
These figures are made by first computing $\Delta L_z$ for each
orbit as a function of its initial values of $E$ and $\kappa$.  We
then use kernel density estimation with cross validation
\citep{Silverman:86} to estimate the smoothing kernel.  We increase
and decrease this estimate to ensure that the resulting density field
is not over smoothed.  Using this procedure, Figure \ref{fig:comp1d3d}
compares the change in the $z$ component of the angular momentum after
evolving the phase space with the one-dimensional and
three-dimensional equations of motion.  In the one dimensional case,
each resonance must be computed separately and we use the following
five resonances (listed in order of their appearance in Figure
\ref{fig:comp1d3d}): $(l_1, l_2)=(-1, 2)$ (ILR), $(2,-2)$, $(1, 0)$
(DRR), $(0,2)$ (corotation) and $(1, 2)$ (outer Lindblad).  Figure
\ref{fig:comp1d3d}a is the sum of the five separate one-dimensional
phase space calculations.

The low-order resonances with energies larger than -5 agree within
20\% for the two cases.  However, the ILR ($E\lta-6$) appears at a
somewhat different location and magnitude in the two panels of Figure
\ref{fig:comp1d3d}.  Because the rate of pattern speed change is so
slow for such a weak bar, the orbits linger near the ILR and become
nonlinear.  A similar experiment with the same parameters but a with a
10 times more massive bar\footnote{This bar has 10\% of the
dark-matter mass inside of its radius and is a 20\% force
perturbation at peak.} has the opposite problem: the pattern speed
decreases by 50\% so the orbits are less likely to linger but the
overall amplitude is sufficiently high that the interaction itself is
nonlinear.  As in Figure \ref{fig:comp1d3d}, the resonances at $E>-5$
agree but the one-dimensional integration predicts more torque at ILR
than the three-dimensional integration.  We also tried using the 1\%
bar but artificially forcing the bar pattern speed to decrease at the
rate of the 10\% bar, which resulted in good agreement between the two
calculations at ILR.

\section{Simulating galaxies with resonances using N bodies}
\label{sec:simres}

\subsection{Description of physical and numerical artifacts}
\label{sec:simdesc}

Because Hamiltonian perturbation theory is impractical for complicated
astronomical problems and inappropriate for large perturbations such
as major mergers, in the end one must ``throw caution to the wind''
and resort to exploring the dynamics using N-body simulations.  Using
the development from \S\ref{sec:hamilton}, we can derive the
requirements necessary to correctly simulate resonance dynamics.
There are three requirements on the particle number.  First, we
require a sufficient number of particles near the resonance to produce
the first-order cancellation demanded by the second-order perturbation
mechanism (Criterion 1).  If you like, an N-body simulation does the
phase averaging by Monte-Carlo integration and this criterion ensures
convergence of this integral.  Second, there must be enough particles
so that the time for an orbit to artificially diffuse across the
resonance is long compared to the characteristic time scale of a
closed orbit in the resonance potential. We separately consider two
regimes.  Artificial non-astronomical diffusion is caused by small
scale graininess in the potential owing to the gravitational force
from individual particles (Criterion 2).  Similarly, artificial
diffusion can be caused by potential fluctuations from Poisson noise
at large spatial scales (Criterion 3).  In addition, Criterion 3 can
describe true astronomical noise.  Numerical integration errors can
also give rise to a similar diffusion, and hence the integration must
be performed accurately.

There is a final criterion that does not depend explicitly on particle
number: the potential solver must be able to resolve scales smaller
than the resonance potential (Criterion 0) and, of course, the
realized phase-space distribution must cover this region.  Clearly,
this criterion must be satisfied by construction given the resonance
potential from equation (\ref{eq:v1exp}) for the set of desired
resonances described by $\bl$ and $\Omega_p$.

We investigate each of these issues using the perturbation theory
developed in \S\ref{sec:solve} to estimate the particle number criteria in
\S\ref{sec:criteria}, derive scaling formula in \S\ref{sec:calib} and then
calibrate these against N-body simulations in \S\ref{sec:results}.

\subsection{Using perturbation theory to investigate particle number
  requirements}
\label{sec:criteria}

\subsubsection{Coverage}

\begin{figure}
\subfigure[$l_1=1, l_2=0$]{%
  \includegraphics[width=0.49\textwidth]{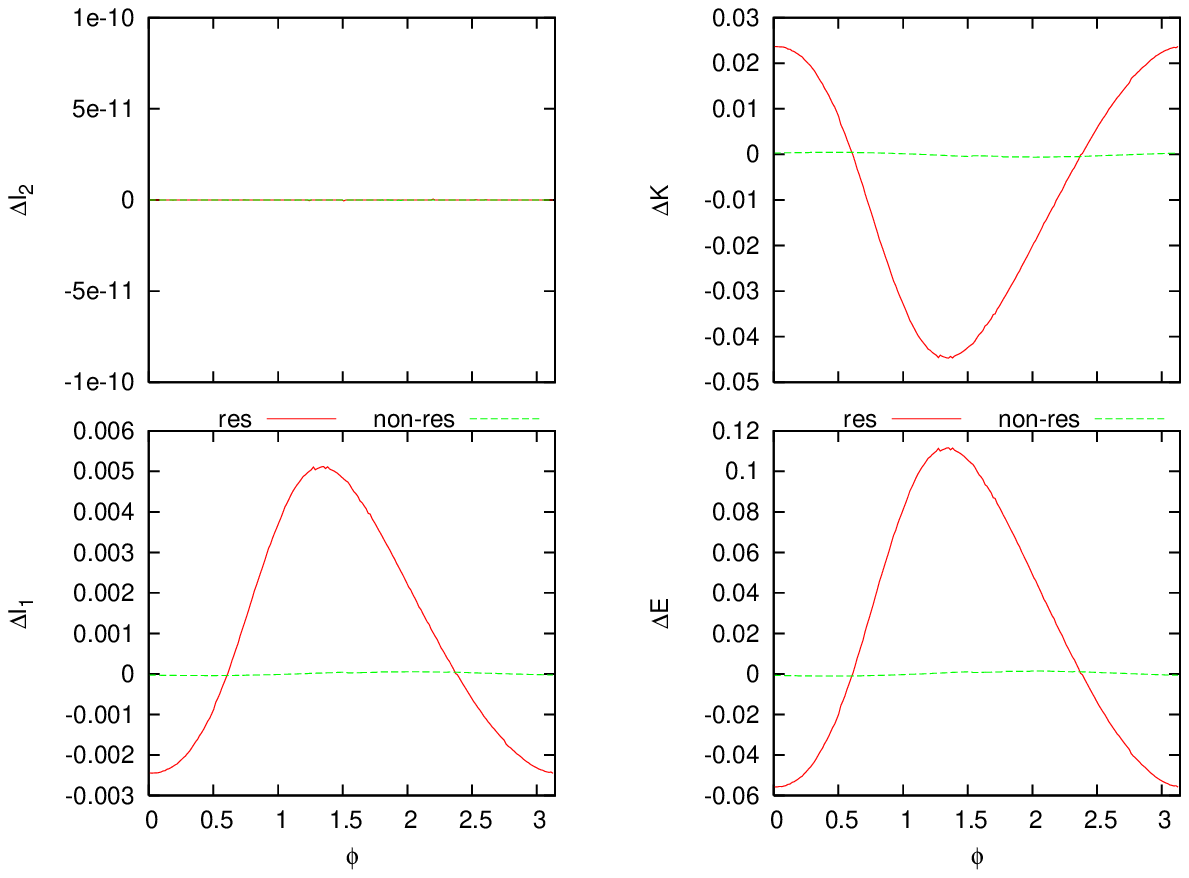}}
\subfigure[$l_1=-1, l_2=2$]{%
  \includegraphics[width=0.49\textwidth]{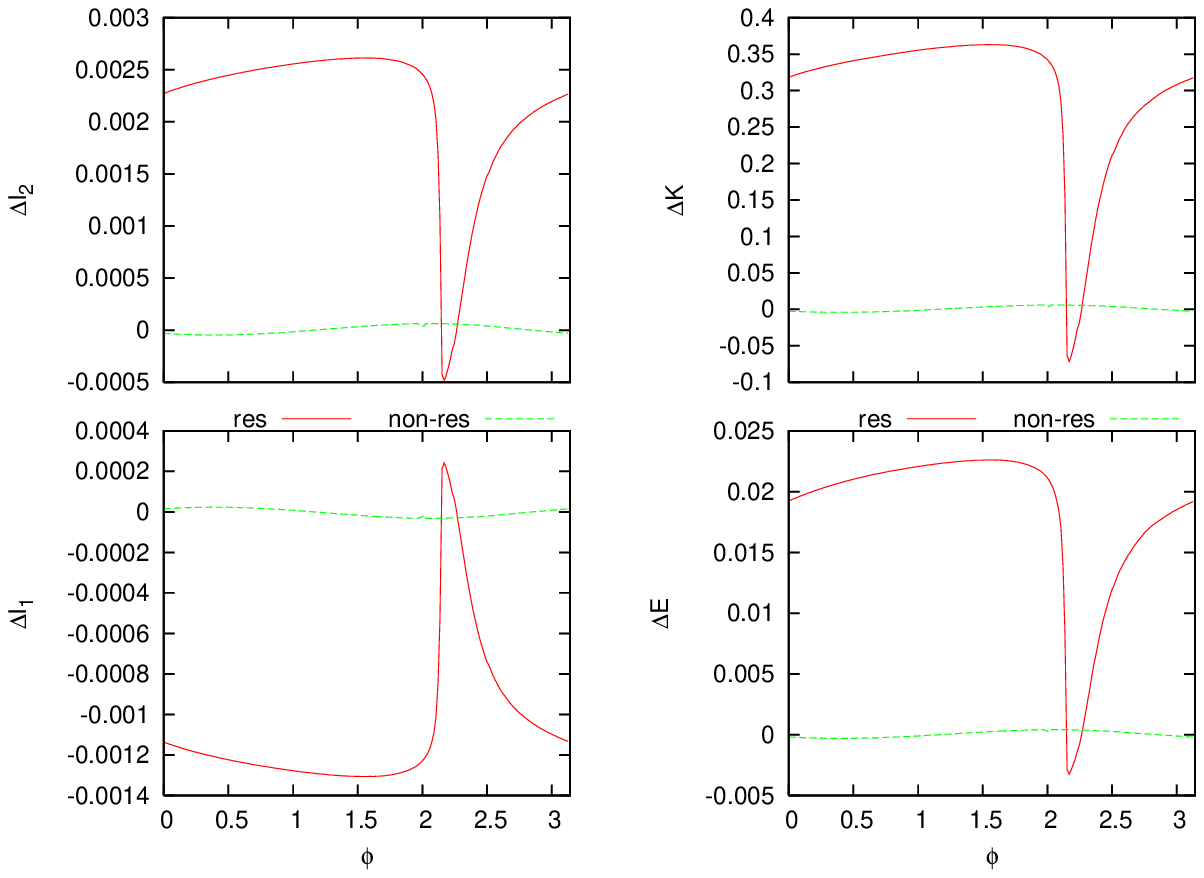}} 
\caption{Variation in conserved-quantity change with bar phase $\phi$
  after evolution in pattern speed for the same orbits shown in
  Figures \protect{\ref{fig:comporb10}} and
  \protect{\ref{fig:comporbILR}}.  The red line is the resonant orbit
  and the green line is the non-resonant orbit.}
\label{fig:vphase}
\end{figure}

The change in the actions $I_1$ and $I_2$ as a function of bar phase
$\phi$ for an ensemble of orbits starting with fixed actions
(e.g. energy, angular momentum, and plane inclination) is shown in
Figure \ref{fig:vphase} for DRR and ILR.  $E$ and $\kappa$ show a
non-sinusoid variation with initial bar phase $\phi$ for orbits that
pass through resonance.  Orbits gain or lose angular momentum and
energy depending on the bar phase at resonance. If one averages over
all the phases there is a net change, as expected.  There is also a
small purely sinusoidal modulation in these quantities for the
non-resonant orbits owing to the rotating bar perturbation, which is
only barely perceptible in these figures.

This example further illustrates the brief statement in
\S\ref{sec:simdesc} that a simulation must have sufficient particles
to cover all the phases shown in Figure \ref{fig:vphase} or one will
not get the correct ensemble average.  The two resonances are excited
by the same bar perturbation but the relative amplitude is much larger
for ILR than for DRR giving ILR a peakier profile.  The peaky profile
requires fewer particles to converge to the mean than the only
slightly asymmetric DRR.  However, the required particle number may
still need to be high to cover ILR adequately owing to the small total
mass at low energies. Clearly, the coverage criterion depends on the
perturbation amplitude of each resonance separately.

\begin{figure}
\subfigure[$N=10^7$, DRR (1,0)]{%
  \includegraphics[width=0.33\textwidth]{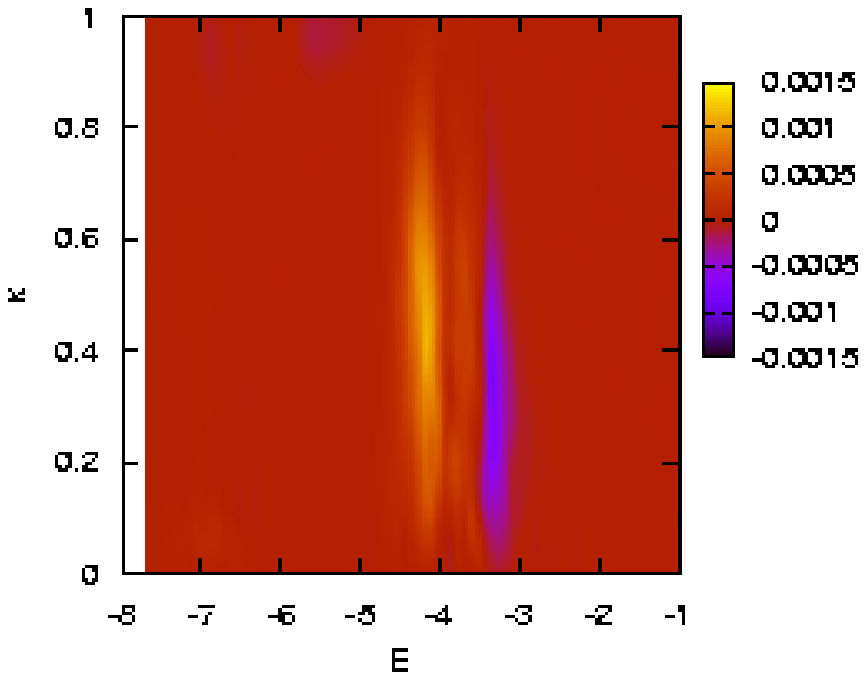}} 
\subfigure[$N=10^6$, DRR (1,0) ]{%
  \includegraphics[width=0.33\textwidth]{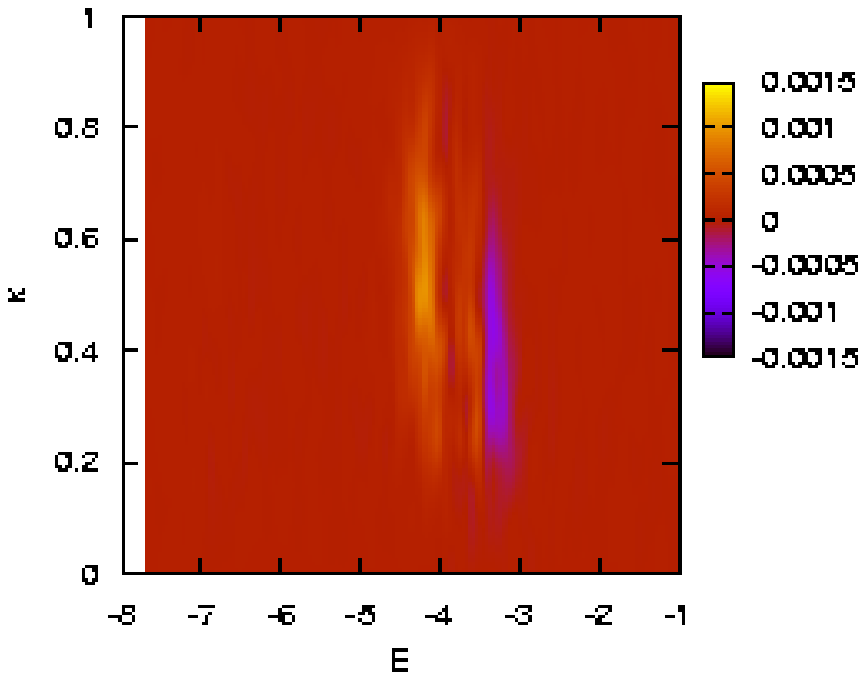}} 
\subfigure[$N=3\times10^5$, DRR (1,0)]{%
  \includegraphics[width=0.33\textwidth]{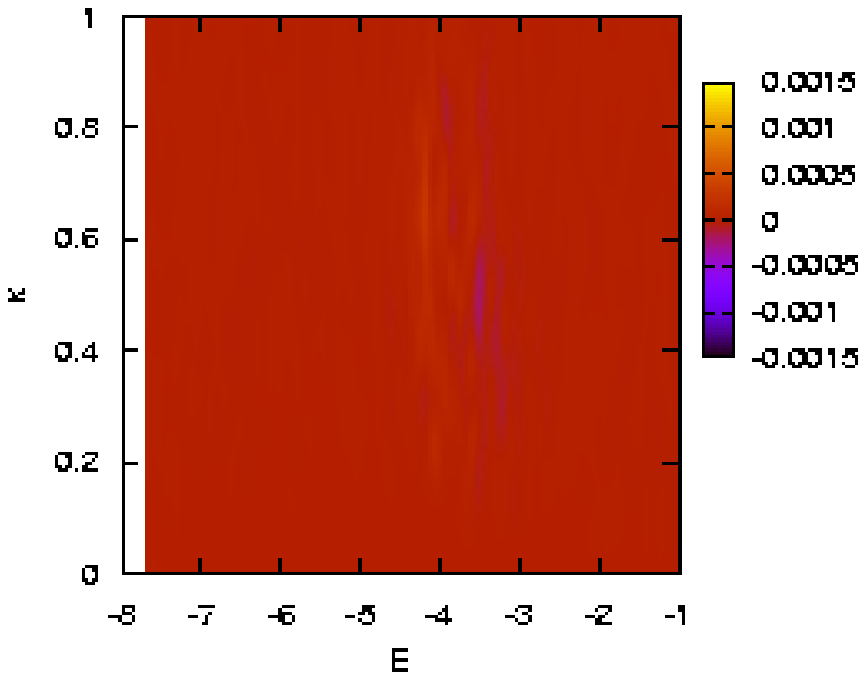}} 
\subfigure[$N=6\times10^7$,ILR (-1,2) ]{%
  \includegraphics[width=0.33\textwidth]{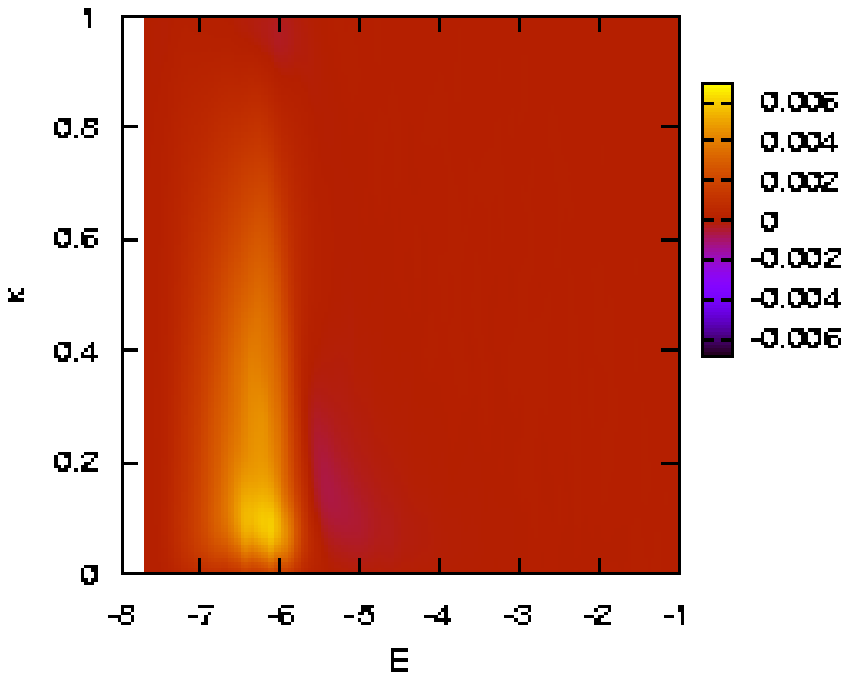}} 
\subfigure[$N=6\times10^6$,ILR (-1,2) ]{%
  \includegraphics[width=0.33\textwidth]{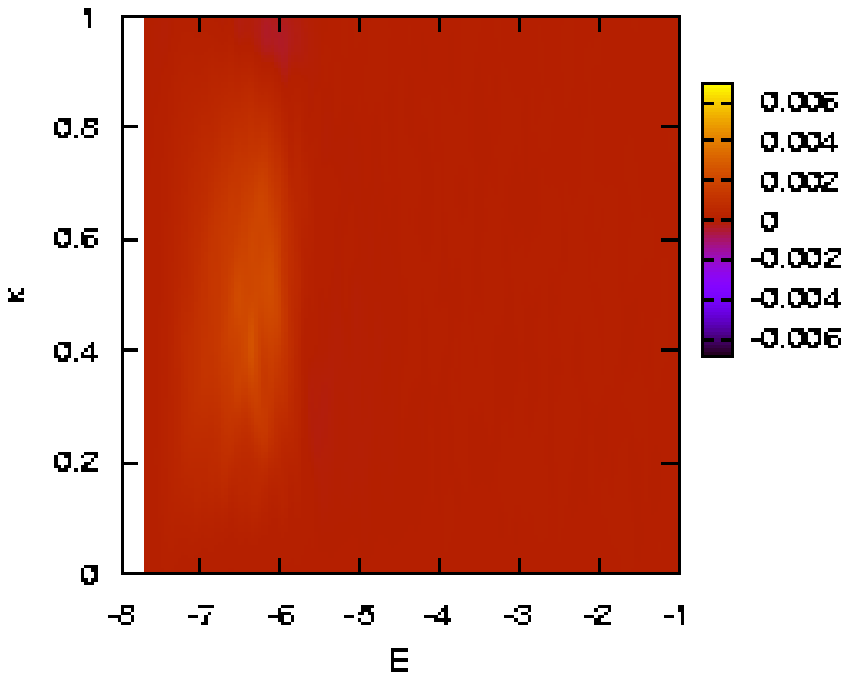}} 
\subfigure[$N=1.8\times10^6$, ILR (-1,2)]{%
  \includegraphics[width=0.33\textwidth]{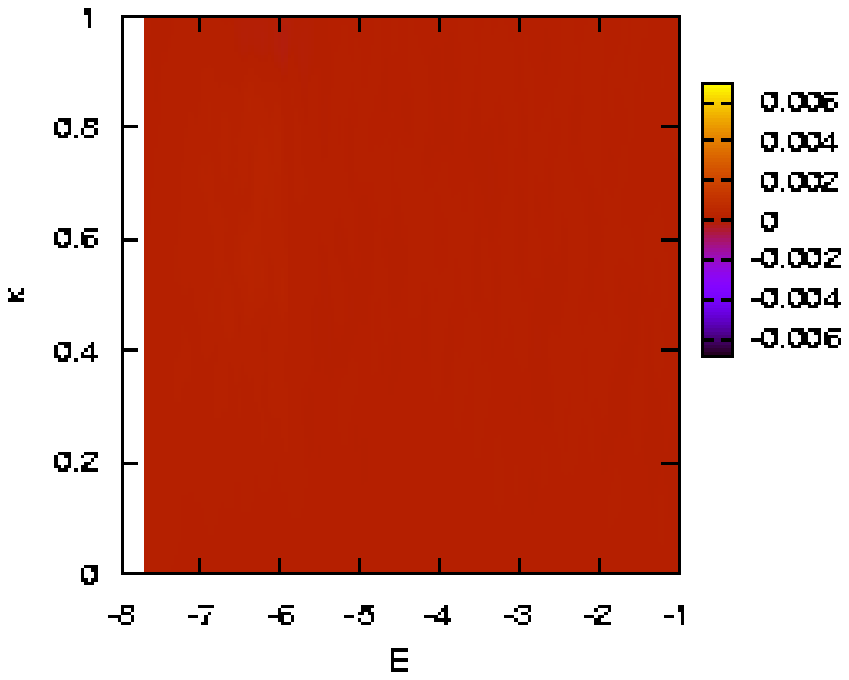}} 
\caption{Distribution of change in the $z$ component of angular
  momentum for a full phase ensemble about the orbits described in
  Figure \protect{\ref{fig:vphase}}.}
\label{fig:distlz}
\end{figure}

In addition, the orbits in Figure \ref{fig:vphase} have identical
initial inclinations near the peak of the contribution for a given
energy $E$ and total angular momentum $J$.  The full ensemble in an
N-body simulation will sample all inclinations and further dilute the
net signal from the resonance.  One can see the full phase space
ensemble result about the initial orbit in Figure \ref{fig:distlz}
where we plot $\Delta L_z$ as a function of initial phase space
coordinate ($E$,$\kappa$) as in Figure \ref{fig:comp1d3d}. The
particle number was increased for both resonances until the amplitude
and location of the resonance features in the $E$--$\kappa$ plane
remained unchanged.  If we decrease the number of particles below this value,
the sampling is insufficient in the vicinity of the resonance to recover
the full amplitude of the resonance.  More than $10^6$ ($6\times10^6$)
equal mass particles within the virial radius are needed for the DRR
(ILR) resonance.  Note that we have chosen a very large bar length to
reduce the required number of particles.  We will see in
\S\ref{sec:results} that ILR for a typical strength scale-length sized
bar requires $> 10^8$ particles!

\subsubsection{Diffusion}
\label{sec:diffu}

We investigate the effects of two-body perturbations on an orbit near
resonance by combining the twist mapping, the solutions to the
one-dimensional, and the three-dimensional equations of motion with a
Monte-Carlo simulation of the Fokker-Planck equation.  The diffusion
coefficients are derived from the isotropic phase-space
distribution for a spherical equilibrium using the standard
prescription \citep[e.g.][]{Spitzer:87,Binney.Tremaine:87}:
\begin{eqnarray}
  \delta v_\parallel &=& \langle\Delta v_\parallel\rangle\delta T + 
  {\cal R}_1(\langle\Delta v_\parallel^2\rangle\delta T)^{1/2}
  \nonumber \\
  \delta v_{\perp\,1} &=& {\cal R}_2(\langle\Delta v_\perp^2\rangle\delta T)^{1/2}
  \cos(2\pi{\cal R}_3) \nonumber \\
  \delta v_{\perp\,2} &=& {\cal R}_2(\langle\Delta
  v_\perp^2\rangle\delta T)^{1/2} \sin(2\pi{\cal R}_3)
  \label{eq:diffmc}
\end{eqnarray}
where $\delta T$ is the time step, ${\cal R}_1$ and ${\cal R}_2$ are
unit-variance zero-mean normally distributed random variates and
${\cal R}_3$ is a random variate uniformly distributed in the unit
interval.  We adopt a constant value of $\ln\Lambda$ with
$\Lambda=b_{max}/b_{min}=1/0.003$.  This conservative value is smaller
than that of state-of-the-art $\Lambda$CDM simulations and corresponds
to a gravitational softening length of about 0.3\% of the virial
radius.  The reader may scale the values of the particle mass $m$ for
any value of $\ln\Lambda$ by keeping the product $m\ln\lambda$ fixed.
The algorithm for applying equation (\ref{eq:diffmc}) is as follows:
(1) at the end of every twist-mapping iteration or every time step for
the symplectic integration of the one-dimensional average equations,
we transform the action-angle variables to Cartesian coordinates; (2)
equation (\ref{eq:diffmc}) is applied to the velocities; and (3) the
Cartesian coordinates are transformed back to actions and angles.

\begin{figure}
  \subfigure[No
  diffusion]{\includegraphics[width=0.5\textwidth]{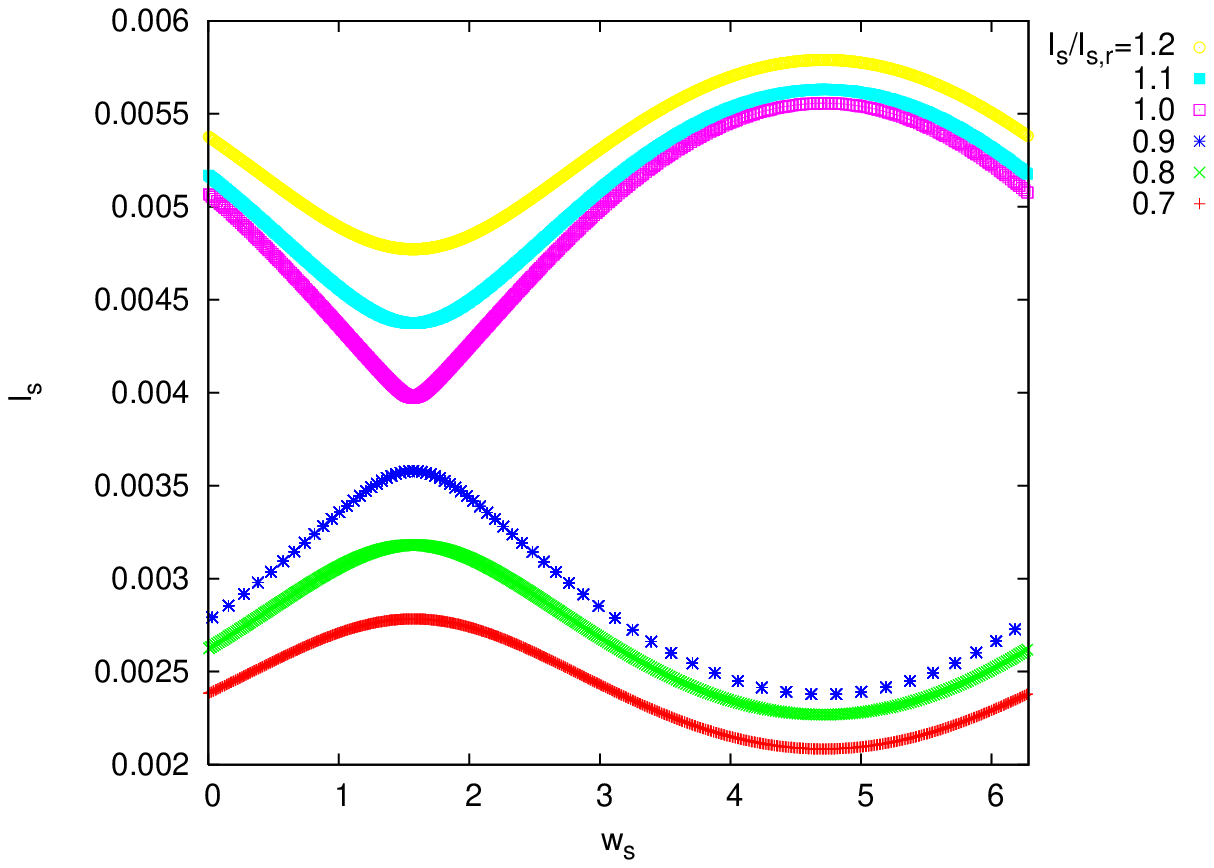}}
  \subfigure[With
  diffusion]{\includegraphics[width=0.5\textwidth]{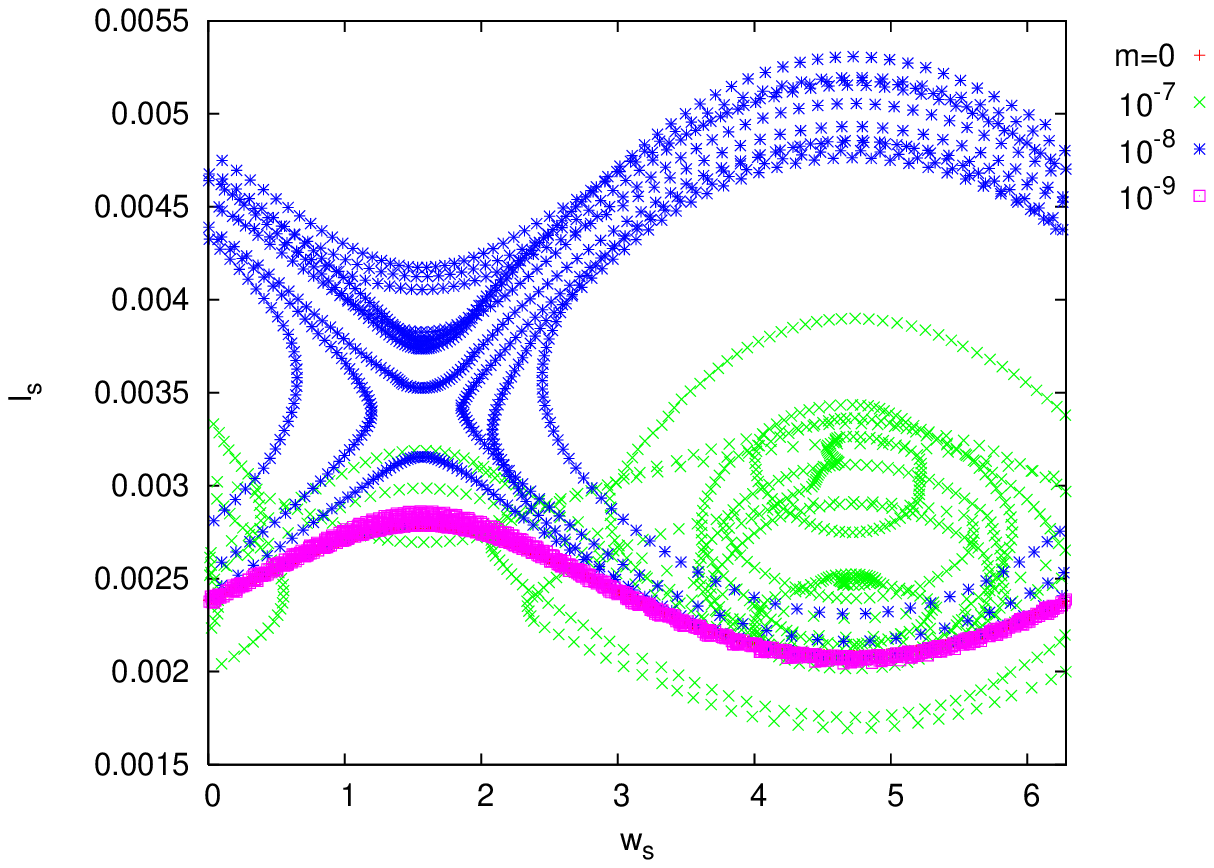}}
  \caption{Panel (a) shows orbits on either side of resonance without
  two-body interactions.  The ratio $I_s/I_s{s,r}$ is the ratio of
  the initial action to the action of the zero-amplitude homoclinic
  trajectory.  Panel (b) shows the orbit from Panel (a) with
  $I_s/I_s{s,r}=0.7$ including two-body interactions
  for a variety of particle masses $m$.}
  \label{fig:twist}
\end{figure}

Figure \ref{fig:twist}a shows the surface of section of the same orbit
near ILR described in the previous figures in the one-dimension slow
variables $I_s$--$w_s$ plane with no small scale diffusion.  The bar
pattern speed is constant.  We label the orbits by $I_s/I_{s,r}$ where
$I_{s,r}$ is the value of the action at resonance for a zero amplitude
bar. Since the amplitude here is not zero, the homoclinic trajectory
is offset slightly from $I_s=I_{s,r}$.  In Figure \ref{fig:twist}a one
can also see how the twist mapping demarcates the expected pendulum
topology. Figure \ref{fig:twist}b begins with the orbit
$I_s=0.7I_{s,r}$ but adds two-body diffusion corresponding to particle
masses of $10^{-7}$, $10^{-8}$, and $10^{-9}$ in units where the total
halo mass is unity.  For $m=10^{-8}$, diffusion becomes very important
and the orbit diffuses to larger $I_s$ and even crosses the homoclinic
trajectory.  For $m=10^{-7}$, the diffusion is so large that the orbit
diffuses into the libration zone.

\begin{figure}
\centering
\subfigure[$m=10^{-6}$]{
  \includegraphics[width=0.49\textwidth]{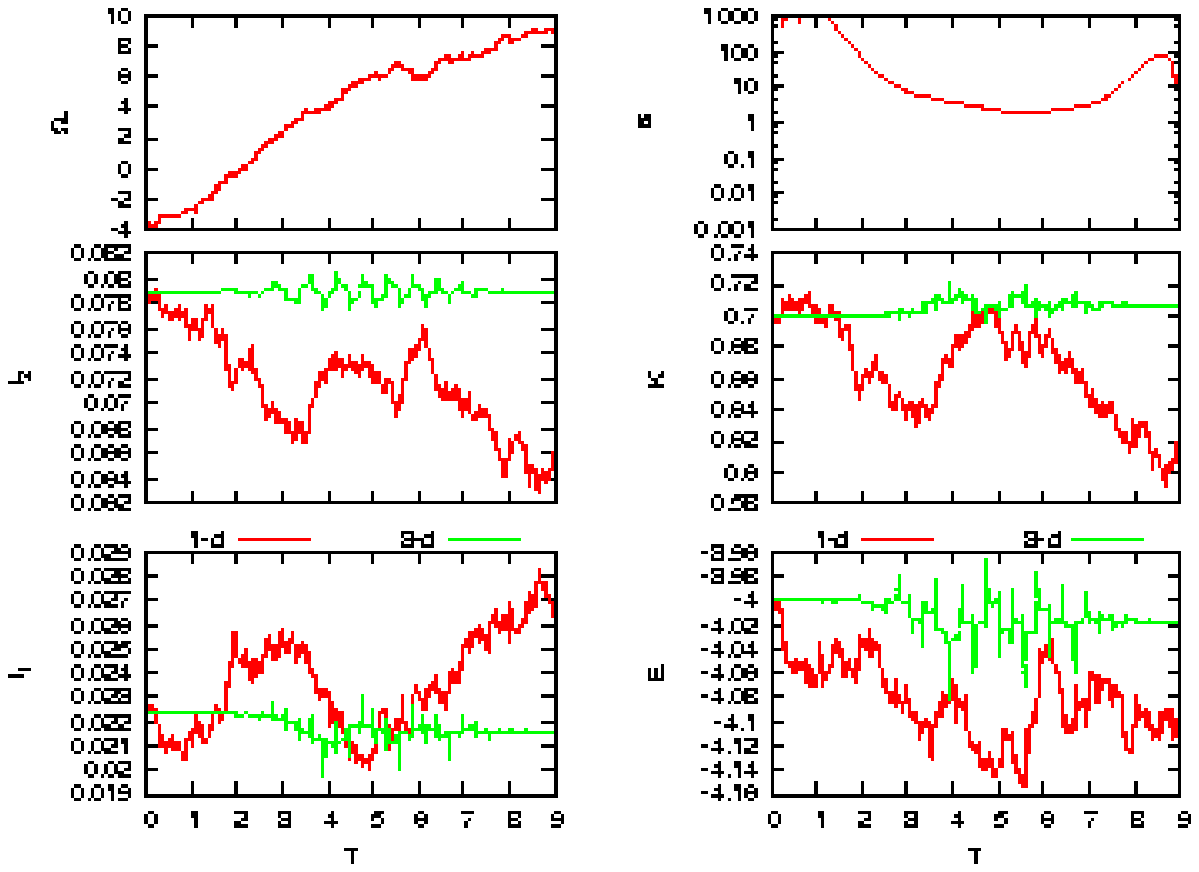}}
\subfigure[$m=10^{-7}$]{
  \includegraphics[width=0.49\textwidth]{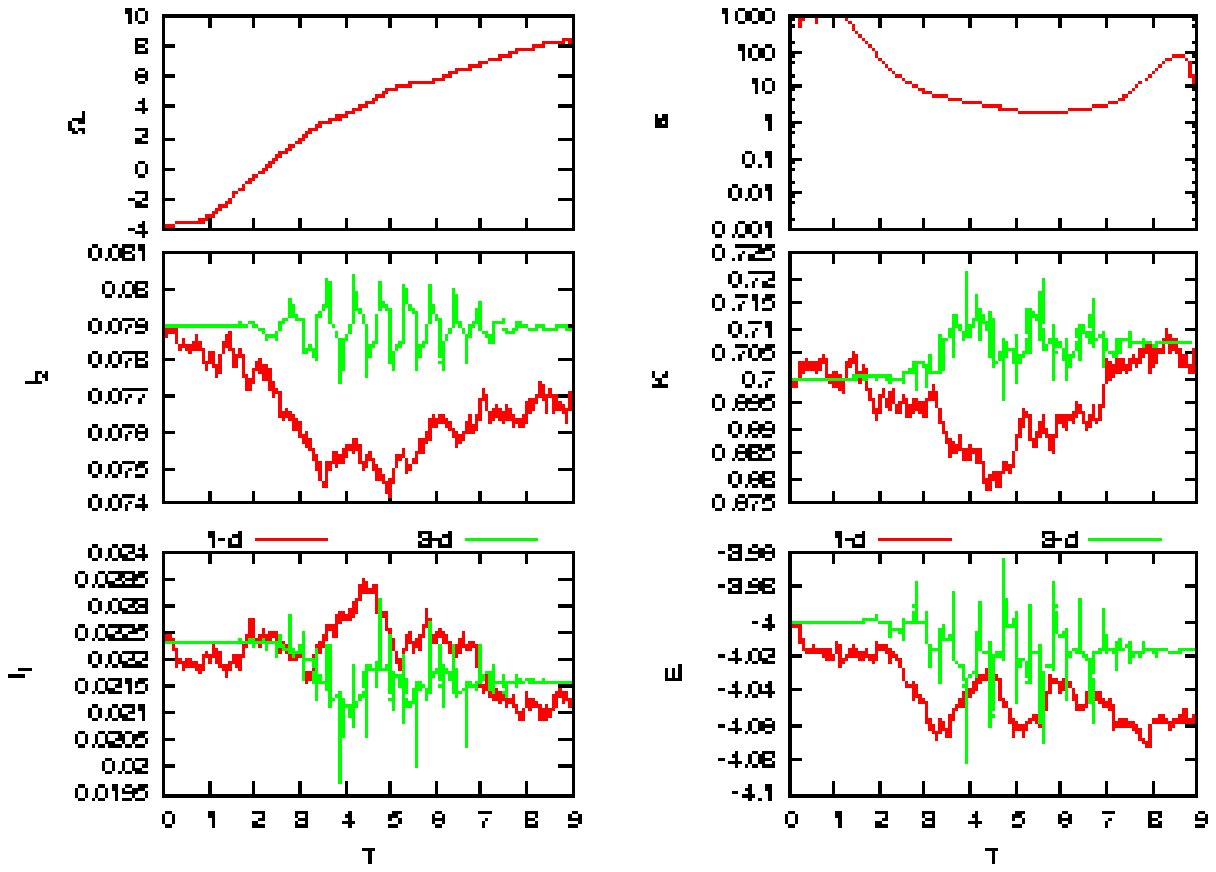}}
\subfigure[$m=10^{-8}$]{
  \includegraphics[width=0.49\textwidth]{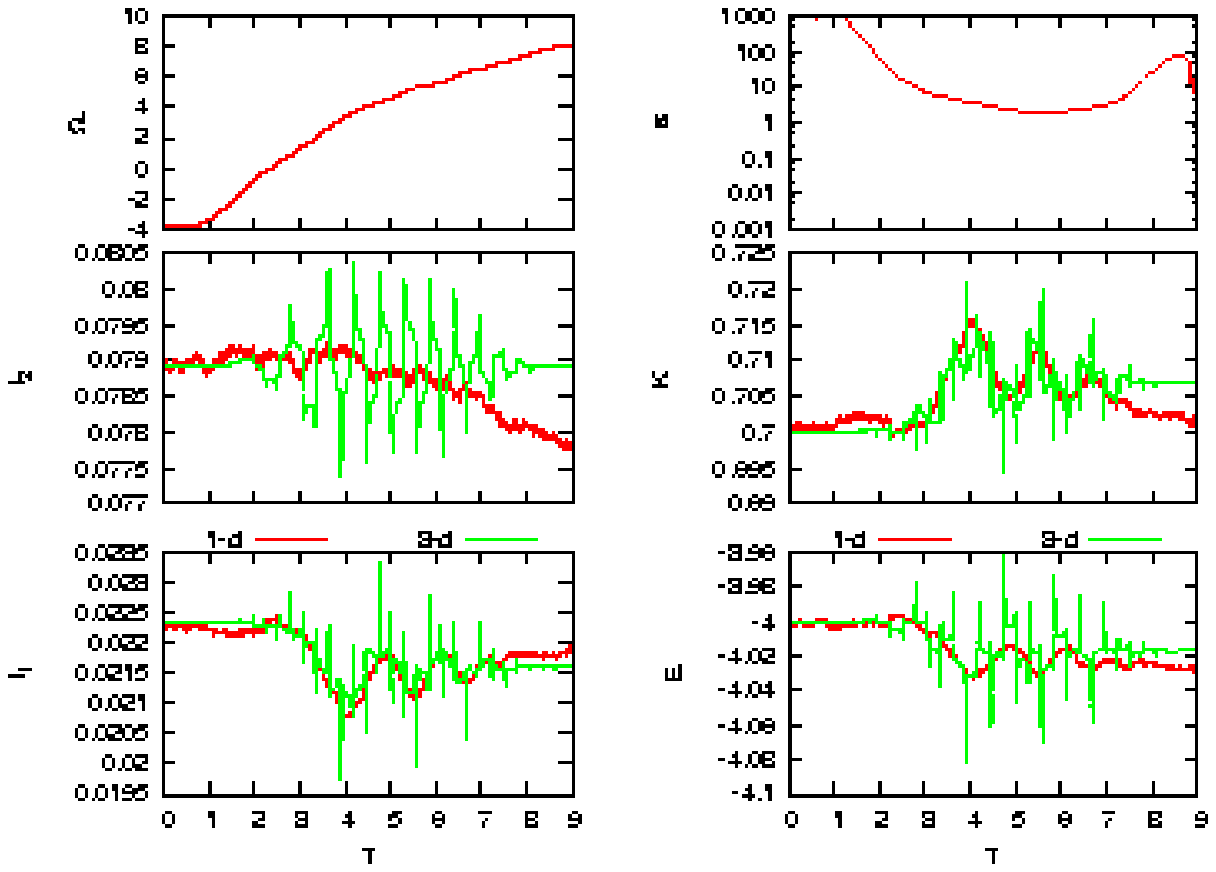}}
\caption{As in Figure \protect{\ref{fig:comporb10}} (DDR) but
  including two-body diffusion for the particle masses indicated in
  units of the total mass.}
\label{fig:tbodyorb10}
\end{figure}

\begin{figure}
\centering
\subfigure[$m=10^{-6}$]{
  \includegraphics[width=0.49\textwidth]{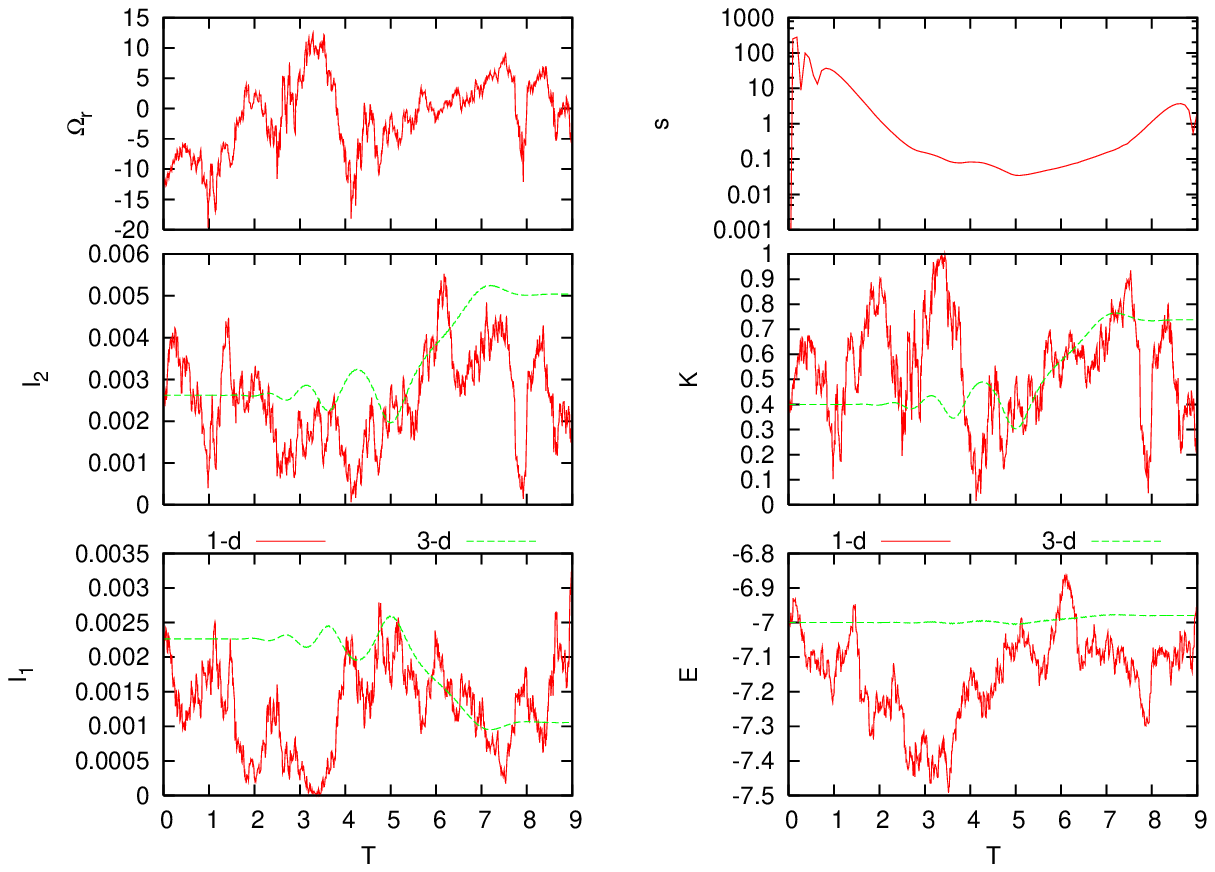}}
\subfigure[$m=10^{-7}$]{
  \includegraphics[width=0.49\textwidth]{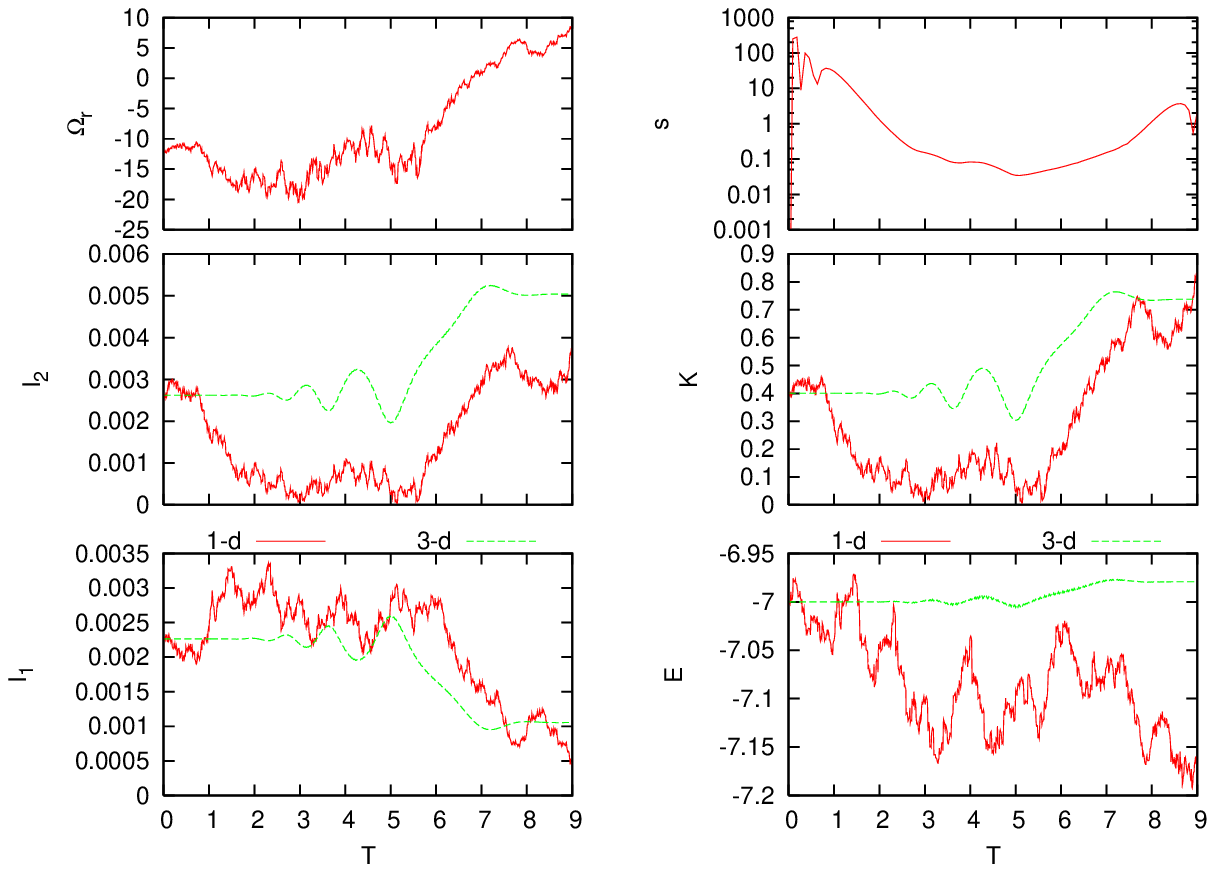}}
\subfigure[$m=10^{-8}$]{
  \includegraphics[width=0.49\textwidth]{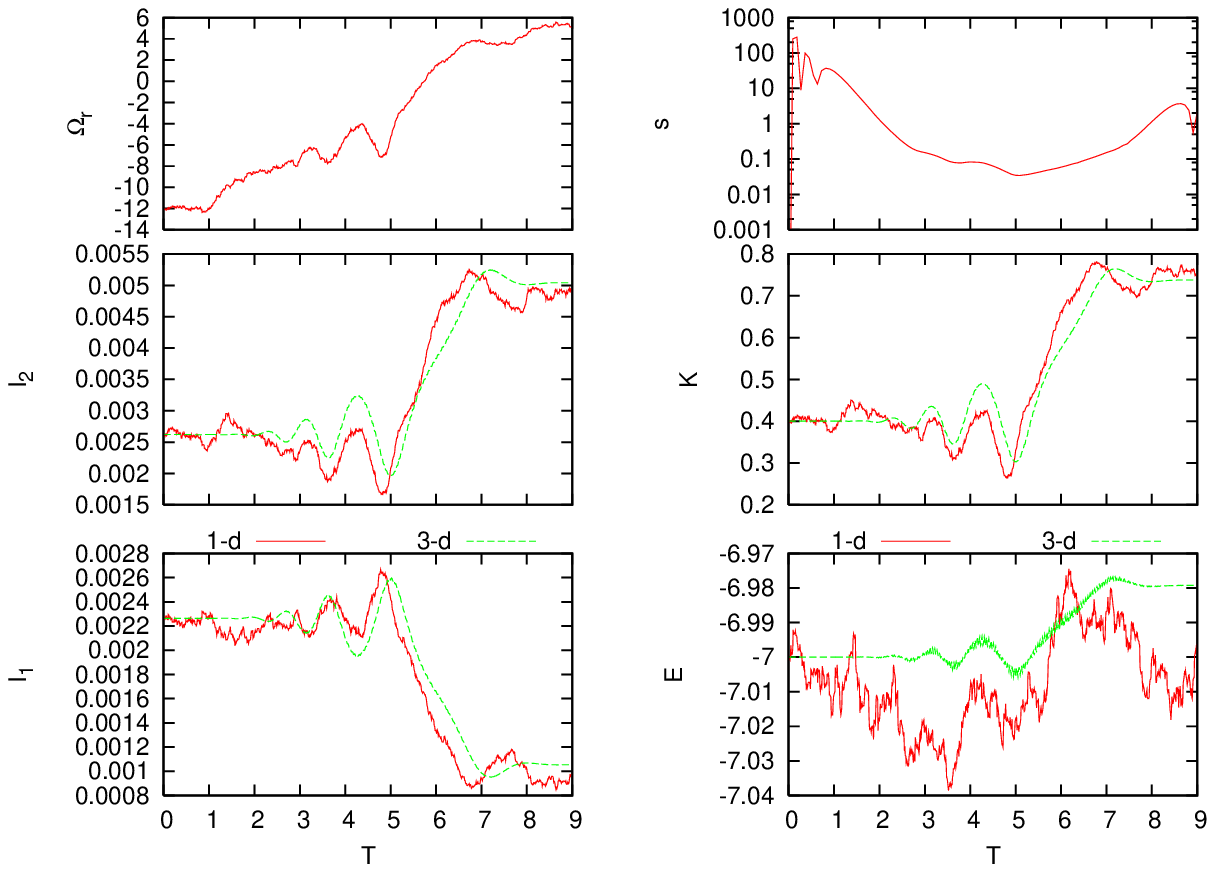}}
\caption{As in Figure \protect{\ref{fig:tbodyorb10}} (ILR) but for the
orbit in Figure \protect{\ref{fig:comporbILR}}.}
\label{fig:tbodyorbILR}
\end{figure}

Similarly, we may solve the one-dimensional equations of motion
including the two-body diffusion for the same test orbits shown in
Figures \ref{fig:comporb10} and \ref{fig:comporbILR}; these are shown
in Figures \ref{fig:tbodyorb10} and \ref{fig:tbodyorbILR},
respectively, for particle masses of $m=10^{-6}$, $10^{-7}$, and
$10^{-8}$.  The two-body fluctuations are only included in the
one-dimensional solution, the three-dimensional solution is left
unchanged for comparison.  Only at $m=10^{-8}$ does the two-body
perturbed solution begin to follow the unperturbed solution.  This is
consistent with our twist mapping results.  In short, {\em values of
$m$ equal to or smaller than those of typical N-body simulations
destroy resonances.}

\begin{figure}
\centering
\subfigure[$m=0$]{
  \includegraphics[width=0.49\textwidth]{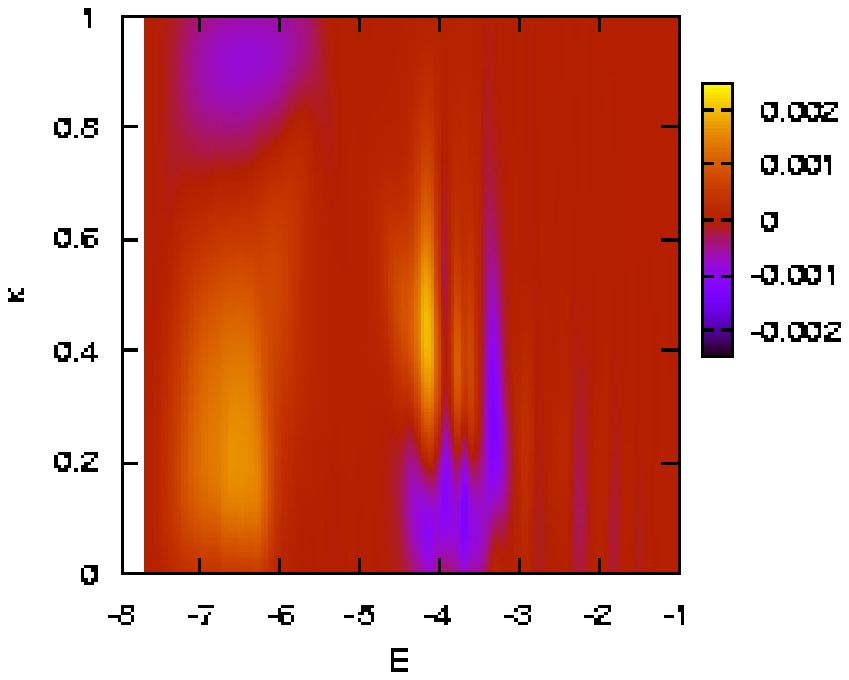}}
\subfigure[$m=10^{-6}$]{
  \includegraphics[width=0.49\textwidth]{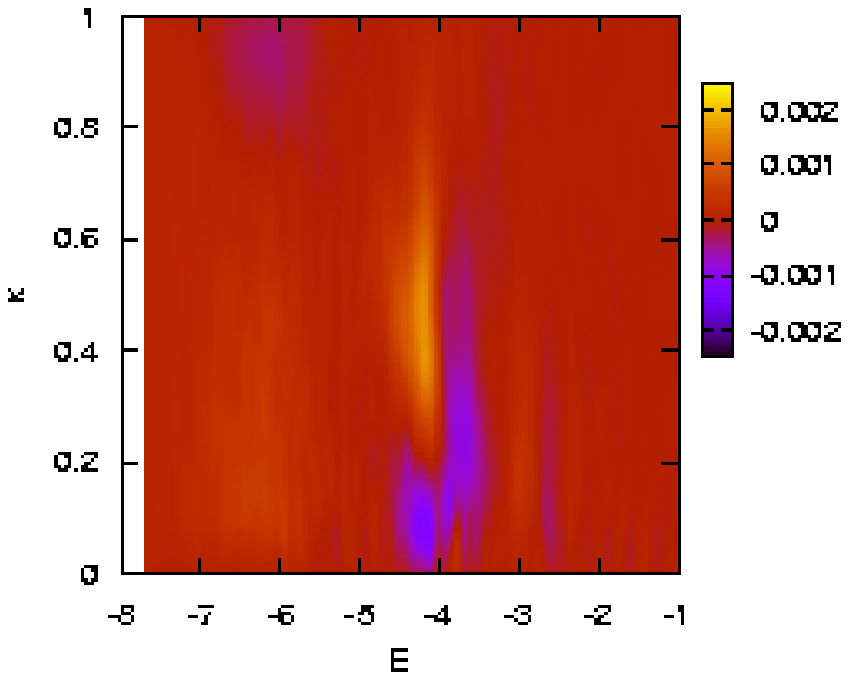}}
\caption{Distribution of $\Delta L_z$ from direct integration results
  of the three-dimensional equations of motion for an evolving
  rotating bar without (a) and with (b) two-body diffusion for
  $m=10^{-6}$.}
\label{fig:tbodyorbTOT}
\end{figure}

Figure \ref{fig:tbodyorbTOT} shows the results of an integration using
the three-dimensional equations of motion for an ensemble
$6\times10^7$ particles in the quadrupole bar perturbation both
without and with two-body diffusion equivalent to $10^6$
particles. This isolates the effects of diffusion from those of
coverage. With two-body diffusion, the inner ILR is significantly
reduced. However, the outer low-order resonances, $l_1=1, l_2=0$ in
particular, still appear at their predicted strength and location.
How can we understand this in the context of the large changes in
orbital structure near resonances we saw in the twist mapping and
one-dimensional solutions?  Even though the random walk in actions
seen in Figures \ref{fig:twist}, \ref{fig:tbodyorb10} and
\ref{fig:tbodyorbILR} may be larger than the width of the resonance,
the pattern speed of the rotating potential will continue to evolve
and sweep past the `walking' orbit at some nearby action whenever
$l_1\Omega_r + l_2\Omega_\phi= m\Omega_p$.  At this point, the
adiabatic invariant corresponding to the slow motion will break and
the orbit will feel a coherent `kick'.  However, orbit cannot linger
near the homoclinic trajectory owing to the random walk and will,
therefore, always be in the {\em fast limit}.  In other words, the
two-body noise diffusion of N-body particle-particle simulations, which is
much larger than that in true dark matter haloes, is sufficient to
cause {\em slow limit} transitions to become {\em fast limit}
transitions.

Note that the speed parameter is fast ($s>1$) for the $l_1=1, l_2=0$
resonance (DDR) (Fig. \ref{fig:comporb10}) but slow ($s<1$) for the
ILR (Fig. \ref{fig:comporbILR}).  The clear difference in magnitude
between the ILR in the slow limit (Fig. \ref{fig:tbodyorbTOT}a) and
when it is forced into the fast limit by noise
(Fig. \ref{fig:tbodyorbTOT}b) is a consequence of the smaller changes
in angular momentum for a fast-limit encounter.  The corotation
resonance ($l_1=0, l_2=2$) appears at $E\approx-3.7$, immediately to
the right of the $l_1=1, l_2=0$ resonance and at lower amplitude.
This resonance has $s\approx1$ and its amplitude is also diminished by
the diffusive effects of two-body encounters.  Only the amplitude of
the DRR is preserved.

\subsubsection{Fluctuations on large spatial scales}
\label{sec:fluct}

\begin{figure}
\centering
\subfigure[$m=10^{-5}$, DRR (1,0)]{
  \includegraphics[width=0.49\textwidth]{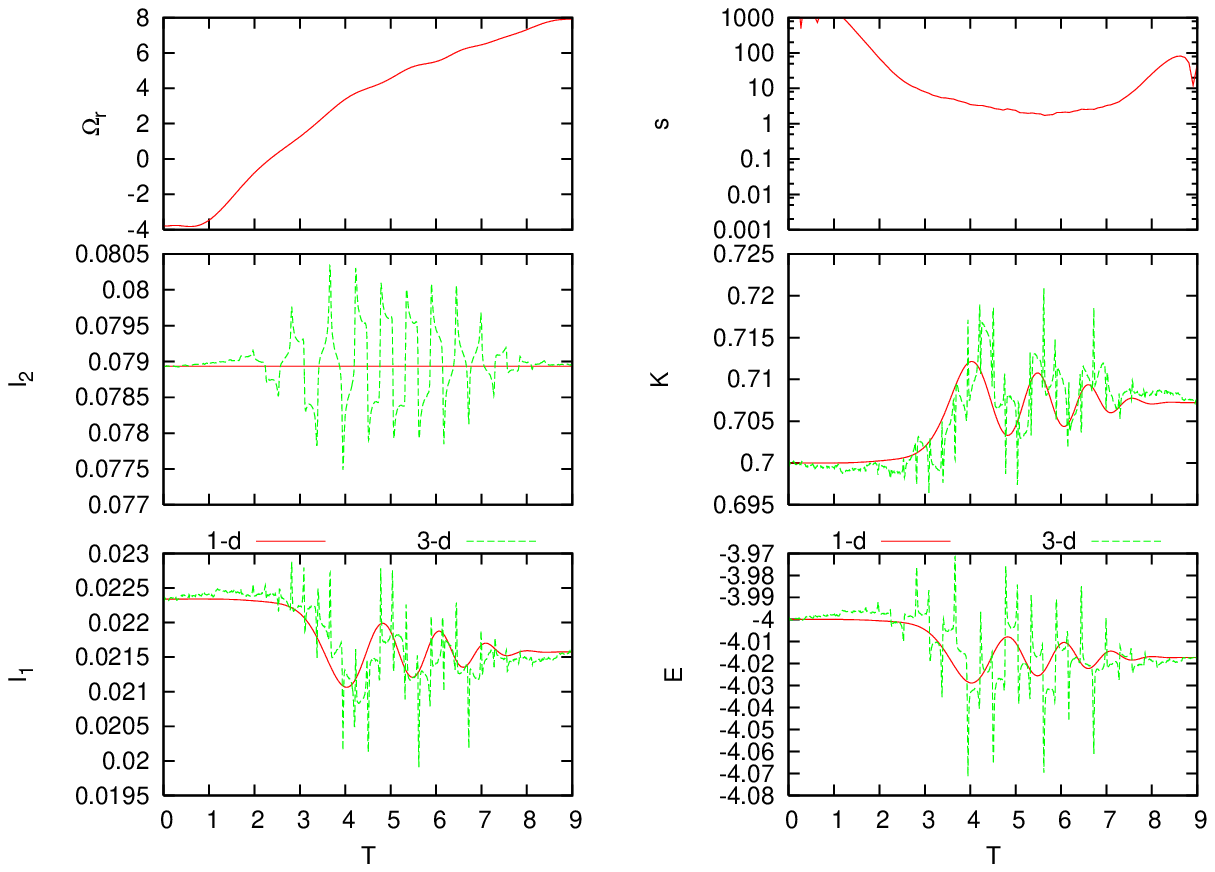}}
\subfigure[$m=10^{-4}$, DRR (1,0)]{
  \includegraphics[width=0.49\textwidth]{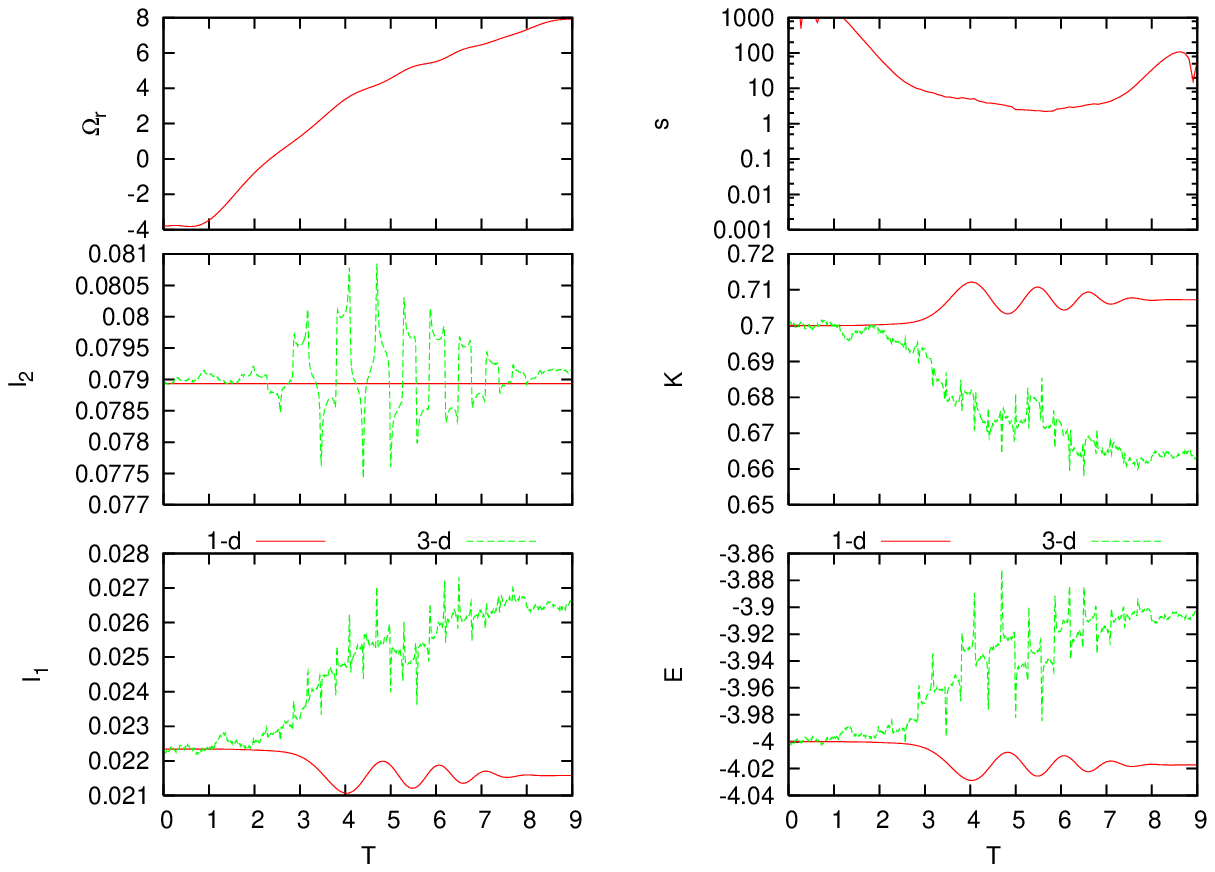}}
\subfigure[$m=10^{-6}$, ILR (-1,2)]{
  \includegraphics[width=0.49\textwidth]{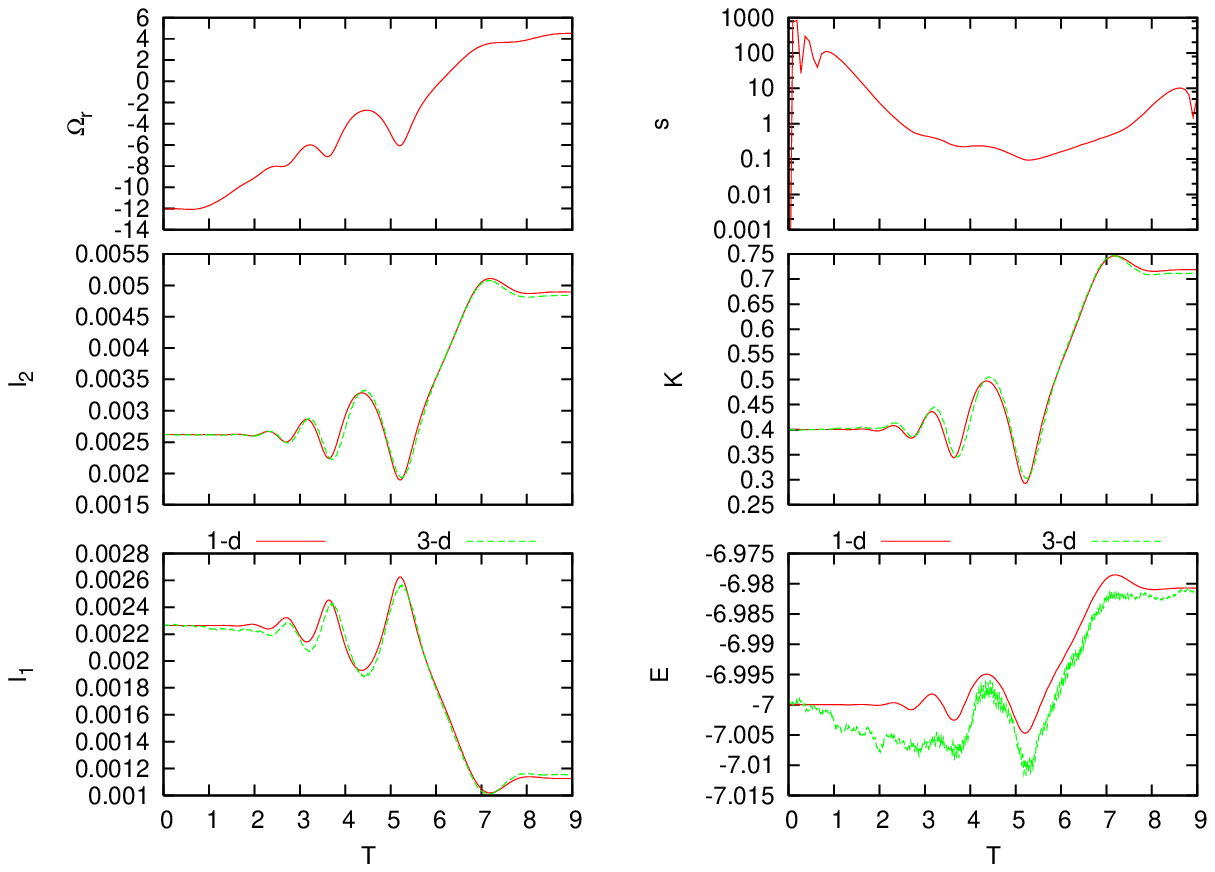}}
\subfigure[$m=10^{-5}$, ILR (-1,2)]{
  \includegraphics[width=0.49\textwidth]{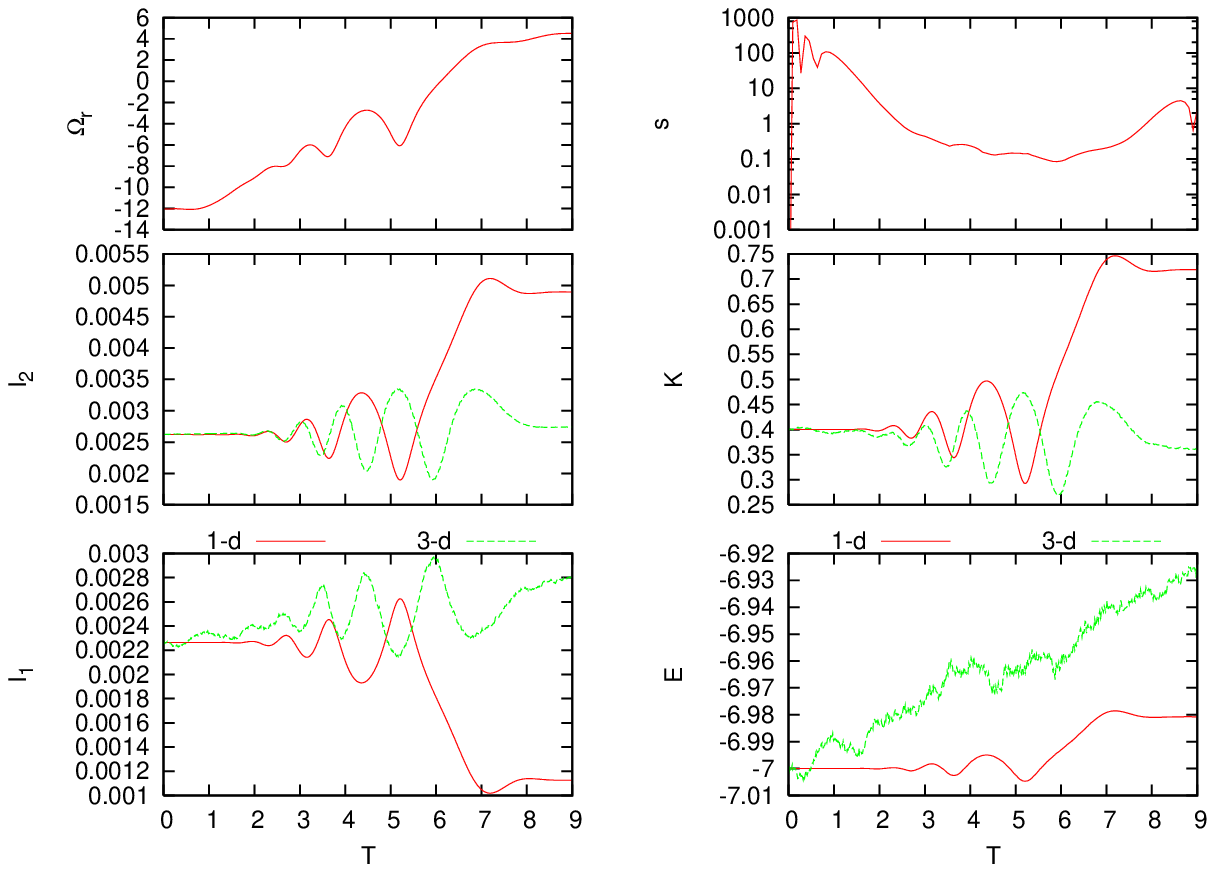}}
\caption{As in Figures \protect{\ref{fig:tbodyorb10}} and
\protect{\ref{fig:tbodyorbILR}} but including
  diffusion caused by large-scale fluctuations for the particle masses
  indicated in units of the total mass.}
\label{fig:fluctorb}
\end{figure}

The calculations in the previous section emphasise the fluctuations in
the gravitational field dominated by two-body encounters on relatively
small scales.  Formally, large scale contributions were also included.
The contribution to the fluctuations on large scales in the
homogeneous approximation leads to a divergence at the upper end of
the Coulombic logarithm $\ln\Lambda$.  Physically, this divergence is
removed by the inhomogeneity of a gravitationally bound galaxy and a
realistic estimate must be computed differently. In this section, we
explicitly address the role of fluctuations owing to noise at large
scales, both in simulations and astronomically owing to dark-matter
substructure.

To appreciate the physical situation, consider the forcing of an
inner-halo orbit by a disturbance (such as orbiting dark-matter
subhaloes) in the outer galaxy. There are two requirements for a
distant disturbance to cause an effect on the inner orbit: 1) the
force must vary over the region
sampled by the orbit, otherwise there can be no work done relative to
the background potential; and 2) the time scale of the disturbance must be
smaller than one of the orbit's natural periods, otherwise the motion
will be adiabatically invariant.  These sorts of considerations are
very similar to those important for the bar--halo interaction
discussed in \S\ref{sec:bardyn} and can be approached similarly.

\cite{Weinberg:01b} derives a formalism for treating the evolution of
orbits including statistical fluctuations caused by perturbers of
various sorts.  Assuming that the perturbations are a first-order
Markovian process, i.e. they do not retain any memory of their
prior state, the evolution equation naturally takes the Fokker-Planck
form \citep{Pawula:67}.  In particular, \cite{Weinberg:01b} works out
analytic expressions for the diffusion coefficients and we will use
them here.  The calculation explicitly includes the spatial and
temporal correlations of any physical perturbation.  The
expression for the coefficients have two parts: 1) a second-order {\em
reinforcement} by the perturbation on the induced distortion on the
orbit, typical of any second-order perturbation theory as described in
\S\ref{sec:hamilton}; and 2) a correlation coefficient that describes
the spatial and temporal correlation of the perturbation.  If we denote the
change in component $j$ of the action at time $t$ after evolving by a
period $\tau$ as $\Delta I_j(t+\tau)$,  then action-space diffusion
coefficients from \cite{Weinberg:01b} are:
\begin{eqnarray}
  D^{(1)}_j(\bI, t) &=& \lim_{\tau\rightarrow0}
  {\left\langle \Delta I_j(t+\tau) \right\rangle \over \tau} \nonumber \\
  &=&
  -{1\over f_0(\bI)}
  l_j{\bf l}\cdot{\partial f_o\over\partial\bI}
  \left|Y_{ll_2}(\pi/2,0)\right|^2 
  \sum_{\mu\nu}\sum_{rs}
  r^l_{l_2m}(\beta)r^{\ast l}_{l_2m}(\beta)
  W^{l_1\,\mu}_{ll_2m}(\bI) 
  W^{\ast l_1\,\nu}_{ll_2m}(\bI) 
  \times \nonumber
  \\
  && \left\{
  (2\pi)^3 \sum_{\bf l} \int d^3I f_o(\bI)
  |Y_{ll_2}(\pi/2,0)|^2 
  r^l_{l_2m}(\beta) r^l_{l_2m}(\beta) 
  \times \right. \nonumber \\ && \left.
  W^{l_1\,r}_{ll_2m}(\bI) W^{\ast l_1\,s}_{ll_2m}(\bI)
  {\cal M}^{lm}_{\mu r}({\bf l}\cdot\bO(\bI))
  {\cal M}^{\ast lm}_{\nu s}({\bf l}\cdot\bO(\bI))
  2\pi\delta\left({\bf l}\cdot\bO(\bI)\right)
  \right\}, \nonumber \\
  \label{eq:mom1expA}
  \\
  D^{(2)}_{jk}(\bI, t) &=& \lim_{\tau\rightarrow0}
  {\left\langle \Delta I_j(t+\tau) \Delta I_k(t+\tau) \right\rangle \over 2\tau} \nonumber \\
  &=&
  {l_jl_k\over 2}
  \left|Y_{ll_2}(\pi/2,0)\right|^2 
  \sum_{\mu\nu}\sum_{rs}
  r^l_{l_2m}(\beta)r^{\ast l}_{l_2m}(\beta)
  W^{l_1\,\mu}_{ll_2m}(\bI) 
  W^{\ast l_1\,\nu}_{ll_2m}(\bI) 
  \times \nonumber
  \\
  && \left\{
  (2\pi)^3 \sum_{\bf l} \int d^3I f_o(\bI)
  |Y_{ll_2}(\pi/2,0)|^2 
  r^l_{l_2m}(\beta) r^l_{l_2m}(\beta) 
  \times \right. \nonumber \\ && \left.
  W^{l_1\,r}_{ll_2m}(\bI) W^{\ast l_1\,s}_{ll_2m}(\bI)
  {\cal M}^{lm}_{\mu r}({\bf l}\cdot\bO(\bI))
  {\cal M}^{\ast lm}_{\nu s}({\bf l}\cdot\bO(\bI))
  2\pi\delta\left({\bf l}\cdot\bO(\bI)\right)
  \right\}. \nonumber \\
\label{eq:mom2expA}
\end{eqnarray}
The spatial component of the response is expanded in a orthogonal
series, the $r^l_{l_2m}(\beta)$ rotate the spherical harmonics, and
the terms $W^{l_1\,r}_{ll_2m}$ describe the coefficients of the
action-angle expansion for the $r^{th}$ basis function.  The operator
${\cal M}^{lm}_{\mu r}(\omega)$ describes the self-gravitating
response owing to an applied frequency $\omega$ in the space spanned by
the basis.  The sums on $\mu,\nu$ and $r,s$ are the sums in the space
of this operator.  In this study, we will ignore self gravity and hence the
${\cal M}^{lm}_{\mu\nu}$ become identity matrices.  The limit
$\tau\rightarrow0$ must be taken in the sense that $\tau$ is small
compared to the evolutionary time scale owing to the fluctuations but
remains large compared to the dynamical time.  The time dependence in
the diffusion coefficients reminds us that the underlying equilibrium
distribution $f_o(\bI)$ changes on an evolutionary time scale but, for
the purposes of the computation, is held fixed on a dynamical time scale.
The lowest-order temporal variation has been explicitly removed by the
limit $\tau\rightarrow0$.  The integrals may be simplified by noting
that $d^3I=dE dJ J d(\cos\beta)/\Omega_1(E,J)$.  We can then do the
integral in $\beta$ using the orthogonality of the rotation matrices
as previously described.  For a given equilibrium distribution
function $f_o(\bI)$, the term in curly brackets in equations
(\ref{eq:mom1expA}) and (\ref{eq:mom2expA}) is the spatial and
temporal correlation of the perturbation and need be computed only once
since it is independent of the local value of the actions.

We now use equations (\ref{eq:mom1expA}) and (\ref{eq:mom2expA})
to compute the fluctuations by random realisation. The procedure
parallels that in \S\ref{sec:diffu}: one generates new values of
$\bI$ from $D^{(1)}_j(\bI, t)$, $D^{(2)}_{jk}(\bI, t)$ and Gaussian
random variates.  Here we do not have the luxury of uncorrelated
parallel and perpendicular motion as in the infinite homogeneous case
but we must diagonalise $D^{(2)}_{jk}(\bI, t)$ (e.g. by a Jacobi
rotation) to generate uncorrelated random variates in $\bI$.

Figure \ref{fig:fluctorb} shows the evolution of an orbit near DRR and
ILR perturbed by large-scale fluctuations
(cf. Figs. \ref{fig:comporb10} and \ref{fig:comporbILR}).  The
particle number requirements are 1000 and 100 times smaller than that
for two-body diffusion, respectively, for the two resonances.  From
the N-body simulation point of view, the two-body relaxation from the
previous section sets a minimum particle number in a particle-particle
code (e.g. direct, tree, mesh, etc.) and the large-scale fluctuations
considered here sets a minimum particle number in an expansion code.
By limiting the spatial scales, the expansion-based solver
dramatically reduces the relaxation and potentially the particle
requirements.

\subsubsection{Implications for dark-matter substructure}

\begin{figure}
\centering \subfigure[${\bar m}f_{ss}=10^{-5}$]{
\includegraphics[width=0.49\textwidth]{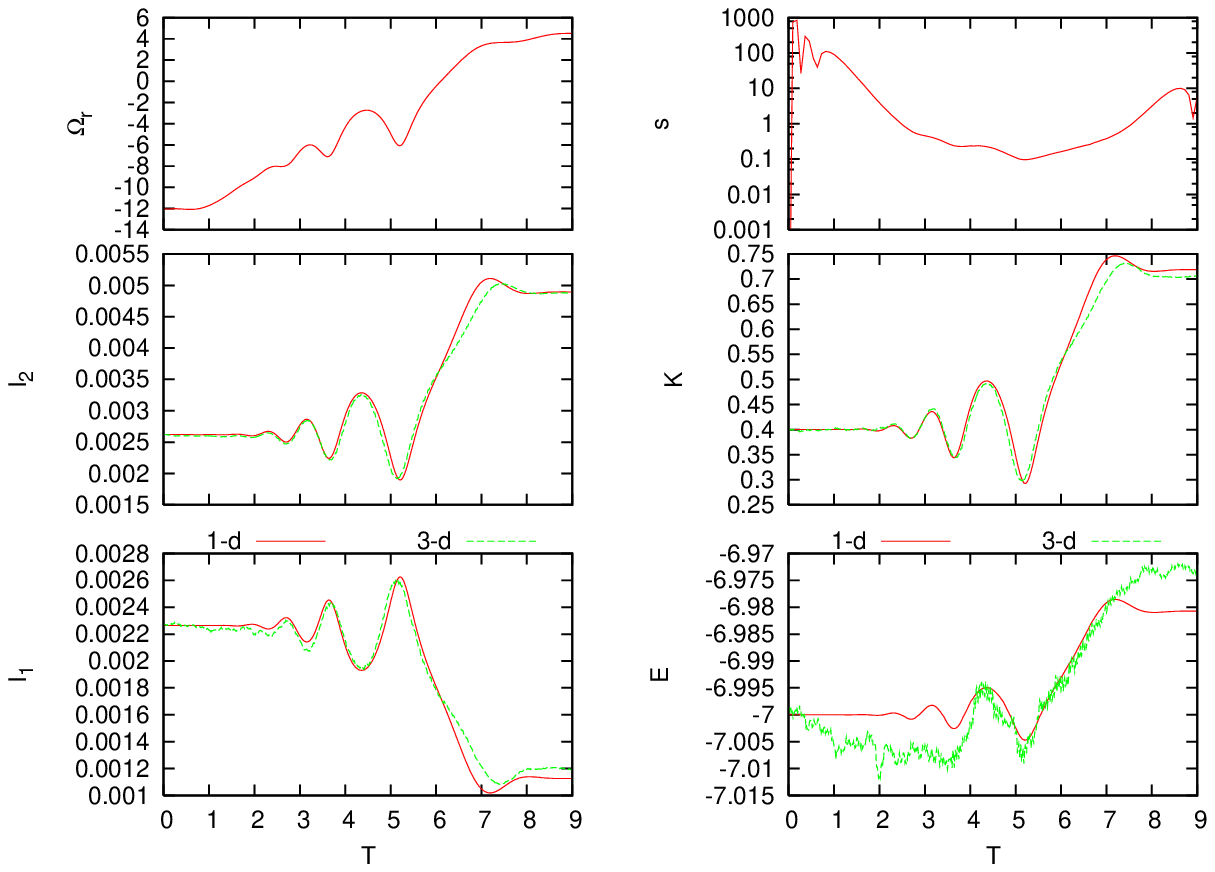}}
\subfigure[${\bar m}f_{ss}=10^{-4}$]{
\includegraphics[width=0.49\textwidth]{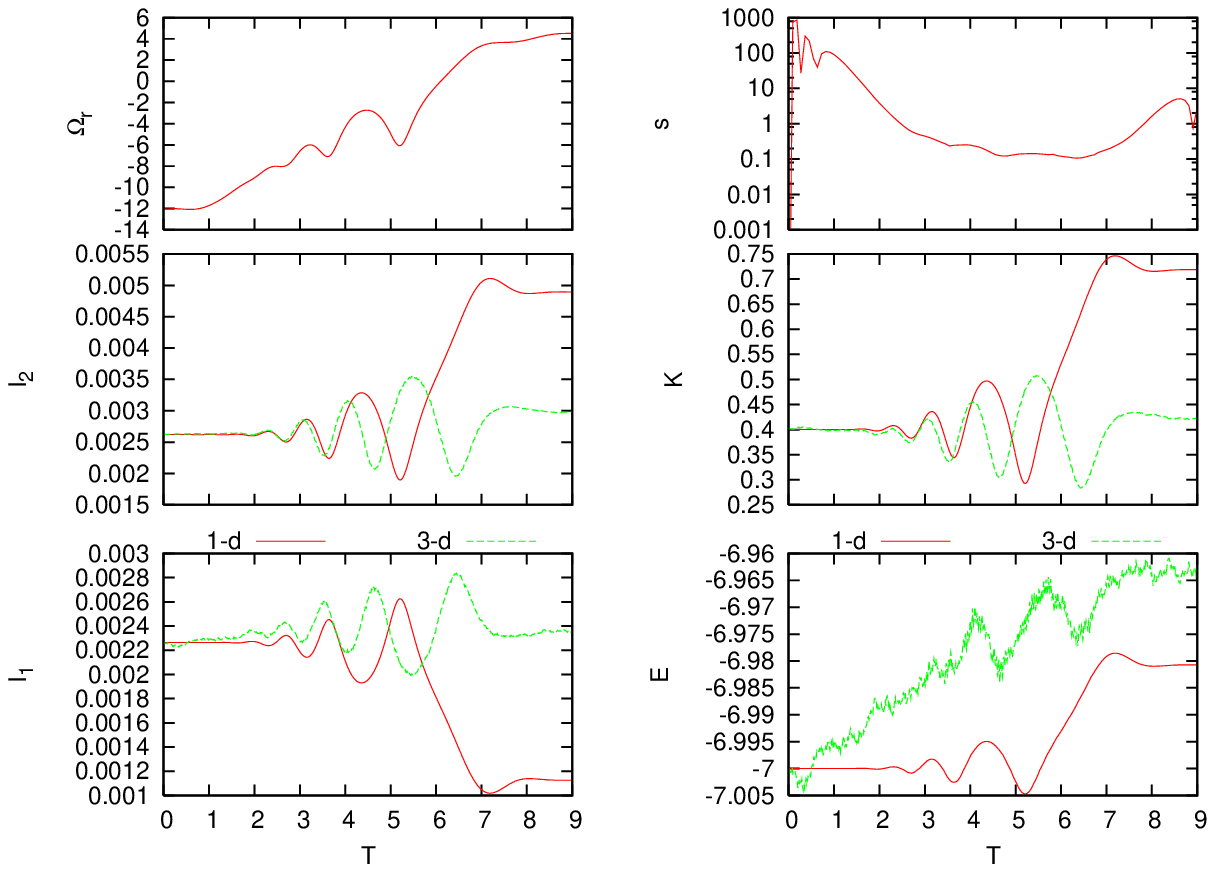}}
\caption{As in Figure \protect{\ref{fig:fluctorb}} for the ILR (-1,2)
including the effects of substructure with orbits restricted to $r>r_s$.}
\label{fig:fluctorbILR}
\end{figure}

We can use these results to estimate the importance of astronomical
noise sources on bar--halo resonances.  In principle, the fluctuations
from any physical process can be computed as described in
\cite{Weinberg:01b}.  The most common and perhaps relevant noise comes
from orbiting substructure.  Ignoring orbital decay, the calculation
is identical to that discussed in the previous section.
$\Lambda$CDM simulations provide both a spatial density distribution
and mass function for dark-matter substructure.
\cite{Oguri.Lee:04} show that these distributions can be 
modelled using a Press-Schechter approach \citep{Press.Schechter:74}
extended to include tidal stripping and orbital decay.  Oguri \& Lee
find that the satellite masses have the cumulative distribution
$N(>m)\propto m^{-1}$ for $m/M_{vir}$ from $10^{-6}$ to $0.3$. The
spatial distribution of substructure is shallower than the overall
dark matter distribution with shallower profiles for larger values of
$m$.  We will assume, for simplicity, that the maximum substructure
mass, $m_{max}/M_{vir}$, is 0.3 and the smallest substructure mass,
$m_{min}/M_{vir}$ is $10^{-6}$. This implies that the
mean mass (in virial mass units) is
\begin{equation}
{\bar m} = {\log(m_{max}/m_{min}) \over 
  \left(1/m_{min} - 1/m_{max}\right)} \approx 1.3\times 10^{-5}.
\label{eq:meff}
\end{equation}
Let the fraction of the dark matter in substructure with $m>m_{min}$
be $f_{ss}$.  Then, the diffusion coefficients from equations
(\ref{eq:mom1expA}) and (\ref{eq:mom2expA}) for particle mass $m$ may
be scaled to $m={\bar m}f_{ss}$.  Assuming that the fraction of the mass
in substructure $f_{ss}\approx 0.1$
\citep{Gao.White.ea:04} and using equation (\ref{eq:meff}) we have
${\bar m}f_{ss}\approx 10^{-6}$.  To implement the effect of tidal
stripping, we assume that the density distribution of substructure
within the dark halo takes the same NFW form as the dark matter but
with no substructure inside of $r_s=1/15$ (units $R_{vir}=1$).  This
requires restricting the phase-space integral inside the curly
brackets in both equations to orbits with $r>r_s$.

Figure \ref{fig:fluctorbILR} shows the now familiar ILR orbit
perturbed by large scale fluctuations from orbiting substructure
restricted to orbits with $r>r_s$.  Now, ${\bar m}f_{ss}$ must be
larger than about $10^{-5}$ to destroy the
resonance, which is an order of magnitude larger than the actual
substructure noise.

\subsection{Calibration of particle number criteria: scaling formulae}
\label{sec:calib}

All of the particle number criteria described in \S\ref{sec:criteria} have
natural scalings in terms of physical quantities: properties of the equilibrium
gravitational potential and the perturbation such as bar strength, bar
shape, bar pattern speed, etc.  We derive simple scaling relations in this
section and calibrate them using the results of \S\ref{sec:simdesc}
and additional simulations.

\subsubsection{Method}

To calibrate the scaling estimates for the particle number requirements
defined in \S\ref{sec:criteria}, we evolve phase-space ensembles as
described in \S\ref{sec:comp1d3d}.  We can use both the
one-dimensional and three-dimensional equations of motion to study
each of the three criteria from \S\ref{sec:criteria}.

\subsubsection{Coverage}

The coverage criterion demands that one samples phase space
sufficiently densely in the vicinity of a resonance to ensure the
correct ensemble average.  The {\em resonance potential} is defined by
the one-dimensional pendulum problem from equation (\ref{eq:hamprt})
and is simply the Fourier action coefficient corresponding to the
commensurability $\bl$.  We define the half-width of the resonance
potential as the maximum extent of the infinite-period trajectory in
the one-dimensional phase space:
\begin{equation}
  \delta I_s = \sqrt{2M^2H_{1\,\bl}}.
  \label{eq:reswid}
\end{equation}
This is called the {\em resonance width} in the celestial mechanics
literature.\footnote{NB: This has nothing to do with the width in
frequency space.}  Hence, we can use perturbation theory to estimate a
characteristic {\em width} or {\em volume} in phase space associated
with each resonance.  Because it depends on both the Fourier
action-angle expansion coefficient and the change in slow frequency
with slow action, this width depends on both the order of the
resonance and the amplitude of the perturbation.

As the galaxy halo slowly evolves, the resonance defined by the
commensurability $\ldo-m\Omega_p=0$ sweeps through phase space as
described in \S\ref{sec:bardyn}.  A secular torque occurs only if
there are enough particles so that the first-order sinusoidal
oscillations cancel to leave the second-order changes. We, therefore,
require sufficient particles, $n_p$, in a fraction $\epsilon_w$ of the
resonance width to obtain a mean cancellation of the first-order
forced response.  We use the resonance width to estimate the number of
particles needed to resolve this phase space volume.  The
commensurability $\bl\cdot\Omega=m\Omega_p$ defines a track in the $E$
and $J$ (or $I_r$ and $I_\phi$) phase-space plane.  The resonance
width in equation (\ref{eq:reswid}) takes values along this locus.
For a spherical isotropic stellar system, most resonance tracks
defining the commensurability have only a small variation with
$\kappa\equiv J/J_{max}(E)$ for fixed $E$.  We optimistically assume
that an ensemble average over angles for orbits over a large range in
$\kappa$ for the same resonance cancels the first-order oscillation.
Clearly more careful approximations that explicitly compute the full
phase-space volume for coherent contributions are possible.  In
addition, we assume that the phase-space distribution is isotropic.
The phase-space fraction within an energy width $\Delta E$ for an
isotropic phase-space distribution function $f(E)$ is:
\begin{equation}
  dF = \int d\bw\,d\bI\, f(E) \delta[E-E(\bI)] = p(E)f(E) dE = n(E) dE
\end{equation}
where
\[
p(E) = 2(2\pi)^3 J_{max}^2(E) \int^1_0 d\kappa\,\kappa/\Omega_1(E,\kappa)
\]
with normalisation
\[
1 = \int dF = \int dE p(E) f(E).
\]
We then use the width in $E$, corresponding to $\delta I_s$ defined in
equation (\ref{eq:reswid}), to estimate the fraction of phase space
$f_{crit}$ that we require to be populated with at least $n_p$
particles as follows:
\begin{eqnarray}
  f_{crit} &\approx& \epsilon_w  p(E)f(E){\partial E\over\partial
    I_s} (\delta I_s) \nonumber \\ 
  &=& \epsilon_w {n(E) (\bl\cdot\Omega)
  \sqrt{2 H_{1\,\bl}} \over \left|\partial^2H_0(I_s)/\partial
  I_s^2\right|^{1/2}_{I_{s,r}}}
  \label{eq:fcrit}
\end{eqnarray}
where ${I_{s,r}}$ is the value of the slow action at the resonance.
This expression for $f_{crit}$ is independent of the choice for $w_f$,
as it must be, but depends of course on the resonance $\bl$.

We can now use $f_{crit}$ to estimate the number of particles needed
to resolve a resonance. Explicitly, our requirement $n_p\le Nf_{crit}$
implies that the minimum critical particle number for resonance $\bl$
is $N_{req,\bl} = n_p/f_{crit}$.  For example, if we demand that
$n_p=10$ particles should span one tenth of the resonance width for
$\epsilon_w=0.1$ to obtain good first-order cancellation, we require
the simulation to contain the following minimum number of particles:
\begin{equation}
  N_{req,\bl} = {n_p\over \epsilon_w} 
  {\left|\partial^2H_0(I_s)/\partial I_s^2\right|^{1/2}
  \over n(E) (\bl\cdot\Omega) \sqrt{2 H_{1\,\bl}}}
   = 100 {\left|\partial^2H_0(I_s)/\partial I_s^2\right|^{1/2}
  \over n(E) (\bl\cdot\Omega) \sqrt{2 H_{1\,\bl}}}.
\label{eq:Ncrit}
\end{equation}
This factor of 100 is only a crude guess but tests in
\S\ref{sec:results} suggest that this is approximately the correct value.

\subsubsection{Noise}
\label{sec:diffusion}

As we have seen from \S\ref{sec:diffu}, fluctuations on very small
spatial scales (two-body relaxation) causes diffusion in the
conserved quantities of an orbit. Recall that in linear perturbation
theory, torque is transferred to orbits whose apses are very slowly
precessing in the bar frame.  This implies that the orbits must remain
quasi-periodic for at least several periods in the bar frame.  The
period of the slowly precessing orbit is best characterised by the
period of the closed resonant orbit.  For our spherical system, this
period is
\begin{equation}
  P_{res} = \max\left(
  {2\pi\over\Omega_1}|l_2|,
  {2\pi\over\Omega_2}|l_1|
  \right).
  \label{eq:pres}
\end{equation}
We therefore require that the diffusion length in the conserved
quantities over some number of $P_{res}$ be smaller than the resonance
width.  Let the diffusion coefficient, the mean-squared rate of spread
of an ensemble of particles in the slow action $I_s$, be denoted as
$D_{I_s I_s}$.  Using equation (\ref{eq:reswid}), we can then express
this criterion as
\begin{equation}
  \tau \equiv {
    \epsilon_r (\delta I_s)^2   \over 
    D_{I_s I_s} P_{res}
  } > 1
  \label{eq:diffureq}
\end{equation}
where $1/\epsilon_r$ is the desired number of resonance periods over
which the orbit must remain stable.  We estimate that plausible values
of $\epsilon_r$ range from 0.1 to 0.3  and use $\epsilon_r=0.3$
in the estimates given below. We test this choice in \S\ref{sec:results}.
For a fixed phase-space distribution with unit mass, the
diffusion coefficient scales as the mass per particle or
inversely with the number of particles $N$.  Now, if we derive $D_{I_s
I_s}$ for a unit-mass particle then the number of particles
required to satisfy the criterion in equation (\ref{eq:diffureq}) is
\begin{equation}
  N_{req} = 1/\tau = 
  {
    D_{I_s I_s}|_{{\bar m}=1} P_{res}   \over 
    \epsilon_r (\delta I_s)^2
  }.
  \label{eq:Nreq}
\end{equation}

Direct-summation N-body simulations (including tree-algorithm based
simulations) follow individual particle motions with a resolution
approximately equal to the interparticle softening scale.  We express
the relaxation present in these codes at small scales using two-body
relaxation as in \S\ref{sec:criteria}.  The standard expressions are
given in terms of energy $E$ and angular momentum $J$.  The diffusion
coefficient $D_{x_j x_k}$ transforms to new variables
$x^\prime_\mu=x^\prime_\mu(\{x_j\})$ as
\begin{equation}
  D_{x^\prime_\mu, x^\prime_\nu} = \sum_{jk} 
  {\partial x^\prime_\mu \over\partial x_j}
  {\partial x^\prime_\nu \over\partial x_k} D_{x_j x_k}
\label{eq:dtrans}
\end{equation}
\citep[e.g.][]{Risken:89}.  Depending on the value of $l_j$, we are
free to choose either $I_1\equiv l_1 I_s$ or $I_2\equiv l_2 I_s$.  We
may then derive $D_{I_s I_s}$ from the homogeneous diffusion
coefficients $D_{E\,E}, D_{E\,J}, D_{J\,J}$ and equation
(\ref{eq:dtrans}).

In the same way, the large-scale expansion-based diffusion
coefficients in equations (\ref{eq:diffureq}) and (\ref{eq:Nreq})
together with equation (\ref{eq:dtrans}) yields $D_{I_s I_s}$ and
particle number criteria for noise on large scales.  We use the same
expansion from our Poisson solver for the expansion in these diffusion
coefficients although any biorthogonal expansion would suffice.  This
approach also facilitates a direct comparison with the simulations.

\subsubsection{Time steps}
\label{sec:timestep}

Comparison between the one- and three-dimensional solutions for
individual orbits allows us to investigate the time step criteria
necessary to obtain a correct transition through resonance.  This is
more stringent than the requirement that the energy or actions be
conserved for an equilibrium orbit.  For an orbit with energy E, we
find that a time step with approximately 1/100 of the period of
the circular orbit with the same energy is necessary.

\subsection{Results}
\label{sec:results}

\begin{table}
  \caption{Scaling of particle number requirements.}
  \begin{tabular}{lll}
    Requirement & Scaling & Description \\ \hline
    $N_{req,\bl}$ & $M_p^{-1/2}$ & coverage \\
    $N_{req,Ch}$ & $M_p^{-1}$ & small-scale noise \\
    $N_{req,SCF}$ & $M_p^{-1}$ & large-scale noise \\ \hline
  \end{tabular}
  \label{tab:scaling}
\end{table}
    
\begin{figure}
  \centering
  \subfigure[ILR (-1,2)]{\includegraphics[width=0.32\linewidth]{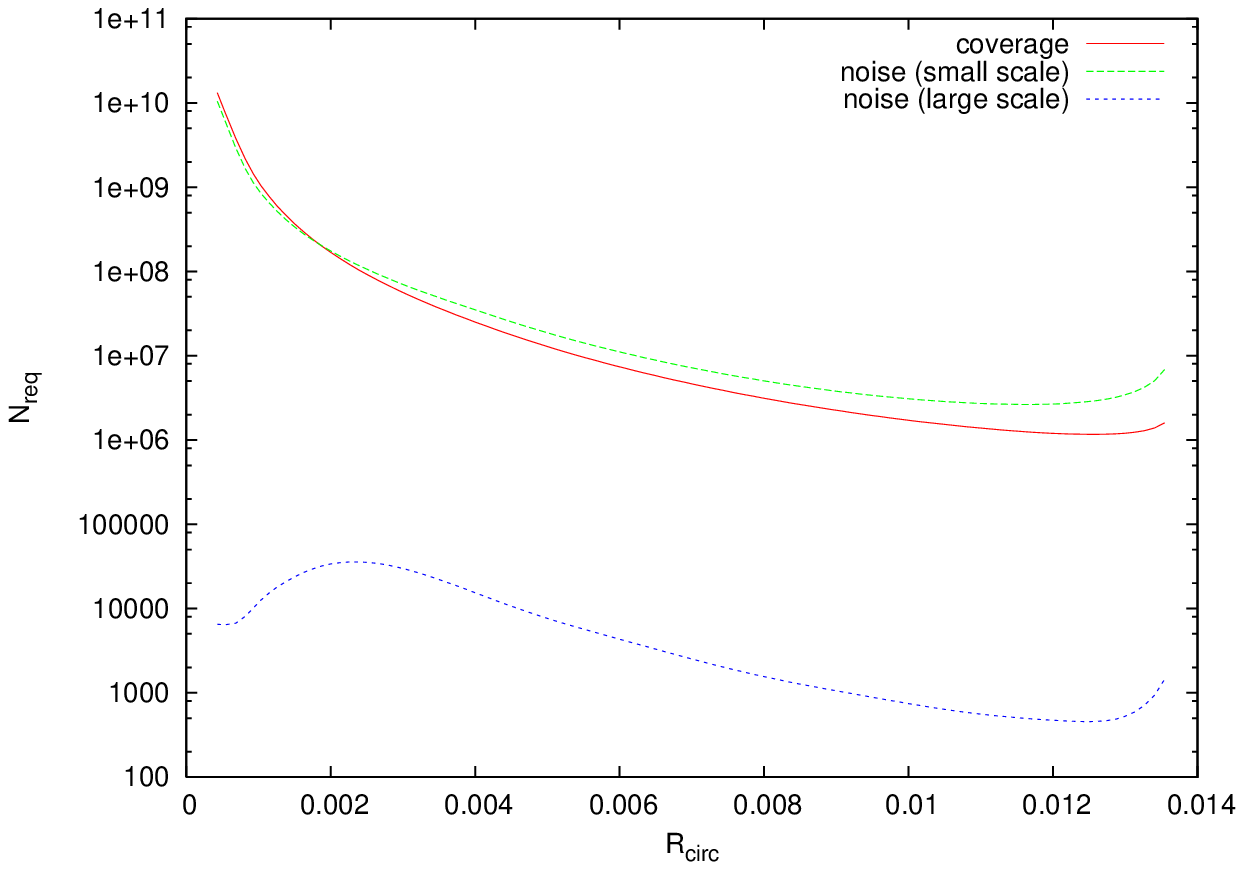}}
  \subfigure[corot (0,2)]{\includegraphics[width=0.32\linewidth]{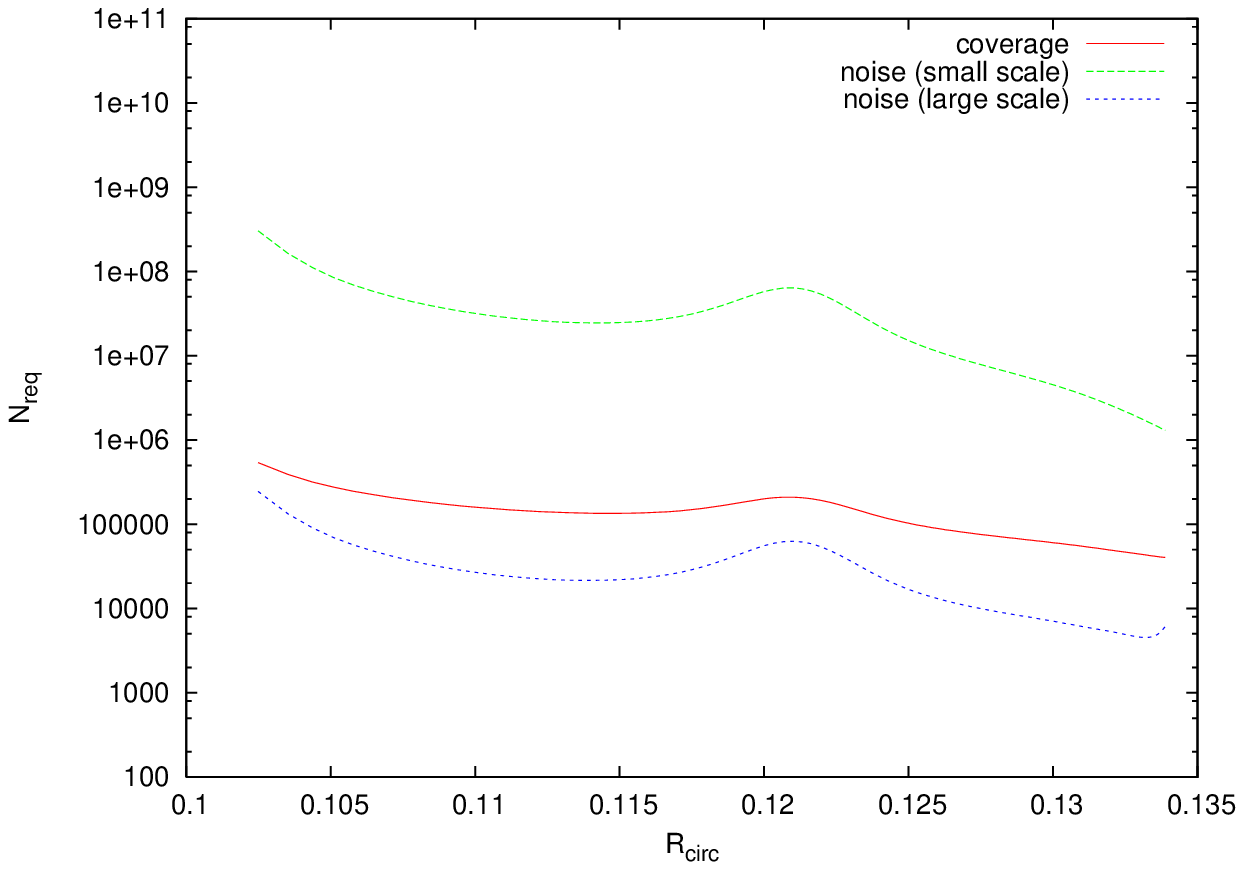}}
  \subfigure[DRR (1,0)]{\includegraphics[width=0.32\linewidth]{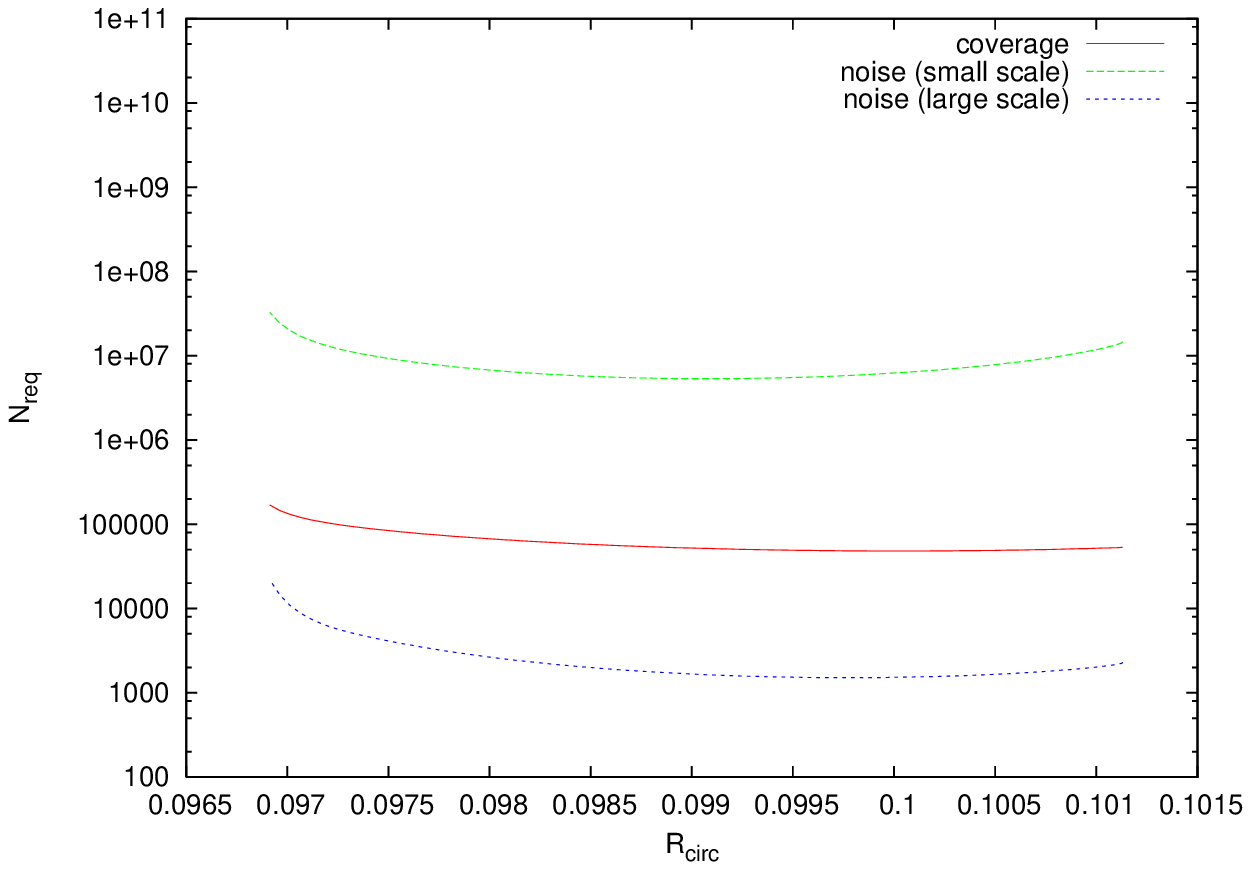}}
  \caption{The three particle number criteria---coverage, and
    small and large scale noise---for the indicated resonances for a large weak
    bar versus radius.}
  \label{fig:DNbigbar}
\end{figure}

We can use the formula derived in \S\ref{sec:calib} to determine the
particle number requirements, i.e. the number of equal mass particles
within the virial radius.  We start with a halo-scale-length bar, i.e.
bar radius $=r_s=0.067$, whose mass is 1\% of the dark matter mass
within that radius and a pattern speed at one-half of its corotation
value.  We plot the results for each of the three criteria for three
of the most important resonances, ILR (-1,2), corotation (0,2), and
DRR (1,0) in Figure \ref{fig:DNbigbar}. Table \ref{tab:scaling}
describes how each of these criteria scale with the mass of the
perturbing bar, $M_b$.  We choose such a weak bar to facilitate
comparison with perturbation theory but the reader may scale these
estimates to any bar mass using these relations. Since both the energy
and angular momentum change along the track of each resonance, as seen
in previous figures (e.g. Fig. \ref{fig:distlz}), we plot the
requirement for each criteria as a function of the radius of a
circular orbit with the same energy, $r_{circ}$.  For this
interaction, the most important resonances are the ILR and DRR.  We
show the corotation resonance (0,2) for comparison.  Note the radial
location of the resonances: the corotation resonance and DRR are
around the end of the bar, 0.067, while the ILR begins at
approximately one quarter of the bar radius, 0.014, and extends inward
to very small radii, at least as small as 0.02 times the bar radius.
It is important for the Poisson solver to resolve down to this scale,
approximately $10^{-3}$ times the virial radius in this case. Hence,
in this figure we assume a gravitational softening length of this size
when calculating the small-scale noise requirement.

For all resonances, the small-scale noise criteria is the most severe
but it is comparable to the coverage criterion for ILR.  Remember
that, as they are usually used, the small scale noise criterion only
applies to particle-particle codes, e.g. direct, tree, and grid based,
and not to expansion SCF codes.  As described in
\S\ref{sec:diffusion}, small-scale noise will strongly affect which
orbits enter resonance and the regime of the resonance.  However,
coverage will dominate the small-scale noise for a stronger, i.e. more
massive, bar at ILR according to the scalings in Table
\ref{tab:scaling}.  The estimates for coverage are consistent with
the convergence of $\Delta L_z$ in phase space with particle number for coverage
alone computed using one-dimensional averaging as shown in Figure
\ref{fig:distlz}.  Similarly, comparison of Figure \ref{fig:DNbigbar}
and Figures \ref{fig:tbodyorb10} and \ref{fig:tbodyorbILR} show
consistency: the scaling relations require more than $10^7$ particles
for DRR and considerably more for ILR.

\begin{figure}
  \centering
  \subfigure[ILR (-1,2)]{\includegraphics[width=0.32\linewidth]{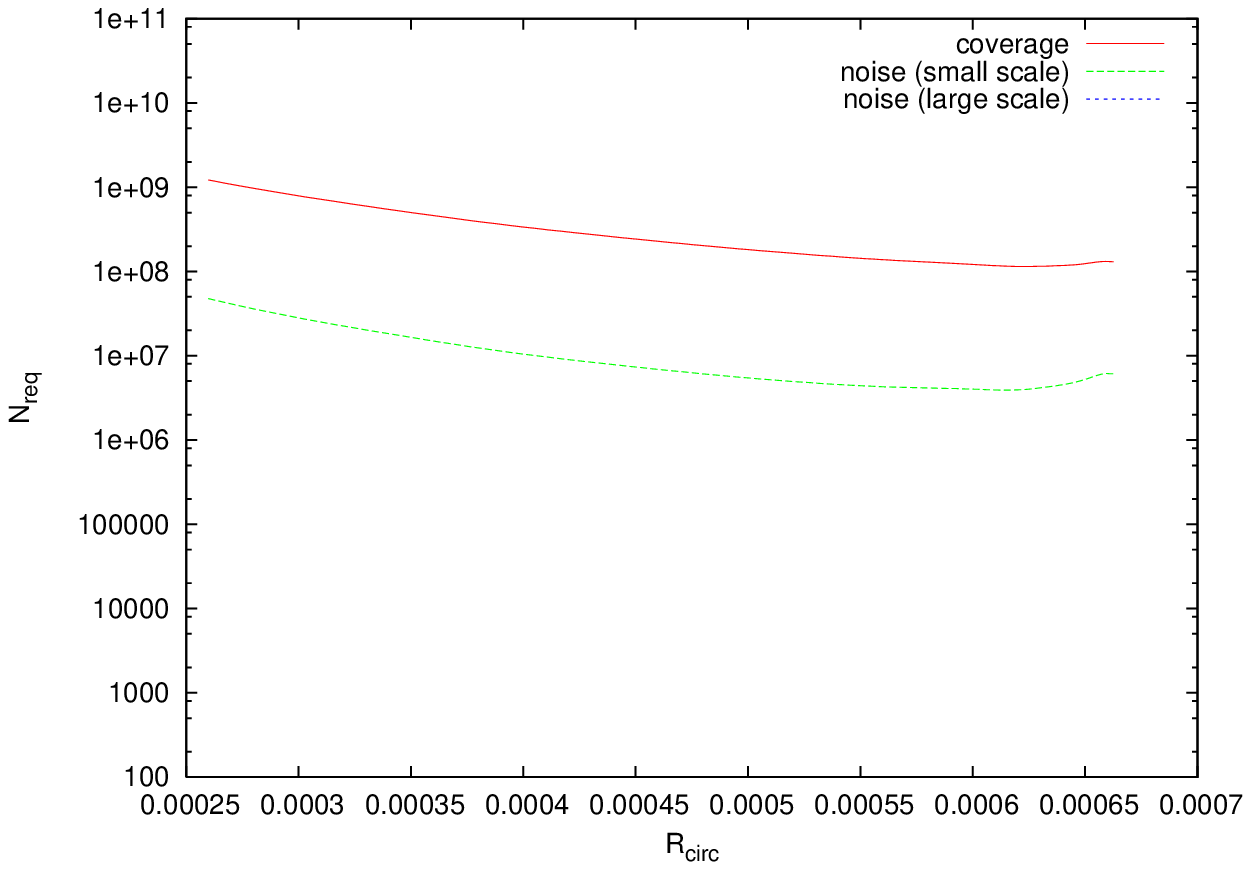}}
  \subfigure[corot (0,2)]{\includegraphics[width=0.32\linewidth]{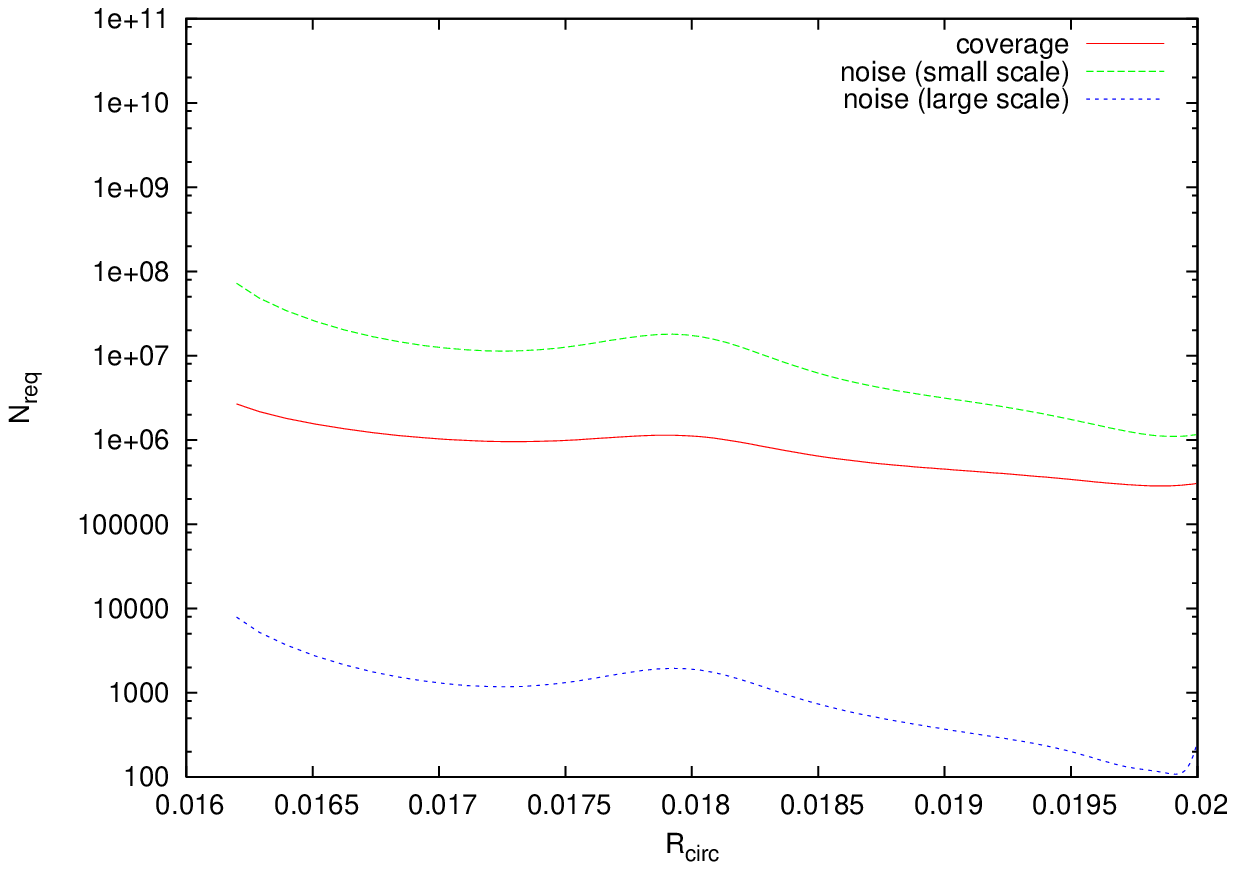}}
  \subfigure[DRR (1,))]{\includegraphics[width=0.32\linewidth]{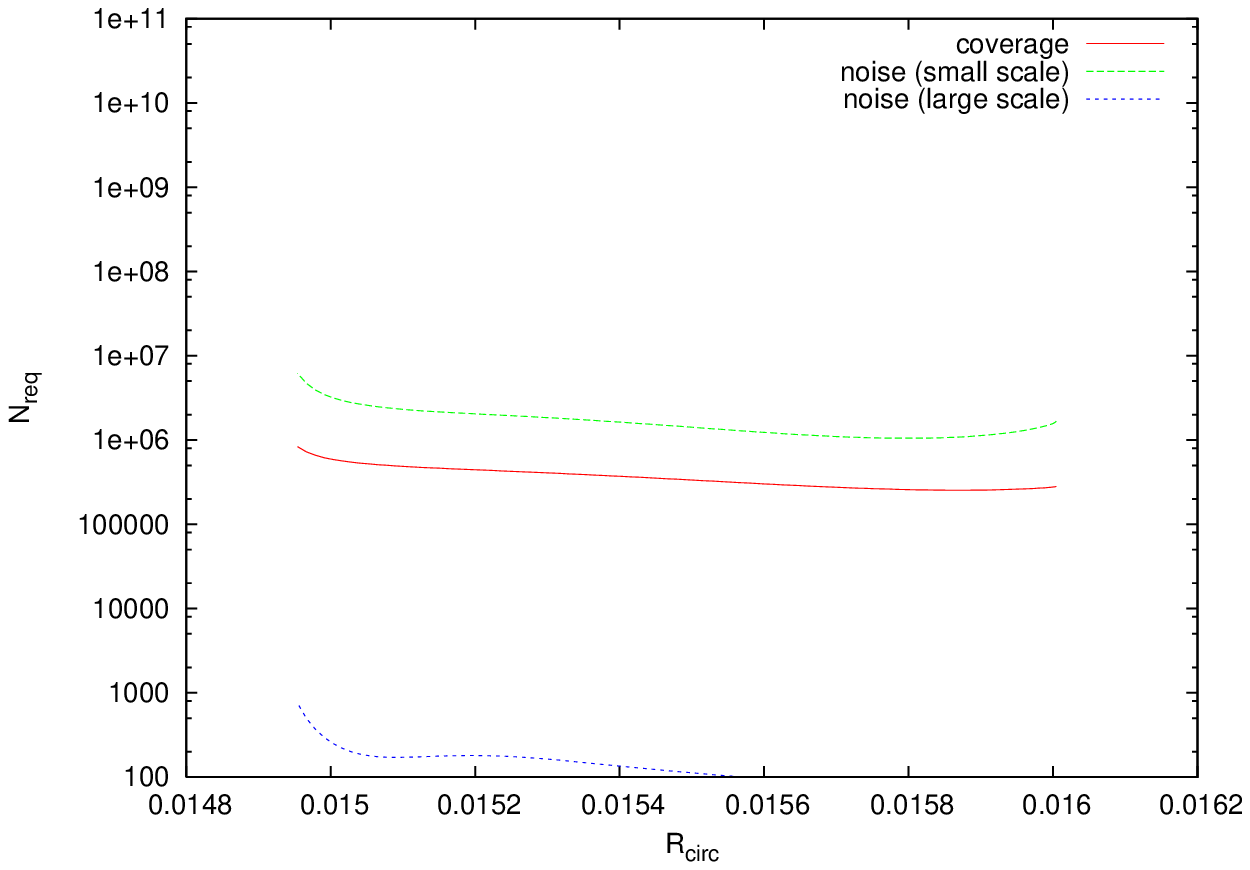}}
  \caption{The three particle criteria---coverage, and
    small and large scale noise---for the indicated resonances for a
    disk-scale-length, moderate strength bar.}
  \label{fig:DNsmlbar}
\end{figure}

Figure \ref{fig:DNsmlbar} shows the same predictions for an
approximately scale-length sized moderate strength bar, i.e.  bar
radius$=r_s/6.7=0.01$.  The bar mass is 10\% of the enclosed dark
matter mass, making the maximum relative amplitude of the asymmetric
to axisymmetric force 30\%.  This bar requires a force resolution of
$2\times 10^{-4}$ times the virial radius to fully resolve the ILR,
which we use in our calculation of the small-scale noise.  The
coverage criterion demands a very large particle number to correctly
couple to the bar--halo ILR: approaching $10^9$ particles. The trends
for the {\em outer} resonances are similar to those for the large bar
but the coverage criterion is relatively more important.
Approximately $5\times10^7$ particles are needed for these resonances.
Because the ILR for a scale-length bar occupies very low energies in
the cusp, we can check the criteria in Figure \ref{fig:DNsmlbar}a for
by realizing only a small subset of the full phase space.  By varying
the energy range, we may use between $10^6$ and $10^7$ actual
particles to achieve a suite of realizations logarithmically spaced
between $10^6$ and $10^9$ equivalent particle numbers.  The ILR
resonance only reaches its full amplitude for $N\gta3\times10^8$
particles consistent with the prediction in Figure \ref{fig:DNsmlbar}.

As described in \S\ref{sec:diffusion}, the diffusion will cause a
drift in the orbits that pass through resonance and cause all resonant
interactions to be in the fast limit, even if they would be in the
slow limit without noise.  Therefore, one may relax the small-scale
noise criteria if one knows {\em a priori} that all resonances are
fast and possibly still obtain the correct overall torque.  This
coincidence explains the convergence of the total torque
in simulations that do not meet the small-scale noise criterion.  For
the bar--halo interaction emphasised here, slow limit interactions
primarily affect the ILR resonance.  The net torque scales as
$M_p^{1/2}$ in the slow limit and $M_P^2$ in the fast limit.
Therefore, the noise-induced fast limit may diminish this interaction
by an order of magnitude for bars of typical strength.

\section{Discussion}
\label{sec:disc}

\subsection{The Chandrasekhar dynamical friction formula}
\label{sec:chandra}

The bar--halo interaction and dynamical friction are governed by the
same dynamics \citep{Weinberg:85,Weinberg:86}.  The traditional
Chandrasekhar dynamical friction is the sum of single scattering
events and, therefore, works without resonance.  However, in the case
of a bound orbit responding to a periodic perturbation such as a bar
(or orbiting satellite), the individual scatterings become repetitive
interactions near a commensurability as described in
\S\ref{sec:bardyn}.  For quasi-periodic systems, the angular momentum
transfer occurs at resonances (and can be understood as the
superposition of many second-order secular Hamiltonian perturbation
theory problems, one for each resonance).  Hence, dynamical friction
works in quasi-periodic systems through resonances.  The second-order
quasi-periodic analogue to dynamical friction obtains for fast-limit
encounters.  Slow-limit encounters have a different scaling.  In
general, one cannot replace the details of resonant dynamics with that
of scattering (Chandrasekhar formula).  For example, the lack of a few
low-order resonances owing to differences in pattern speed or the halo
profile can cause large changes in the torque.  We will see examples
of this in Paper II.

In \S\ref{sec:diffu}, we argued that scattering owing to small scale
fluctuations in the potential can be large enough, even for particle
numbers typical of state-of-the-art N-body simulations, that the
location and dynamics of the resonance change.  If potential
fluctuations rapidly scatter orbits through the resonances, the orbits
have no memory that they have been affected by the same perturbation
during previous periods.  However, one can observe dynamical friction
in a poor simulation even if the quasiperiodic nature of the orbits
are lost but it is in the wrong, scattering regime.  In this case it is
possible that the total torque can be larger for small $N$.  If this
torque changes the actions of the orbits that drive the structural
evolution, then the evolution will be even faster.

\subsection{Relationship to the LBK formula}
\label{sec:LBK}

Although the first-order perturbation may dominate the instantaneous
changes to phase space, the second-order changes describe the net
changes that will persist after many dynamical times.  Most often,
these results are computed assuming that the perturbation began
infinitely long in the past, the {\em time-asymptotic limit}.  LBK
derived a formula describing the exchange of angular momentum between
a spiral pattern and the rest of the disk in this
limit,  which can be straightforwardly applied to a bar interacting
with a halo \citep{Weinberg:85}.  However, not only is the number of
characteristic dynamical times in a galactic age modest but the growth of
a bar is most likely only a small fraction of the galactic age.  In the
previous section, we argued that the finite time effects yield
different angular momentum distributions in the halo than for a
time-asymptotic system.  \citet{Weinberg:04} shows that the LBK
formula does not give an accurate description of the bar slow down for
the same reasons and presents the following generalisation for an
arbitrary time-dependent perturbation:
\begin{eqnarray}
  \left\langle\left\langle {dL_z\over dt}\right\rangle\right\rangle 
  &=&
  -(2\pi)^3\sum_\bl\int d\bI\, l_3\bl\cdot{\partial f_o\over\partial\bI}
  e^{-i\ldo t^\prime} \left\{\int^t_0 dt^\prime e^{i\ldo t^\prime}
    H_{1\,\bl}(\bI, t^\prime) \right\} H_{1\,\bl}^\ast(\bI, t).
  \label{eq:LBK1}
\end{eqnarray}
This expression reduces to the LBK formula for a perturbation with a
constant pattern speed in the $t\rightarrow\infty$ limit \citep[for
additional details, see][]{Weinberg:04}.

\begin{figure}
\centering 
\includegraphics[width=0.49\linewidth]{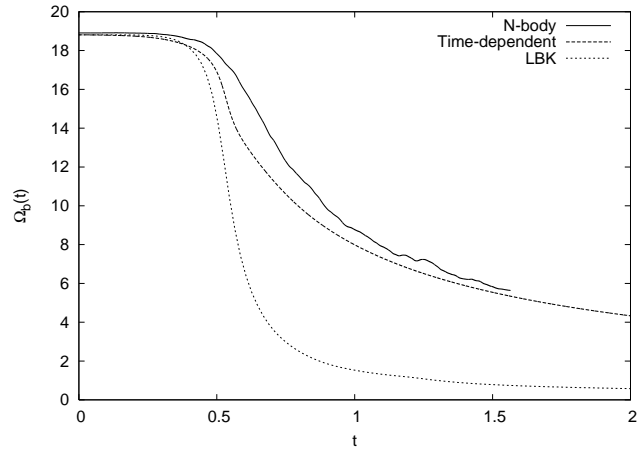}
\caption{Comparison of pattern speed in an N-body simulation with
predictions from second-order perturbation theory in the
time-dependent and LBK limits.}
\label{fig:pattern}
\end{figure}

The simulations described here assume a fixed bar profile with a
changing pattern speed.  Because angular momentum is conserved, the
net change in halo angular momentum is equal and opposite to the bar
angular momentum.  Therefore ${\dot L}_b = I_b{\dot\Omega}_b = -
\langle\langle dL_z/dt\rangle\rangle$.  Figure \ref{fig:pattern}
compares the evolution of the pattern speed $\Omega_b(t)$ in an N-body
simulation, the time-dependent torque from equation (\ref{eq:LBK1}),
and the time-asymptotic LBK limit.  The details of the simulation will
be described in Paper II.  For now, note that the time-dependent
perturbation theory gives fair agreement while the LBK limit predicts
a much more rapid slow down than observed in the simulation.  This is
caused by a variety of resonances that do not contribute in the
time-asymptotic limit but contribute appreciable torque (of both
signs) over a finite-time at a sufficient magnitude so that the
overall evolution is affected.  Clearly, time-dependent theory is
necessary to describe this interaction and we use it for all
subsequent predictions.

\subsection{Comparison with a N-body simulation}
\label{sec:nbody}

\begin{figure}
  \centering
  \subfigure[$M_b=0.1$]{\includegraphics[width=0.32\linewidth]{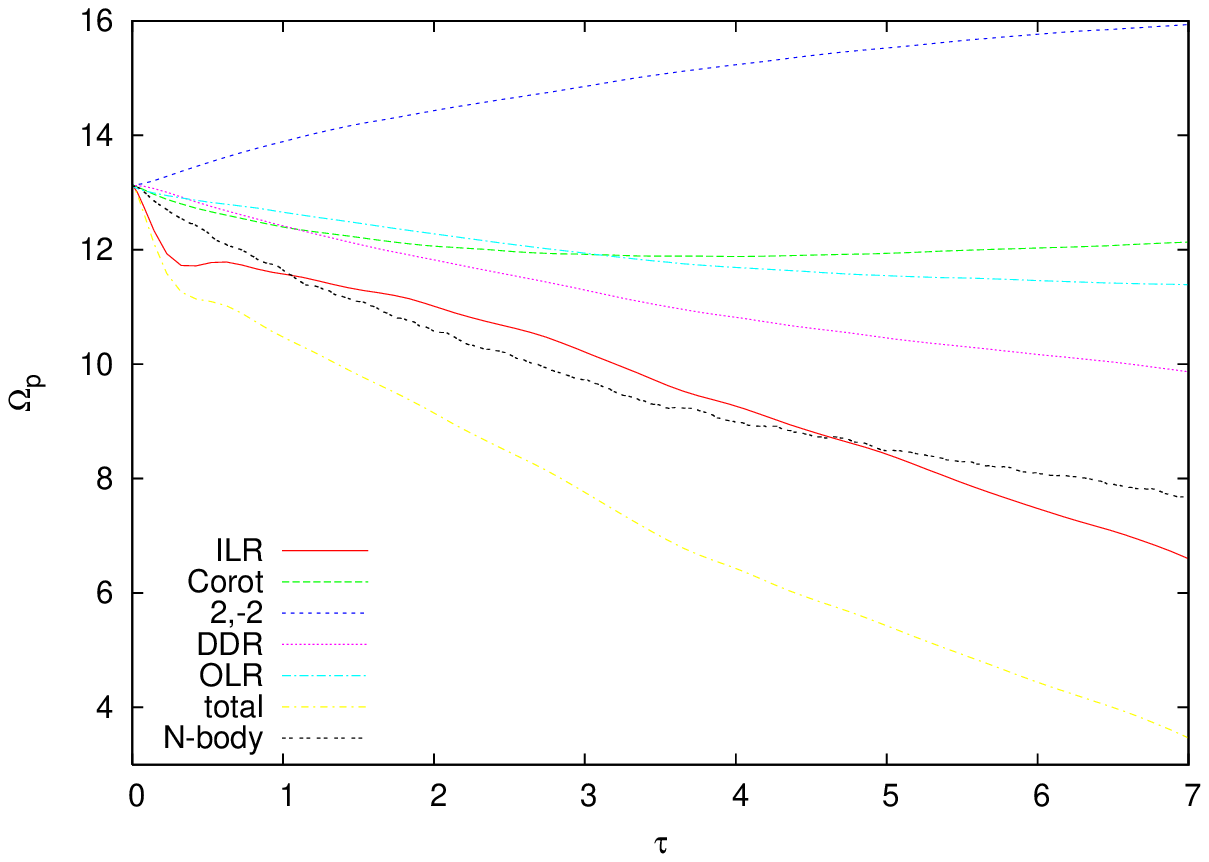}}
  \subfigure[$M_b=0.1$, 1/2 ILR]{\includegraphics[width=0.32\linewidth]{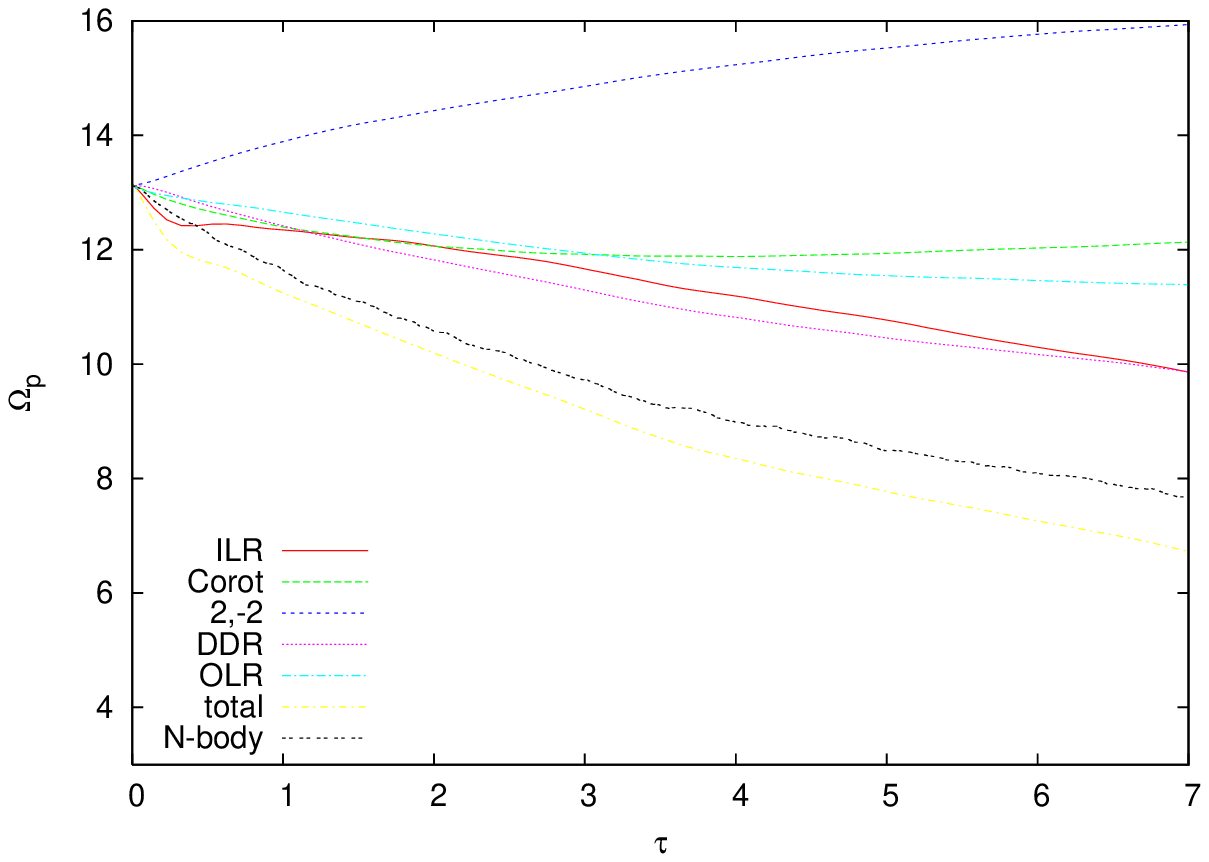}}
  \subfigure[$M_b=0.01$]{\includegraphics[width=0.32\linewidth]{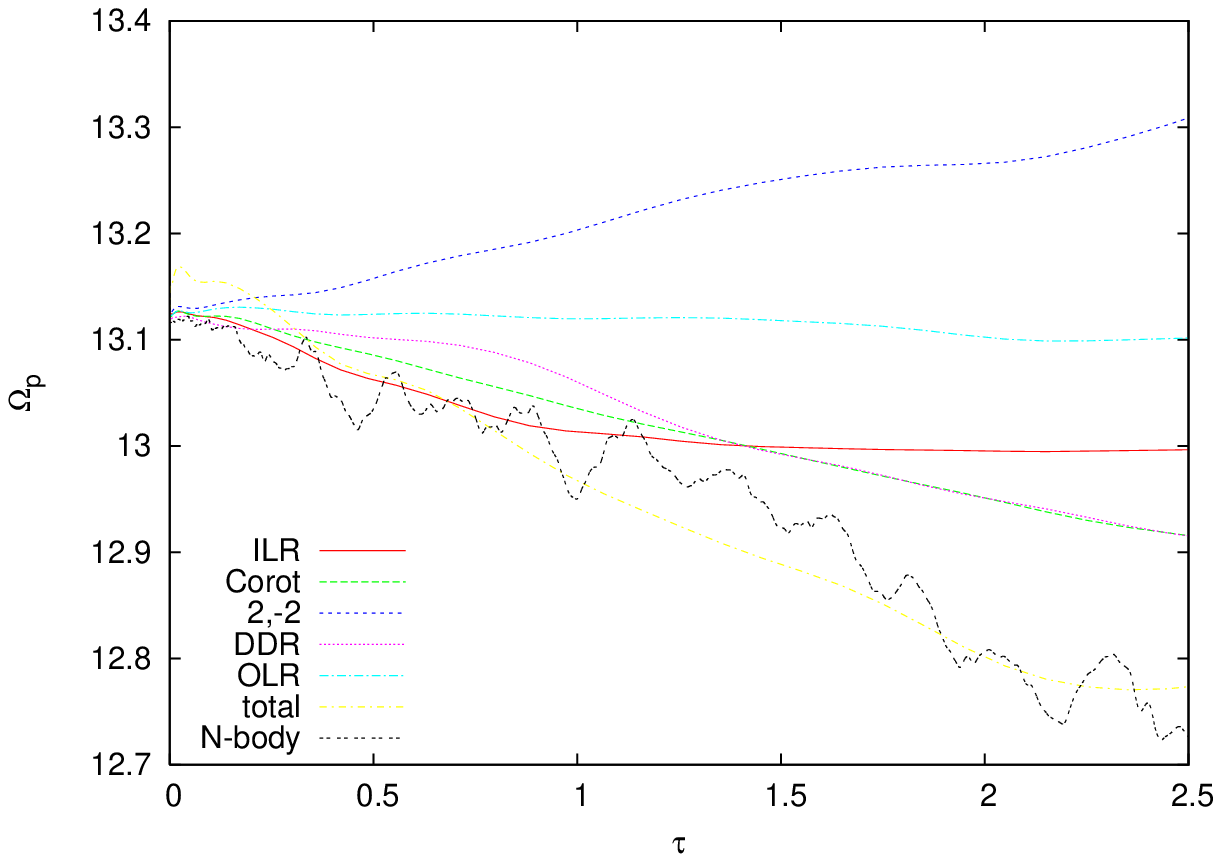}}
  \caption{The change in pattern speed $\Omega_p$ with scaled time
    $\tau$ as predicted by the one-dimensional evolution and N-body
    simulations.  The contributions from five resonances
    ILR, $(2, -2)$, DRR, corotation and OLR are shown along with the
    total.}
  \label{fig:ocontrib}
\end{figure}

We compare a N-body self-consistent simulation of a bar that forms in
a live disk\citep[see][for details]{Holley-Bockelmann.Weinberg.ea:05}
to the one-dimensional perturbation theory calculations from
\S\ref{sec:numavg}. For these simulations, both the expansion centre
and the quadrupole centre are at the origin.  We eliminate the $l=1$
term from the expansion and apply only the quadrupole part of the bar
perturbation.  This eliminates issues of centring and combined
halo-disk equilibria.  We will treat these in paper II.  The evolution
of the pattern speed is computed by enforcing total conservation of
angular momentum.  The bar has a mass that is 10\% of the dark matter
enclosed within the bar radius and corotates at 3/2 of the bar radius.
The bar is slowly turned on and off with the function $A(t)$ defined
in \S\ref{sec:barpert}. In Figure \ref{fig:ocontrib}, we compare the
evolution of the pattern speed in the N-body simulation to the
prediction from the one-dimensional averaged perturbation theory. We
also plot the contributions from the five most important resonances
separately, calculated using the one-dimensional perturbation theory.
Naively, the rate of secular evolution should be proportional to the
perturbation amplitude.  However, because of the slow and fast limits
this simple scaling is broken.  Nonetheless, we can remove some of the
dependence on the turn on and turn off by plotting the pattern speed
evolution in Figure \ref{fig:ocontrib} using the scaled time
\begin{equation}
  \tau\equiv\int_0^t dt^\prime A(t^\prime).
\end{equation}

\begin{figure}
  \centering
  \includegraphics[width=0.49\linewidth]{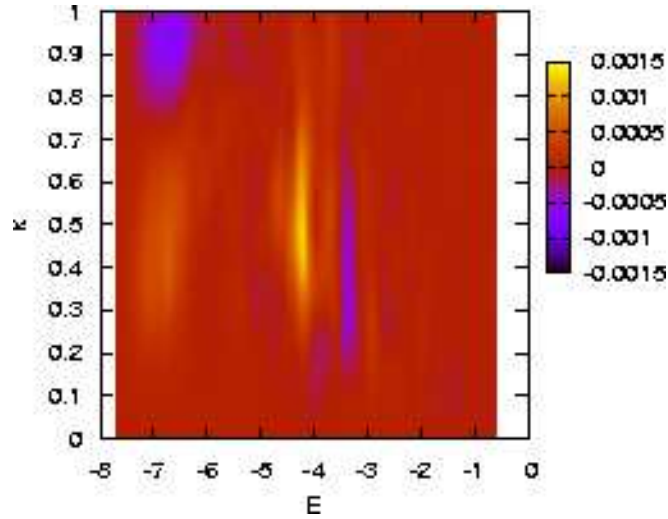}
  \caption{$\Delta L_z$ plot for the N-body simulation in 
  Figure \protect{\ref{fig:ocontrib}}.  Compare this with
    Fig. \protect{\ref{fig:distlz}}.}
  \label{fig:nbodyprof}
\end{figure}

As one can also see from Figure \ref{fig:ocontrib}a, the rate of
pattern speed evolution predicted by the one-dimensional approach exceeds that
in the N-body simulation owing to a stronger ILR contribution.  As
described in \S\ref{sec:comp1d3d}, the ILR is strongly nonlinear and
the amplitude predicted by the one-dimensional solution is roughly a
factor of two larger than for the three-dimensional solution.  Figure
\ref{fig:ocontrib}b shows the same run but with the prediction for ILR
decreased, in an ad-hoc way, by a factor of two, which gives a good
correspondence between the theory and the simulation.  Even with this
decrease, we see that the torque is still dominated by the ILR and
DRR.  To check this interpretation, we decreased the bar amplitude by
a factor of 10 to $M_b=0.01$.  In this case, the ILR is not strongly
nonlinear and we have a good correspondence between the prediction and
the simulation without any ad hoc adjustment.  Figure
\ref{fig:nbodyprof} shows the distribution of angular momentum change
in phase-space as a result of the bar interaction. It looks very
similar to the perturbation theory prediction in Figure
\ref{fig:distlz} with the expected differences in the ILR.

\subsection{A numerical testing ground}
\label{sec:angmom}

For testing ones own N-body code's ability to properly evolve the
bar-halo problem, we recommend reproducing some of the angular
momentum exchange experiments performed here.  For the bar--halo
interaction, most researchers only check the evolution of the pattern
speed with time.  A more sensitive test, and one that is crucial to
understanding the component--component interaction during the overall
secular evolution, is to verify that each resonance is numerically
converged.  Comparing the distribution of the change in angular
momentum in phase space is an effective tool to accomplish this goal
(e.g. Figs.  \ref{fig:distlz} and \ref{fig:comp1d3d}).  These figures
may be constructed as follows.  For a spherical halo, the phase-space
distribution function will be a function of energy and total angular
momentum, $E$ and $J$.  To make this plane rectangular, use the
normalised angular momentum variable $\kappa\equiv J/J_{max}(E)$ so
that $\kappa\in[0,1]$. For a spherical equilibrium, $J_{max}$ is the
angular momentum of the circular orbit with energy $E$.  This can be
determined by finding the radius of the circular orbit $r_c$ by solving
$E=v_c^2(r)/2 + V(r)$ implicitly for $r$, where $v_c(r)\equiv
\sqrt{r\,dV/dr}$ and then $J_max(E)= r_c v_c(r_c)$.  One then computes
the change in angular momentum in some time interval on this plane.
If resonant angular momentum transfer dominates the evolution, the
change in angular momentum will follow the curves in $E$ and $\kappa$
defined by the commensurabilities.  If scattering or more general
non-resonant interaction dominates, the change in angular momentum
will populate a broad area in this plane.  This is only a rough guide;
the ILR, for example, is broad $E$ and $\kappa$ owing to its special
degeneracy (see Fig. \ref{fig:comp1d3d}).  A $\Delta L_z$ plot can be
made from an N-body simulation using the following procedure:
\begin{enumerate}
\item Save the state of the N-body phase space separated by
  a time interval that is small but of order of or larger than the
  orbital time scale.  For the three-dimensional calculation (N-body
  simulation with fixed background potential), the evolution of the
  pattern speed may be compared with Figure \ref{fig:barpert} in
  \S\ref{sec:barpert}.

\item For each orbit, compute the change in the $z$-component of the
  angular momentum, $\Delta L_z$.  For figures in this paper, we
  compute $\Delta L_z$ for the entire simulation from before the bar
  is turned on until after it is turned off using $A(t)$
  (eq. \ref{eq:turnonoff}).

\item Use a density estimator to produce the distribution of $\Delta
  L_z$ over the phase space in the plane defined by $E$ and $\kappa$.
  Simple histograms will suffice but some tuning of bin sizes will be
  necessary to resolve the features.
\end{enumerate}

Using this graphical tool, one may straightforwardly reproduce some of
the examples presented here.  Here are some recommended tests:
\begin{enumerate}
\item Compute the $\Delta L_z$ plots for the large bar used in
  \S\S\ref{sec:criteria}--\ref{sec:calib}.  The details of the bar
  perturbation are given in \S\ref{sec:barpert}.  The $\Delta L_z$
  diagrams should be compared to Figures \ref{fig:distlz} and
  \ref{fig:comp1d3d}.  One should attempt to identify each of the
  resonances.
\item The bar used for these examples is much larger than bars in
  Nature.  This experiment should be repeated for a scale-length sized
  bar of moderate strength as described in \S\ref{sec:results}.
  Ideally, check that their position and amplitude is converged by
  drastically increasing the particle number.
\end{enumerate}

Since it is not possible to make particle number criteria that are
appropriate for all problems, it is necessary to determine the
criteria for each class of problem individually using the procedure
that we outlined above.  One should check these other scenarios
against perturbation theory whenever possible. This is not always
possible so an idealised experiment, i.e. one which imposes an
approximate form of the perturbation and its time dependence, should
suffice to verify the particle number criteria.  Sometimes it is
necessary to use a weaker form of the perturbation and in this case
the derived particle number criteria can be scaled to the real
problem.  For these other perturbations, one can repeat the one- and
three-dimensional computations outlined in \S\ref{sec:bardyn} and
compare them with N-body simulations.  The perturbation theory allows
one to separate the effects of the individual resonances and compute
their locations and importance (e.g. Fig. \ref{fig:ocontrib}).
Numerical perturbation theory, i.e. the integration of the reduced
one-dimensional equations of motion, bypasses most of the
complications associated with solving the CBE while retaining all of
the dynamics of the resonance interaction typical in canonical
perturbation theory.  Furthermore, this approach automatically
includes any desired time dependence.  All of the necessary details
including a symplectic integration scheme are provided in
\S\ref{sec:numavg}.  One may also compare individual orbits as in
Figures \ref{fig:comporb10} and \ref{fig:comporbILR} both to check of
one's perturbation theory calculation and to verify the choice of time
step (see \S\ref{sec:timestep}).  Finally, a three-dimensional
ensemble integration permits sampling a specific energy range in phase
space with more particles than is possible with a full N-body
simulation, as a test of convergence.

\section{Summary}
\label{sec:summ}

Secular evolution in galaxies is caused by intercomponent interactions
possibly triggered by environmental disturbances.  These interactions
manifest themselves as large-scale asymmetric features such as spiral
arms, bars, ovals, and lopsidedness.  The dynamical mechanisms
mediating the secular evolution in the collisionless components are
resonantly driven.

The linear and nonlinear dynamics of a single particle in resonance
and the features of orbit families in model potentials have been well
understood using perturbation theory or numerical solutions of the
equations of motion.  However, an N-body simulation adds two
additional complications.  First, a simulation provides the ensemble
result but the effect of a resonance on a particular orbit depends
sensitively on initial conditions.  The resulting angular momentum
exchange in phase space at resonance has contributions of both signs
depending on the orbits' initial conditions.  The ongoing secular
evolution imposes a directionality that breaks the symmetry in the
initial conditions. However, the simulation must have a sufficient
number of particles in the vicinity of the resonance to obtain the
correct net torque.  Second, representation of the dark matter and
stellar components by an unnaturally small number of particles leads
to fluctuations in the gravitational potential.  For modern
simulations, the magnitude of these fluctuations yield a very long
relaxation time but we have shown that the noise is sufficient to
cause orbits to random walk through resonances.  Of course, if some
orbits walk out of the resonance, others walk in.  However for some
resonances, ILR in particular, orbits would like to {\em linger} near
the resonance for many rotation periods.  The noise prevents this and
in doing so changes both the amplitude of the net torque and the
location of the orbits in phase space receiving the torque.  In this
paper, we explored the details of the resonance mechanism and its
manifestation in an N-body simulation from a rigorous kinetic theory
point of view.  We presented particle number criteria for each type of
discreteness error and checked them with numerical experiments for the
bar--halo interaction.

Both the coverage and noise criteria require a very large number of
particles for important resonances.  In the case of coverage, the net
torque from such a resonance may not be discernible until the critical
particle number is reached. In the case of noise, assuming that the
coverage criterion is met, diffusion may force the resonant
interaction into a different regime until the noise is reduced to the
correct threshold by using a sufficient number of particles.  In both
cases, checking for convergence by (e.g.) doubling the number of
particles may not be a good indicator.  For example, Figure
\ref{fig:tbodyorbTOT} shows that the ILR for a very large, albeit
weak, bar is reduced by a factor of 3 and the angular momentum is
transferred to orbits with higher energy (and therefore with large
radii) for noise equivalent to $10^6$ particles.  Scaling to a
moderate strength disk-scale-length-sized bar would demand between
$10^8$ and $10^9$ particles to attain the correct dynamical regime.

Although larger $N$ at fixed spatial resolution will reduce the
fluctuations, many N-body practitioners choose to use their particle
number to optimise the resolution since the cost of a simulation is
driven by $N$.  Unfortunately, this results in maximising the noise.
For simulations of secular evolution, it is worth decreasing the
spatial resolution and, thereby, use some of one's particle resources to
reduce noise.  Alternatively, an expansion code allows us to directly
control the spatial resolution and efficiently eliminate the
high-spatial frequencies that dominate the noise.  We have shown that
this Poisson solver reduces the required number of particles to
satisfy the noise criterion by several of orders of magnitude, in
practise leaving only the coverage criteria.  For this reason, we
recommend this Poisson solver for treating problems of secular
evolution and use it in Paper II to study the self-consistent
evolution of the bar-perturbed halo. Of course, this method also has
its price: one sacrifices adaptiveness and introduces some bias.
However, in balance we have found it well suited to studying secular
evolution.

The work presented here was the product of a sustained cycle of
perturbation theory predictions followed by numerical experiments, and
we needed many more cycles than anticipated!  Our experience suggests
that it may not be possible to provide a universal set of particle
criteria for an arbitrary dynamical interaction for several reasons.
The strength of the perturbation may change the morphology of the
stable periodic orbits, the rate of secular evolution may change
the resonance topology, and regimes such as fast and slow limit
interactions in resonances may deceive N-body convergence tests.
Rather, each secular mechanism must be studied in
detail to develop a meaningful set of particle number criteria for that
particular problem.  Simulations alone are unlikely to be sufficient
for a full understanding of the dynamics or even ascertaining
convergence.  The important astronomical processes of disk heating and
secular bulge formation are worthwhile topics for this sort of future
investigation.

\section*{Acknowledgments}

Many thanks to Kelly Holley-Bockelmann for suggestions, discussions
and a careful reading of this manuscript and to Julianne Dalcanton for
pleading for a numerical testing ground.  We would also like to
acknowledge many electronic discussions with Jerry Sellwood.  Parts of
this work was completed at the Institute for Advanced Study in
Princeton and MDW thanks his host John Bahcall for his hospitality.  This
work was supported in part by NSF AST-9802568, AST-9988146,
AST-0205969, and NASA ATP, NAG5-12038 \& NAGS-13308.

\bibliographystyle{mn2e}

\appendix

\section{Details of the bar perturbation}
\label{sec:barpert}

Rather than apply a force from a full bar, we may capture the essence
of the nonaxisymmetric interaction by considering the quadrupole
component only.  Laplace's equation demands that the quadrupole part
of the gravitational potential increases as $r^2$ for radii well
within the bar and decreases as $r^{-3}$ far from the bar and,
therefore, comes to a peak somewhere in between.  We find that the
following functional form
\begin{equation} 
  U_{22} = b_1 {r^2\over 1+(r/b_5)^5}
  \label{eq:quad}
\end{equation}
fits the quadrupole radial component for bars quite well.  To be
explicit, our halo is an NFW model with concentration $c=15$.  We
scale all radii and masses by the virial radius and virial mass,
respectively.  We fit equation (\ref{eq:quad}) to a homogeneous
ellipsoid with $b/a=1/5$ and $c/b=1/10$.  The major axis for our main
test case is chosen to be $a=r_s=1/15$.  Our standard tests assume a
bar mass of 1\% of the dark-matter mass enclosed with the bar
semi-major axis.  This leads to the following values: $b_1=-70.4$,
$b_5=0.0262$.  

To compare with perturbation theory, we attempt to minimise transients
by slowly turning on and off the perturbation using the
following function:
\begin{equation}
  A(t) = {1\over4}
  \left[1+\erf\left({t-t_0\over\Delta t}\right)\right]
  \left[1+\erf\left({t_f-t\over\Delta t}\right)\right]
  \label{eq:turnonoff}
\end{equation}
where $t_0$ and $t_f$ are the turn and turn off times, respectively,
and $\Delta$ is the width of the turn and turn off.  The final
perturbing potential is therefore:
\begin{equation}
  H_1({\bf r}, t) = A(t)\Re\left\{{U_{22}(r)Y_{lm}(\theta,
  \phi-\phi_p(t))}\right\}.
  \label{eq:bpert}
\end{equation}

\begin{figure}
  \subfigure[$M_b=0.1, t_0=1, t_f=9, \Delta=0.2$]{
    \includegraphics[width=0.49\textwidth]{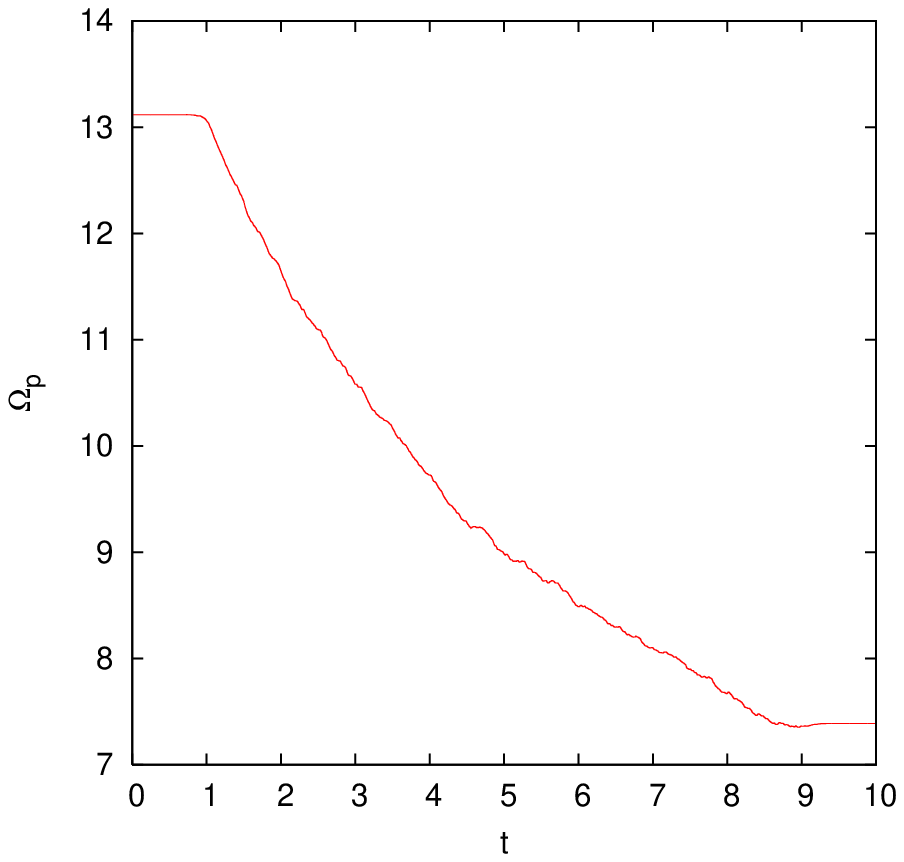}
  }
  \subfigure[$M_b=0.01, t_0=3, t_f=7, \Delta=1$]{
    \includegraphics[width=0.49\textwidth]{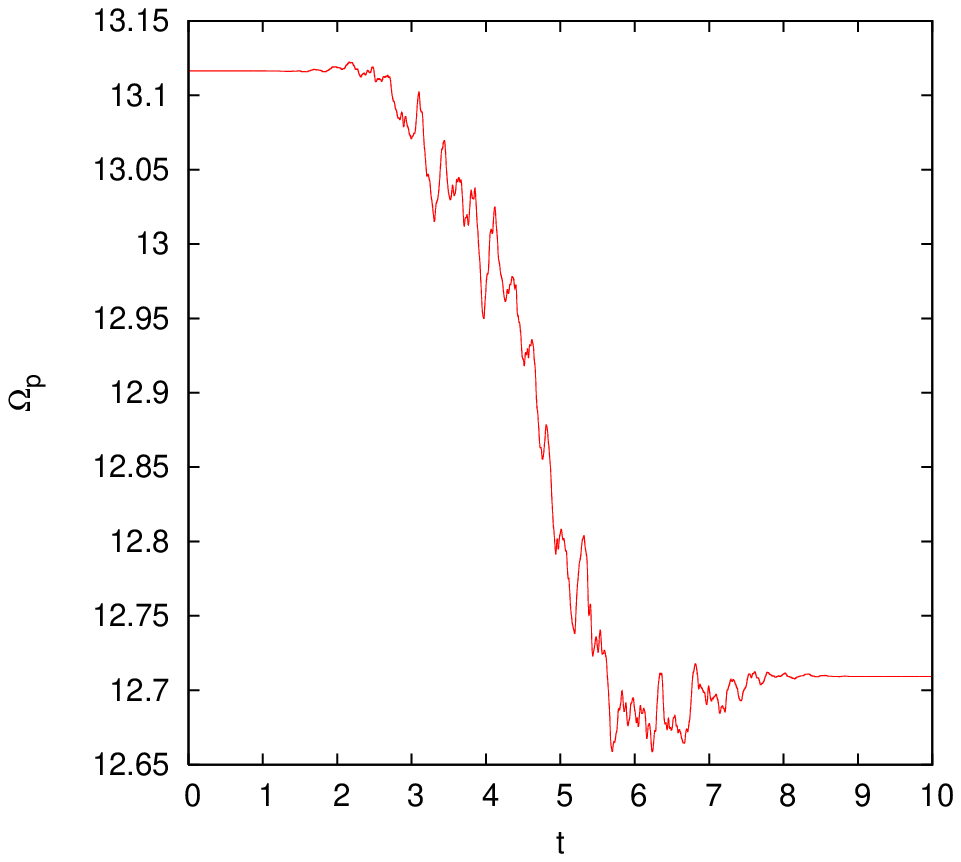}
  }
  \caption{}
  \label{fig:barpert}
\end{figure}

The moment of inertia for the homogeneous ellipsoid about the minor
axis is $I_z=M_b(a_1^2+a^2)/5$ and the angular momentum of the bar is
then $L_{bar} = I_z\Omega_p$.  Let the $z$ component of angular
momentum of the phase space be $L_{zp}$. Then in the three-dimensional
direct integration of a phase-space ensemble, the pattern speed is
derived by enforcing the conservation of momentum $L_{z\,tot}(t) =
I_z\Omega_p(t) + L_{zp}(t) = L_{z\,tot}(0) = L_{bar}(0) + L_{zp}(0)$
to give: $\Omega_p(t) = (L_{bar}(0) + L_{zp}(0) - L_{zp}(t))/I_z$.
Figure \ref{fig:barpert} shows the evolution of the pattern speed for
the two bar examples from \S\ref{sec:criteria}.

\label{lastpage}
  
\end{document}